\documentclass[]{aa} 

\usepackage{natbib}
\bibpunct{(}{)}{;}{a}{}{,} % to follow the A&A style
\usepackage{epsfig}

\newcommand{\beq}{\begin{equation}} 
\newcommand{\eeq}{\end{equation}} 
\newcommand{\beqa}{\begin{eqnarray}} 
\newcommand{\eeqa}{\end{eqnarray}} 
\newcommand{\bea}{\begin{array}} 
\newcommand{\ea}{\end{array}} 

\newcommand{\dd}{{\rm d}}

\newcommand{\lag}{\langle} 
\newcommand{\rag}{\rangle} 

\newcommand{\ii}{{\rm i}}

\newcommand{\Om}{\Omega_{\rm m}}

\newcommand{\bb}{\overline{b}}
\newcommand{\nb}{\overline{n}}
\newcommand{\Nb}{\overline{N}}

\newcommand{\xih}{\xi^{\rm h}}
\newcommand{\zetah}{\zeta^{\rm h}}
\newcommand{\etah}{\eta^{\rm h}}

\newcommand{\vk}{{\bf k}}
\newcommand{\vOm}{{\bf \Omega}}

\newcommand{\vr}{{\bf r}}

\newcommand{\vtheta}{{\vec{\theta}}}

\newcommand{\vx}{{\bf x}}

\newcommand{\tW}{\tilde{W}}

\newcommand{\hn}{\hat{n}}
\newcommand{\hN}{\hat{N}}
\newcommand{\hw}{\hat{w}}
\newcommand{\hwLS}{\hat{w}^{\rm LS}}
\newcommand{\hxi}{\hat{\xi}}
\newcommand{\hxiLS}{\hat{\xi}^{\rm LS}}
\newcommand{\hxic}{\hat{\xi}^{\rm c}}

\newcommand{\cA}{{\cal A}}
\newcommand{\cD}{{\cal D}}

\newcommand{\cI}{{\cal I}}
\newcommand{\cN}{{\cal N}}

\newcommand{\cR}{{\cal R}}
\newcommand{\cV}{{\cal V}}

\newcommand{\xiconzj}{\overline{\xi^{(j)}_{\rm con}}}
\newcommand{\xicylzj}{\overline{\xi^{(j)}_{\rm cyl}}}
\newcommand{\xicyl}{\overline{\xi_{\rm cyl}}}
\newcommand{\xicylalpha}{\overline{\xi_{\rm cyl,(\alpha)}}}
\newcommand{\xir}{\xi^{(r)}}
\newcommand{\xith}{\xi^{(\theta)}}

\newcommand{\QQ}{Q}
\newcommand{\chiim}{\chi_{i,-}}
\newcommand{\chiip}{\chi_{i,+}}
\newcommand{\zim}{z_{i,-}}
\newcommand{\zip}{z_{i,+}}
\newcommand{\sn}{({\rm s.n.})}
\newcommand{\sv}{({\rm s.v.})}
\newcommand{\Rim}{R_{i,-}}
\newcommand{\Rip}{R_{i,+}}

\newcommand{\Rjp}{R_{j,+}}

\newcommand{\Sh}{S^{\rm h}}

\newcommand{\Qw}{{\cal Q}}
\newcommand{\thetaim}{\theta_{i,-}}
\newcommand{\thetaip}{\theta_{i,+}}

\newcommand{\degs}{deg$^{2}$}
\newcommand{\flux}{$\rm erg \, s^{-1} \, cm^{-2}$}

\begin{document}

\title{Covariance matrices for halo number counts and correlation
functions}    
%% ancien titre: Means and covariance matrices for halo number counts and correlation functions, for deep surveys
%% proposition N.C. : Covariance of observables in galaxy clusters deep surveys

%\titlerunning{Means and covariance matrices in galaxy clusters deep surveys}

\author{
P. Valageas\inst{1}
\and N. Clerc\inst{2}
\and F. Pacaud\inst{3} 
\and M. Pierre\inst{2}
}
\institute{
Institut de Physique Th\'eorique, CEA Saclay, 91191 Gif-sur-Yvette, 
France
\and
Laboratoire AIM, CEA/DSM/IRFU/Sap, CEA Saclay, 91191 Gif-sur-Yvette, France
\and
Argelander-Institut f\"ur Astronomie, University of Bonn, Auf dem H\"ugel 71,
53121 Bonn, Germany
}  
\date{Received / Accepted } 
 
\abstract
{}
{We study the mean number counts and two-point correlation functions, along with 
their covariance matrices, of cosmological surveys such as for clusters. In particular,
we consider correlation functions averaged over finite redshift intervals, which
are well suited to cluster surveys or populations of rare objects, where one needs
to integrate over nonzero redshift bins to accumulate enough statistics.}
{We develop an analytical formalism to obtain explicit expressions of all
contributions to these means and covariance matrices, taking into account both
shot-noise and sample-variance effects. We compute low-order as well
as high-order (including non-Gaussian) terms.}
{We derive expressions for the number counts per redshift bins both for the
general case and for the small window approximation. We estimate the range
of validity of Limber's approximation and the amount of correlation between
different redshift bins.
We also obtain explicit expressions for the integrated 3D correlation
function and the 2D angular correlation. We compare the relative importance of
shot-noise and sample-variance contributions, and of low-order and high-order
terms. We check the validity of our analytical results through a comparison with
the Horizon full-sky numerical simulations, and we obtain forecasts for several
future cluster surveys.}
{}


\maketitle

\section{Introduction} 
\label{Introduction}

The large-scale structure of the Universe is a key test of modern cosmological
scenarios. Indeed, according to the standard cosmological model, the
large-scale structures of the present Universe have formed through the
amplification by gravitational instability of small almost-Gaussian primordial
fluctuations \citep{Peebles1980}. Then, from observations of the recent
Universe, such as galaxy surveys \citep{Cole2005,Tegmark2006},
cluster surveys \citep{Evrard1989,Oukbir1992,Pacaud2007},
weak-lensing studies \citep{Massey2007,Munshi2008}, or
measures of baryon acoustic oscillations \citep{Eisenstein1998,Eisenstein2005},
one can derive constraints on the cosmological parameters (e.g., the mean matter
and dark energy contents) and on the properties of the initial perturbations
(e.g., possible deviations from Gaussianity).
Moreover, one can check whether these structures have really formed through
this gravitational instability process.

In the case of well-defined astrophysical objects, such as galaxies or X-ray
clusters, which form a discrete population, standard probes are the abundance
of these objects, that is, ``number counts'', and their low-order correlation
functions. Galaxies are governed by complex gas physics and star formation
processes, which makes it difficult to relate their abundance as a function
of optical luminosity to theoretical predictions. However, since they are
rather common objects (one can reach almost $10^6$ galaxies in current surveys,
e.g. \citet{Abazajian2009}) it is possible to reconstruct halo density fields 
using subsamples of luminous red galaxies and to compare their power
spectrum (i.e. the Fourier transform of their two-point correlation) with
theory to derive constraints on cosmology \citep{Reid2010}.
In contrast, as the largest nonlinear objects in the present Universe,
galaxy clusters are much rarer (current cluster samples have a density
of about ten per deg$^2$ at most, e.g. \citet{Adami2011}),
but their relationship with dark matter halo mass is controlled better.
This means that their abundance is very sensitive to cosmological parameters
(especially $\Om$ and $\sigma_8$, \citet{Evrard1989,Oukbir1992}), and
has already been used to derive constraints on cosmology, but their clustering
has not yet provided much cosmological information (because of low statistics).
However, upcoming cluster surveys should allow the use of both number counts
and spatial clustering to derive cosmological constraints \citep{Pierre2011}.

To compare observations with theory, one needs either to relate the observed
properties of the objects (e.g., optical galaxy luminosity, X-ray cluster luminosity
or temperature) to the quantities that are predicted by theoretical models
(e.g., virialized halo mass) or to use semi-analytical models that attempt to
build mock catalogs \citep{Harker2007}.
For clusters, one can use scaling laws between luminosity, temperature, and
mass, calibrated on observations of the local universe 
\citep{Arnaud2005,Arnaud2010}.
Then, one must take the selection function of the survey being considered
into account, since the probability of detecting the objects is usually more
complex than a sharp cutoff on mass or luminosity \citep{Pierre2011}.
Finally, one needs to estimate the error bars of the statistical quantities
that are measured, in order to derive meaningful constraints on cosmology.
In addition to the uncertainties associated with the relationship between
observed quantities (e.g., luminosity) and theoretical quantities (e.g., halo mass)
discussed above, two unavoidable sources of uncertainty are the ``shot-noise''
effects of the discrete character of the population and the
``sample variance'' due to the limited size of the survey.
In the case of a full-sky survey, the latter is also known as the ``cosmic
variance'', due to the fact that we only observe ``one sky'' so that there is
only a limited number of low-$k$ modes to be measured. 

In practice \citep{Benoist1996,Maller2005,Norberg2009}, one often estimates
error bars from
the data itself by subsampling the data and by computing the scatter between
the means measured within each subsample (e.g., jackknife resampling).
However, if one studies rare objects (e.g., clusters) or large scales, it is not
possible to obtain reliable estimates from such subsamplings (because of
low-quality statistics). Moreover, one often wishes to estimate the
signal-to-noise ratio of future surveys, even before they have been approved
by research agencies, in order to evaluate their scientific possibilities and to
compare the efficiency of different probes. Then, one must use numerical
simulations \citep{Pierre2011,Croton2004,Kazin2010}
or analytic methods. The former have the advantage of greater
power (in the sense that one may explicitly introduce complex recipes for the
formation of the objects, such as cooling processes and feedback, or intricate
survey geometry), but are
limited by finite resolution on large scales and for rare objects. 
Analytical approaches allow one to describe a wider range of scales and halo
masses, and usually provide faster computations. Hence they remain a useful
complementary method, which we investigate in this paper.

Thus, in this paper, we present a general analytical formalism for computing the
shot-noise and sample-variance error bars of estimators of number counts and 
real-space two-point correlations for deep surveys that cover
a significant range of redshifts. We consider both the 3D correlation
function and the 2D angular correlation function on the sky.
As explained above, this study is motivated by the need for such covariance
matrices to compare any survey with theory. This extends over
some previous works, which only considered the sample variance of number
counts \citep{Hu2003} or neglected high-order or non-Gaussian terms in the
sample variance of estimators for two-point correlations or power spectra
\citep{Feldman1994,Majumdar2004,Eisenstein2005,Cohn2006,Crocce2011}.
Indeed, while the sample variance of number counts (i.e., the mean number
of objects per unit volume) only involves the two-point correlation of the objects,
the sample variance of estimators of the two-point correlation itself also involves
the three- and four-point correlations (and so on for estimators of higher 
order correlation functions) \citep{Bernstein1994}.

Some previous studies have already included the contributions of such
higher order correlations to covariance matrices
\citep{Szapudi1996,Meiksin1999,Scoccimarro1999,Eisenstein2001,Smith2009},
mostly in the context of galaxy surveys.
However, we extend these works by comparing
the various contributions to expected error bars in detail, including all shot-noise
and sample-variance terms, as well as high-order contributions, and by
studying real-space two-point correlation functions instead of Fourier-space
power spectra.
Moreover, since we have the application to cluster surveys in mind,
and more generally to deep surveys of rare objects, we consider quantities
that are defined by integration over finite redshift bins.
For instance, number counts may be associated with bins $\Delta z=0.1$
while the two-point correlation functions are integrated over a significant
redshift interval, such as $0<z<1$, to accumulate enough statistics.
Then, the statistical quantities that we consider involve integrations along the
line-of-sight, rather than local power spectra or two-point correlations in
a small box at a given redshift.
Moreover, for number counts we consider arbitrary angular scales, from small
angles, where the Limber approximation applies, to full-sky surveys.
In these various respects, our study fills a gap in published works.

In view of the application to cluster surveys, we consider
a population of objects defined by their mass $M$, and focusing on the
case of dark matter halos, we use the halo mass function and bias measured
in previous numerical studies for our numerical computations. 
To estimate the three- and four-point correlation functions
(needed for the covariance of the two-point estimator), we
use a simple hierarchical model \citep{Peebles1980,Bernstein1994}, which
writes these higher order correlations as products of the two-point
correlation. 
We give the explicit expressions of our results, which we also compare with
numerical simulations, and we provide several realistic illustrations.
In particular, we take advantage of our analytical formalism to compare
the various contributions to the error bars and derive approximate scalings.
Although we eventually apply our results to several future cluster surveys, 
our method is more general and could be applied to other objects
(e.g., galaxies or quasars), defined by other quantities (e.g., luminosity),
provided one has a model for their multiplicity function and their two-point
correlation, and the three- and four-point correlation functions can be
described reasonably well by a hierarchical model in the regime where they
are relevant.

This paper is organized as follows. In Sect.~\ref{Halo-density} we first briefly 
describe the analytic models that we use to estimate
the means and covariance matrices of numbers counts and halo correlations,
as well as the numerical simulations that we use to check the accuracy of our
results. Then, we study the halo number counts per redshift bins in
Sect.~\ref{Number-density}. This allows us to introduce on
a simple example our approach to evaluate the mean and the covariance matrix of
various estimators. We consider the cases of both small angular windows and
arbitrary angular windows (including full-sky surveys), and we estimate the
accuracy of small-angle approximations and the decay in correlations between
distant redshift bins.
Next, we study the real-space 3D halo correlation function in
Sect.~\ref{Two-point-correlation}. We consider both the Peebles \& Hauser
and the Landy \& Szalay estimators and compare their
covariance matrices. We also discuss the relative importance of different contributions
to these covariance matrices (shot noise/sample variance,
low-order/high-order terms).
Then, we investigate the halo angular correlation in Sect.~\ref{Angular-correlation},
using the same approach.
Finally, we apply our formalism to several real survey cases in
Sect.~\ref{Applications} and we conclude in Sect.~\ref{Conclusion}.

We give details of our calculations in several appendices. We discuss shot-noise
terms in App.~\ref{Method}, in the simple case of number counts, and finite-size
effects in App.~\ref{Finite-size}. Then, we describe our computation of the
mean and covariance of the estimators of the 3D halo correlation in
Apps.~\ref{Computation-of-the-mean-of-the-estimator}
and \ref{Computation-of-the-covariance-of-the-estimator}, for the Peebles \& Hauser
estimator, and in App.~\ref{Computation-Landy-Szalay} for the Landy \& Szalay
estimator. We give further details on high-order contributions to the covariance
matrices in App.~\ref{Computation-high-order-terms}, for the 3D correlation,
and App.~\ref{Computation-high-order-terms-w}, for the angular correlation.

\section{Halo density fields}
\label{Halo-density}

Before we describe our analysis of the covariance matrices for halo number
counts and correlation functions, we present in this section the analytic
models that we use for the underlying halo distributions (mass and bias function,
etc.) and the numerical simulations that we use to validate our results.

\subsection{Analytic models}
\label{Analytic-models}

\subsubsection{Halo mass function and correlation}
\label{Halo-mass-function}

To be consistent with the numerical simulations, in
Sects.~\ref{Mean-number-counts-in-redshift-bins} to \ref{Angular-correlation},
where we develop our formalism and compare our results with simulations,
 we use the WMAP3 cosmology
\citep{Spergel2007}, that is, $\Omega_{\rm m}=0.24$, $\Omega_{\rm de}=0.76$,
$\Omega_b=0.042$, $h=0.73$, $\sigma_8=0.77$, $n_s=0.958$, and
 $w_{\rm de}=-1$.
In Sect.~\ref{Applications}, where we apply our formalism to obtain forecasts
for current and future surveys, we use the more recent WMAP7 cosmology
\citep{Komatsu2010}, that is, $\Omega_{\rm m}=0.274$, $\Omega_{\rm de}=0.726$,
$\Omega_b=0.046$, $h=0.702$, $\sigma_8=0.816$, $n_s=0.968$, and
 $w_{\rm de}=-1$.
 
In this paper, keeping in mind the study of X-ray clusters, we consider the
number counts and correlations of dark matter halos defined by the
nonlinear density contrast $\delta=200$. These halos are fully characterized
by their mass, and we do not investigate the relationship between this mass 
and cluster properties such as the gas temperature and X-ray luminosity.
These scaling laws can be added to our formalism to derive the cluster
number counts and correlations, depending on the quantities that are actually
measured, but we keep a more general setting in this paper.

We use the halo mass function, $\dd n/\dd \ln M$, of \citet{Tinker2008}, 
and the halo bias of \citet{Tinker2010}. Thus, the two-point correlation function
$\xih_{i,j}$ between two halos labeled ``$i$'' and ``$j$'' can be 
factored\footnote{A weaker hypothesis would be to write
$\xih_{i,j} = b^2_{i,j} \xi(x_{ij};z)$, where the dependence on $M_i$ and $M_j$
does not factor in (i.e., $b^2_{i,j} \neq b_i b_j$), but keeping the
factorization with respect to the relative distance $x_{ij}$. This only slightly
modifies our expressions in a straightforward manner.}
in as
\beq
\xih_{i,j} = b_i b_j \, \xi(|\vx_i-\vx_j|;z) ,
\label{xij-bb}
\eeq
where $\xi$ is the matter density correlation, and the bias factors $b_i$ and
$b_j$ do not depend on scale,
\beq
b_i = b(M_i,z_i) .
\label{b-def}
\eeq
This approximation of scale-independent halo bias is valid to better than $10\%$
on scales $20<r<130 h^{-1}$ Mpc \citep{2009arXiv0912.0446M},
with a small feature on the baryon acoustic scale ($r \sim 100 h^{-1}$Mpc)
of amplitude of $5\%$ \citep{2010PhRvD..82j3529D}.

Throughout most of this paper we assume that correlations are negligible
over cosmological distances (of order $c/H_0$), so that the redshift $z$ on the
right-hand side of Eq.(\ref{xij-bb}) can be taken at will as $z_i$ or $z_j$
(or the mean $(z_i+z_j)/2$).
In Sect.~\ref{Large-angular-windows}, where we consider the case of large
angular windows (and go beyond the flat sky and Limber's approximations),
we do not use this approximation but replace the matter density correlation
by its linear approximation. This yields the alternative factorization
$\xih_{i,j}  \propto  b(M_i,z_i)  b(M_j,z_j) D_+(z_i) D_+(z_j) \xi_{L0}(\vx_i,\vx_j)$
that allows one to handle arbitrary redshifts $z_i$ and $z_j$.

For the nonlinear matter correlation $\xi(x;z)$, and the Fourier-space nonlinear
power spectrum $P(k)$. defined by
\beq
\xi(x;z) = \int \dd\vk \, e^{\ii\vk\cdot\vx} \, P(k;z) ,
\label{xi-Pk}
\eeq
we use the popular fitting formula to numerical simulations of
\citet{Smith2003}.

Throughout this paper, all angular number densities are in units of deg$^{-2}$.

\subsubsection{Three-point and four-point halo correlations}
\label{three-point}

The covariance matrices of the estimators $\hxi$ for the halo two-point
correlation $\xih$ also involve the halo three-point and four-point correlation
functions, $\zetah$ and $\etah$, so we must define a model for these
quantities.
On large scales, for Gaussian initial conditions, the three- and four-point
correlation functions of the matter density field behave as $\zeta \sim \xi^2$
and $\eta\sim\xi^3$ at lowest order over $\xi$
\citep{Bernardeau2002,Goroff1986}.
On small scales, these scaling laws remain a reasonable approximation
\citep{Colombi1996}, but with numerical prefactors that are different from the
large-scale ones (and may slightly vary with scale).
On the other hand, for rare massive clusters, using the standard approach
of \citet{Kaiser1984} where virialized objects are identified with overdense
regions in the linear density field, \citet{Politzer1984} obtain
$1+\xih(\vx_1,..,\vx_N) = \prod_{i>j} [1+\xih(\vx_i,\vx_j)]$. 
Since our goal is only to estimate the magnitude of high-order contributions
we consider in this article a simple ``hierarchical clustering ansatz'',
where the $N-$point correlation function can be expressed in terms
of products of $(N-1)$ two-point correlation functions through tree diagrams
\citep{Groth1977,Peebles1980}. This is the simplest
model\footnote{An alternative would be to use a halo model \citep{Cooray2002},
coupled to
perturbation theory predictions on large scales. This could provide more
accurate estimates, since such an approach is able to describe low-order
correlation functions well from very large to small scales
\citep{Scoccimarro2001,Giocoli2010,Valageas2011,Valageas2011a}.
However, this would introduce several contributions associated with ``1-halo''
up to ``4-halo'' terms, and would require additional parameters such as halo
occupation functions and high-order bias parameters, depending on the objects
that one considers.
Therefore, we do not investigate this approach in this paper (although it would
certainly deserve further attention), and we restrict ourselves to the simpler
hierarchical models, as described in Eqs.(\ref{zeta-def})-(\ref{eta-def}).}
that is in qualitative
agreement with large-scale theoretical predictions and small-scale numerical
results, as well as with observations.

Through a comparison with numerical
simulations, we check that the accuracy of this model is sufficient for our purpose,
which is to estimate signal-to-noise ratios and compare different survey strategies
(while we would require a higher accuracy for the computation of the means
themselves, that is, the number counts and two-point correlations that we wish
to measure). The advantage of this simple model is that it describes all scales,
through the scalings recalled above, and does not require additional free
parameters.

\begin{figure}
\begin{center}
\epsfxsize=8 cm \epsfysize=2.5 cm {\epsfbox{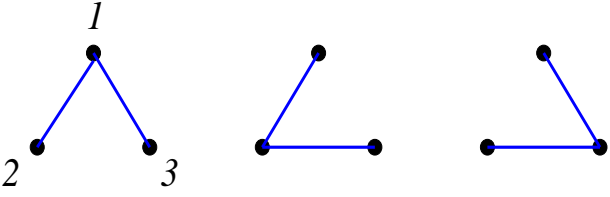}}
\end{center}
\caption{The ``hierarchical clustering ansatz'' for the three-point correlation
function $\zetah_{1,2,3}$ of Eq.(\ref{zeta-def}). Each solid line corresponds to
a two-point correlation $\xi$, and $\zetah$ is written as the sum of these
three diagrams, with a multiplicative factor $b_1b_2b_3 S_3/3$.}
\label{fig-zeta}
\end{figure}

\begin{figure}
\begin{center}
\epsfxsize=8 cm \epsfysize=2.6 cm {\epsfbox{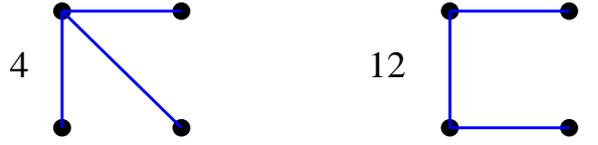}}
\end{center}
\caption{The two topologies of the four-point diagrams associated with the
``hierarchical clustering ansatz'' for the four-point correlation, as in
Eq.(\ref{eta-def}). The numbers are the multiplicity factors of each diagram.}
\label{fig-eta}
\end{figure}

Thus, as in \citet{Bernstein1994}, \citet{Szapudi1996}, \citet{Meiksin1999},
we write the three-point halo correlation function as
\beq
\zetah_{1,2,3} = b_1 b_2 b_3 \; \frac{S_3}{3} \; \left[ \xi_{1,2} \xi_{1,3} 
+ \xi_{2,1} \xi_{2,3} + \xi_{3,1} \xi_{3,2} \right] ,
\label{zeta-def}
\eeq
where we sum over all three possible configurations over the three halos
labeled ``1'', ``2'', and ``3''. This corresponds to the
three tree-diagrams shown in Fig.~\ref{fig-zeta}.
We use a linear bias model\footnote{Within a local bias model, one writes
the halo density field as $\delta_{\rm h}= \sum_k b_k \delta^k/k!$,
where $\delta$ is the matter density
contrast smoothed on large scales. Then, the halo many-body correlations
also depend on the higher order bias coefficients $b_k$ \citep{Fry1993},
but for simplicity we only consider a linear bias model here (i.e. $b_k=0$ for
$k\geq 2$).},
as in Eq.(\ref{xij-bb}), and for the matter density
normalization factor $S_3$, we take its large-scale limit, which is obtained by 
perturbation theory \citep{Peebles1980,Fry1984a,Bernardeau2002},
\beq
S_3 = \frac{34}{7} - (n+3) ,
\label{S3-def}
\eeq
where $n$ is the slope of the linear power spectrum at the scale of interest.
Within the same ``hierarchical clustering ansatz'', the four-point correlation
function is expressed in terms of products of three two-point
functions, as shown in Fig.~\ref{fig-eta}. We have two possible topologies,
and sixteen different diagrams for four distinct halos.  For simplicity, in this work we
give the same weight  to all sixteen diagrams, independently of their topology,
as
\beqa
\etah_{1,2,3,4} & = & b_1 b_2 b_3 b_4 \; \frac{S_4}{16} \;
\left[ \xi_{1,2} \xi_{1,3} \xi_{1,4} + 3 \, {\rm cyc.} \right. \nonumber \\
&& \left. + \xi_{1,2} \xi_{2,3} \xi_{3,4} + 11 \, {\rm cyc.} \right] ,
\label{eta-def}
\eeqa
where ``3 cyc.'' and ``11 cyc.'' stand for three and eleven terms that are
obtained from the previous one by permutations over the labels ``1,2,3,4''
of the four halos.
Again we take for $S_4$ its large-scale
limit,
\beq
S_4 = \frac{60712}{1323} - \frac{62}{3} (n+3) + \frac{7}{3} (n+3)^2 .
\label{S4-def}
\eeq

This is the simplest possible model, where the angular dependence only comes
from the decomposition over the terms in brackets in Eqs.(\ref{zeta-def})
and (\ref{eta-def}).
More complex models and exact computations at lowest order of perturbation
theory would introduce homogeneous kernels $Q_N(\vx_1,..,\vx_N)$
(in place of the numbers $S_N$)
that also depend on the angles between the vectors $\vx_{i}$ and $\vx_{j}$
\citep{Scoccimarro1999}.
Observations of galaxy clustering show, for instance, that $Q_3$ displays a weak
dependence on the triangle shape, while remaining close to unity
\citep{Gaztanaga2005,Kulkarni2007}.
Here we simply take the constant value $Q_3=S_3/3$.

In terms of the halo two-point correlation, Eqs.(\ref{zeta-def}) and (\ref{eta-def})
imply halo coefficients $\Sh_N$ that behave as $\Sh_N \sim b^{2-N} S_N$.
For $b\sim 1$ and $n\simeq -1$ and from Eqs.(\ref{S3-def}) and
(\ref{S4-def}), this gives the values $\Sh_3 \sim S_3 \simeq 2.9$ and
$\Sh_4 \sim S_4 \simeq 13.9$, which roughly agree with observations of
galaxy clustering \citep{Szapudi2001,Croton2004,Ross2006,Marin2008}.

\subsubsection{Flat-sky and Limber's approximations}
\label{Flat-sky-approximation}

In this paper we often encounter quantities, such as the mean
matter density correlation over a redshift bin $j$, $z_{j,-}<z'<z_{j,+}$, 
with respect to some redshift $z$ in a second bin $i$, integrated over some angular
window of area $(\Delta\Omega)$,
\beq
\xiconzj(z) = \int_{\chi_{j,-}}^{\chi_{j,+}} \frac{\dd\chi'}{\cD(z)} \int
\frac{\dd\vOm\dd\vOm'}{(\Delta\Omega)^2} \, \xi(\vx,\vx') ,
\label{xib-ij-def}
\eeq
where $\vx=(\chi,\cD\vOm)$ and $\vx'=(\chi',\cD'\vOm')$.
Here $\chi(z)$ and $\cD(z)$ are the comoving radial and angular distances,
and we introduce the factor $1/\cD(z)$ so that $\xiconzj$ is
dimensionless.
Equation (\ref{xib-ij-def}) is a ``conical'' average, within the observational cone.
However, for small angular windows it is possible to use a flat-sky approximation
and to approximate this ``conical'' average $\xiconzj$ by a ``cylindrical''
average $\xicylzj$.
Thus, using Eq.(\ref{xi-Pk}) and assuming that the correlation $\xi$ is negligible
on cosmological scales, we write for circular angular windows of radius
$\theta_s$, for a redshift $z$ that also belongs to the $j$-bin,
\beqa
z_{j,-} \!<\! z \!<\! z_{j,+} : \;\; \xicylzj(z) & = & 
\int_{\chi_{j,-}}^{\chi_{j,+}} \! \frac{\dd\chi'}{\cD} \, \int 
\! \frac{\dd\vtheta\dd\vtheta'}{(\pi\theta_s^2)^2}  \int \! \dd\vk \,
 \nonumber \\
&& \hspace{-1cm} \times \; e^{\ii k_{\parallel}\cdot(\chi'-\chi)
+\ii \vk_{\perp}\cdot\cD(\vtheta'-\vtheta)} \; P(k;z) ,
\label{Cij-3}
\eeqa
and $\xicylzj(z) = 0$ if $z$ does not belong to the $j$-bin.
Here $k_{\parallel}$ and $\vk_{\perp}$ are the longitudinal and transverse
components of $\vk$, with respect to the line of sight, while $\vtheta$ and
$\vtheta'$ are the 2D transverse angular vectors.

For a redshift binning that is not too small, $\Delta\chi \gg \cD\theta_s$,
longitudinal wavenumbers above $1/(\Delta\chi)$ are suppressed by
integrating $\chi'$
along the line of sight, and the integral is dominated by wavenumbers with
$k \simeq k_{\perp}$ and $k_{\perp} \sim 1/(\cD\theta_s)$.
Thus, using the Fourier form of Limber's approximation \citep{Limber1953},
which is widely used in weak-lensing studies \citep{Kaiser1992,Munshi2008},
the integration over $\chi'$ yields a Dirac term $(2\pi)\delta_D(k_{\parallel})$,
and the integration over $k_{\parallel}$ gives
\beq
\xicylzj(z) \simeq \xicyl(z) ,
\label{flat-Limber}
\eeq
with
\beq
\xicyl(z) = \frac{2\pi}{\cD}  \int \frac{\dd\vtheta \dd\vtheta'}{(\pi\theta_s^2)^2}
\int\! \dd \vk_{\perp} \, e^{\ii \vk_{\perp}\cdot\cD(\vtheta'-\vtheta)} \, 
P(k_{\perp};z) ,
\label{Cij-4}
\eeq
which does not depend on the size of the redshift bin $j$ (because we have
taken the limit of a very large redshift bin).
Introducing the 2D Fourier-space circular window\footnote{For more complicated
angular shapes we can still define a Fourier-space window
$\tW_2(\vk_{\perp}\cD\theta_s)$, normalized by $\tW_2(0)=1$, but it will also
depend on the direction of $\vk_{\perp}$.},
\beq
\tW_2(k_{\perp}\cD\theta_s) = \int \frac{\dd\vtheta}{\pi\theta_s^2} \,
e^{\ii\vk_{\perp}\cdot\cD\vtheta} =
\frac{2J_1(k_{\perp}\cD\theta_s)}{k_{\perp}\cD\theta_s} ,
\label{W-thetas}
\eeq
we obtain
\beq
\xicyl(z) = \pi  \int_0^{\infty} \frac{\dd k}{k} \frac{\Delta^2(k,z)}{\cD k}
\tW_2(k\cD\theta_s)^2 ,
\label{I-thetas-def}
\eeq
where we defined the 3D power per logarithmic wavenumber, $\Delta^2(k,z)$, by
\beq
\Delta^2(k,z) = 4\pi k^3 P(k,z) .
\label{Delta2-def}
\eeq
We will evaluate the accuracy of this approximation, based on the flat-sky
and Limber's approximation, in Sect.~\ref{accuracy-flat-sky-Limber}.

\subsection{Numerical simulations}
\label{simulations}

Our analytical formalism allows us to consider a broad range of scales and
halo masses, from small angular windows up to full-sky surveys, 
and to compare the relative contributions to covariance matrices that arise
from shot-noise and cosmic variance effects, and from low-order and
high-order large-scale correlations.
Numerical simulations do not easily allow such a detailed analysis; however,
in order to validate our approach, we must check whether it agrees with
estimates from simulations, wherever a comparison is possible.

We use the high-resolution full-sky Horizon simulation \citep{Teyssier2009},
based on the WMAP3 cosmology \citep{Spergel2007}.
This is a 68.7 billion particle N-body simulation, featuring more than 140
billion cells in the AMR grid of the RAMSES code \citep{Teyssier2002}.
The simulation consists in a lightcone spanning the entire sky up to redshift
$\sim 1$, with a mass resolution of $1.1 \times 10^{10} M_{\odot}$.

Halos in the (2 Gpc)$^3$ N-body simulation are found with the
HOP algorithm \citep{EisensteinHut1998}. Their
comoving positions are then converted into sky coordinates,
taking their radial velocity into account when calculating redshifts.
The physical effects of the baryons are neglected and the total mass is given
by the number of particles inside the halo.
We only consider halos at redshifts $z\leq 0.8$,
up to which the simulation is complete towards all directions.

We design simulated surveys by extracting rectangular 
fields in angular coordinates.
To minimize the effect of intrinsic sample correlations, we impose
a 10 (resp. 20) deg gap between consecutive fields when computing number
counts (resp. correlation functions), which yields 138 (resp. 34) nonoverlapping
fields that can be cut out in the simulation.

For the purpose of clustering analysis, auxiliary random fields are constructed by
shuffling the angular coordinates of halos in the data fields,
thus preserving the halo mass and redshift distributions. To gain
in computational efficiency, the number of halos per random field
is ten times the average number of halos in data fields.
The Landy \& Szalay estimator is then scaled accordingly to the ratios
of pair numbers in data and random fields.

The mean and covariance of all quantities of interest are estimated by
sample averaging over the extracted surveys.
Because of the uniqueness of the simulation, a residual noise is expected whenever
the area of individual fields becomes large and their number diminishes. 
Thus, there are about $41253 {\rm deg}^2/(\Delta\Omega)$
nonoverlapping fields of area $(\Delta\Omega)$. (For instance, we cut
41 fields of 400 deg$^2$ for the analysis of the angular correlation function.)

\section{Number density of halos}
\label{Number-density}

\subsection{Mean number counts in redshift bins}
\label{Mean-number-counts-in-redshift-bins}

We consider a population of objects defined by some property, such as their
mass $M$, with a mean comoving number density per logarithmic interval of
$M$ written as $\dd n/\dd\ln M$. Then, the mean number of objects
in the redshift interval $[z,z+\dd z]$, within the solid angle $\dd\vOm$
on the sky, with a mass in the range $[M,M+\dd M]$, reads as
\beq
\dd\vOm \, \dd N =  \dd z \, \dd\vOm \, \left|\frac{\dd V}{\dd z\dd\vOm}\right|\!(z)
\, \frac{\dd M}{M} \, \frac{\dd n}{\dd \ln M}(M,z) ,
\label{dndz}
\eeq
where $|\dd V/\dd z\dd\vOm|$ is the cosmological volume factor, which is given
by
\beq
\left|\frac{\dd V}{\dd z\dd\vOm}\right|(z) = \cD(z)^2 \; \frac{\dd\chi}{\dd z} ,
\label{V-chi}
\eeq
and $\chi(z)$ and $\cD(z)$ are the comoving radial and angular distances.
In Eq.(\ref{dndz}) and in the following we define $N$ as the number density
of objects per unit area on the sky, instead of the total number of objects within
a given window $(\Delta\Omega)$. This choice is more convenient for practical
purposes because it allows a simpler comparison between different surveys
that have different total areas.

We can split the interval of mass\footnote{Although $M$ stands for the mass
of the objects, as for the mass function of
clusters of galaxies, or of galaxies themselves, it can also represent any
other quantity, such as temperature, luminosity, or a vector made of several
such quantities.} over several bins ``$\alpha$'',
$[M_{\alpha,-},M_{\alpha,+}]$ and the observational
cone over nonoverlapping redshift intervals ``$i$'', $[\zim,\zip]$ with
$\zip \leq z_{i+1,-}$.
In practice, one usually takes $\zip = z_{i+1,-}$ so as to cover a continuous
range of redshifts.
Then, the number of objects per unit area, in the bin $(i,\alpha)$, reads as
\beq
\hN_{i,\alpha} = \int_{\zim}^{\zip} \!\! \dd z \, \frac{\dd\chi}{\dd z} \, \cD^2
\int_{\Delta\Omega} \! \frac{\dd\vOm}{(\Delta\Omega)}
\int_{M_{\alpha,-}(z)}^{M_{\alpha,+}(z)} \! \frac{\dd M}{M} \, \frac{\dd\hn}{\dd\ln M}  ,
\label{Ni-1}
\eeq
where $\dd\hn/\dd\ln M$ is the observed density of objects.
Here and in the following, we note observed
quantities by a hat (i.e. in one realization of the sky) to distinguish them from mean
quantities, such as the comoving number density of Eq.(\ref{dndz}), 
that correspond to expectation values over many realizations.
In practice, assuming ergodicity, these expectation averages are assumed to
be identical to volume averages (in the case of statistically homogeneous
and isotropic cosmologies).

To simplify notations we define the mean cumulative number density
of objects observed at a given redshift, within the mass bin $\alpha$
(with boundaries that may depend on $z$), 
\beq
\nb_{\alpha}(z) =  \int_{M_{\alpha,-}(z)}^{M_{\alpha,+}(z)} \frac{\dd M}{M} \, 
\frac{\dd n}{\dd\ln M}(M,z) ,
\label{Nbz-def}
\eeq
and we omit the explicit boundaries on mass in the following.
Then, the mean number of objects per unit area in the redshift and mass bins
$(i,\alpha)$ reads as
\beq
\lag \hN_{i,\alpha} \rag = \int_{\chiim}^{\chiip} \dd\chi \, \cD^2 \, \nb_{\alpha} .
\label{Ni-3}
\eeq

\begin{figure}[htb]
\begin{center}
\epsfxsize=8 cm \epsfysize=6 cm {\epsfbox{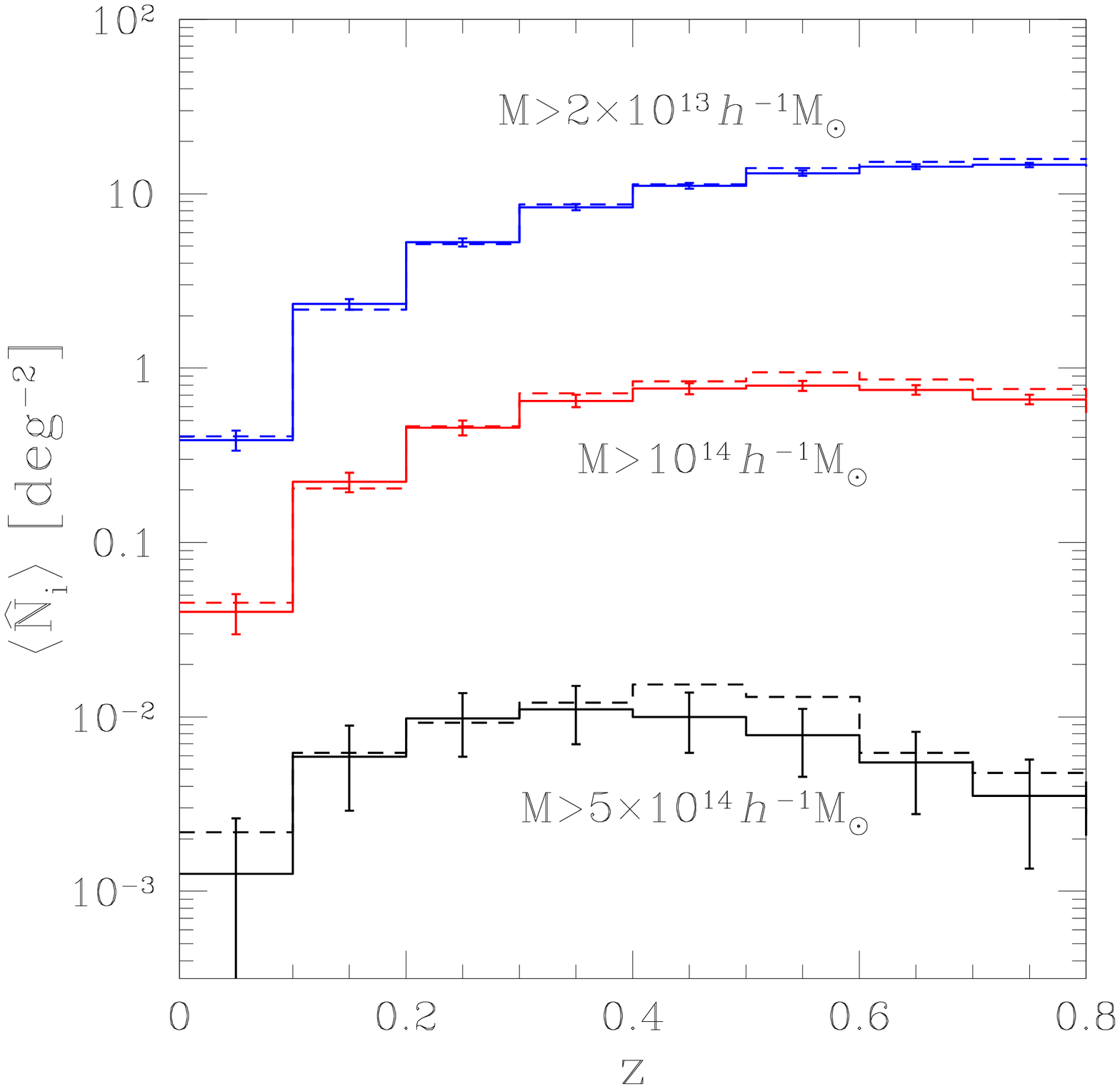}}
\end{center}
\caption{The mean number density of dark matter halos per square degree,
within redshift bins of width $\Delta z= 0.1$. We count all halos
above the thresholds $M_*=2\times 10^{13}, 10^{14},$ and $5\times 10^{14} 
h^{-1} M_{\odot}$, from top down to bottom. We compare our analytical results
(solid lines) with numerical simulations (dashed lines).}
\label{fig_nz_Horizon}
\end{figure}

We plot the mean number counts $\lag \hN_{i,\alpha} \rag$ of Eq.(\ref{Ni-3})
in Fig.~\ref{fig_nz_Horizon}, per square degree, for redshift bins of
width $\Delta z= 0.1$.
Here we select all dark matter halos above a mass threshold $M_*$, with
$M_*=2\times 10^{13}, 10^{14},$ and $5\times 10^{14}h^{-1} M_{\odot}$.
The error bars are the $3-\sigma$ statistical errors obtained from the covariance
matrices derived in Sect.~\ref{small-angles}, for 138 fields of $50$
deg$^2$ as used in the simulations.
We can check that our estimates agree reasonably well with the
numerical results. The small discrepancies are probably due to
the complex relation between theoretical halo masses
and the actual halos found in the simulation box. In particular, it is well known
that using different algorithms, such as spherical-overdensity, HOP or
friends-of-friends algorithms, can lead to slightly different results 
(e.g. \citet{EisensteinHut1998,Tinker2008}).
However, this point is beyond
the scope of this paper, as we only wish here to check that our analytical
results provide reasonable estimates of the mean and covariance of halo
number counts and correlations.

\subsection{Covariance of number counts}
\label{covariance-counts}

As usual we define the covariance $C_{i,\alpha;j,\beta}$ of the statistical quantities
$\hN_{i,\alpha}$ and $\hN_{j,\beta}$ by
\beq
C_{i,\alpha;j,\beta} = \lag \hN_{i,\alpha} \hN_{j,\beta} \rag - \lag \hN_{i,\alpha} 
\rag \lag \hN_{j,\beta} \rag .
\label{C-ij-def}
\eeq
As recalled in App.~\ref{Method}, following \citet{Peebles1980},
it can be decomposed over ``shot-noise'' and ``sample-variance''
contributions,
\beq
C_{i,\alpha;j,\beta} = C_{i,\alpha;j,\beta}^{\sn} + C_{i,\alpha;j,\beta}^{\sv} ,
\label{C-NiNj-sn-sv}
\eeq
which write from Eq.(\ref{hni-hnj}) as
\beq
C_{i,\alpha;j,\beta}^{\sn} = \delta_{i,j} \, \delta_{\alpha,\beta} \; 
\frac{\lag \hN_{i,\alpha} \rag}{(\Delta\Omega)} ,
\label{Cij-sn}
\eeq
(for nonoverlapping mass binning), and
\beqa
C_{i,\alpha;j,\beta}^{\sv} & = & \int_i\dd \chi \, \cD^2
\int\frac{\dd\vOm}{(\Delta\Omega)} \int_{\alpha} 
\frac{\dd M}{M} \, \frac{\dd n}{\dd\ln M} \nonumber \\
&& \hspace{0cm} \times \int_j\dd \chi' \, \cD'^2
\int\frac{\dd\vOm'}{(\Delta\Omega)} \int_{\beta}\frac{\dd M'}{M'} \,
\frac{\dd n}{\dd\ln M'}  \; \xih .
\label{Cij-xi}
\eeqa
Here we denote $\int_i$ and $\int_{\alpha}$ as the integrals over the redshift
and mass bins $i$ and $\alpha$. 
The superscript ``$h$'' refers to the ``halo'' correlation function,
which depends on the two redshifts, angular directions, and masses (or
temperatures, etc.), $\xih= \xih(M,\vx;M',\vx')$, with $\vx=(\chi,\cD\vOm)$.

The first term $C_{i,\alpha;j,\beta}^{\sn}$ is the shot-noise contribution and
vanishes for nonoverlapping bins. As expected it decreases with the
survey size as $1/(\Delta\Omega)$. (We recall that $\hN$ is the angular number
density.) 
The second term $C_{i,\alpha;j,\beta}^{\sv}$ is due to the ``sample-variance''
cross-correlation $\xih$ between distant objects \citep{Hu2003}.
Using the approximation (\ref{xij-bb}), that is, the factorization of the dependence
on mass and distance of $\xih$, we define the mean bias $\bb_{\alpha}$ at
redshift $z$, for the mass bin $\alpha$, through 
\beq
\bb_{\alpha}(z) \, \nb_{\alpha}(z) = \int_{\alpha} \frac{\dd M}{M} \, b(M,z) \,
\frac{\dd n}{\dd\ln M}(M,z) ,
\label{bb-def}
\eeq
where $\nb_{\alpha}$ was defined in Eq.(\ref{Nbz-def}),
so that Eq.(\ref{Cij-xi}) also writes as
\beqa
C_{i,\alpha;j,\beta}^{\sv} & = & \int_i\dd\chi \, \cD^2 \, \bb_{\alpha} \nb_{\alpha} 
\int_j \dd\chi' \, \cD'^2 \, \bb_{\beta}^{\,'} \nb_{\beta}^{\,'} \int\!
\frac{\dd\vOm \dd\vOm'}{(\Delta\Omega)^2} \, \xi(\vx,\vx') . \nonumber \\
&&
\label{Cij-sv-1}
\eeqa

\subsubsection{Small angular windows}
\label{small-angles}

Using the approximation that the correlation function is negligible on cosmological
scales, whence $\cD' \simeq \cD$, $\bb^{\,'} \simeq \bb$ and $\nb^{\,'}\simeq \nb$,
we obtain
\beq
C_{i,\alpha;j,\beta}^{\sv} = \int_i \dd\chi \, \cD^5 \, \bb_{\alpha} \bb_{\beta} \,
\nb_{\alpha} \nb_{\beta} \, \xiconzj ,
\label{Cij-2}
\eeq
where $\xiconzj(z)$ was defined in Eq.(\ref{xib-ij-def}). It is a ``conical'' average,
over objects ``$i$'' and ``$j$'' that are located at unrelated positions $(\chi,\vOm)$ and 
$(\chi',\vOm')$. To recall this ``conical'' integration along
$\chi'$ we have added the subscript ``con''. This helps distinguishing
quantities such as (\ref{xib-ij-def}) from other averages of $\xi$ over 3D
spherical shells, which we encounter in Sect.~\ref{Two-point-correlation}
below.
To further distinguish from the quantities encountered in
Sects.~\ref{Two-point-correlation} and \ref{Angular-correlation}, we put the
label $j$, which refers to a redshift bin, as a superscript, whereas the labels
$i$ or $j$ of the radial or angular bins studied in Sects.~\ref{Two-point-correlation}
and \ref{Angular-correlation} appear as indices. (The parenthesis refer to
the fact that in Limber's approximation of wide redshift bins the dependence
on the boundaries of the bin $j$ disappears, because they are pushed to
infinity, see Eq.(\ref{I-thetas-def}).)

In the case where the nonoverlapping redshift bins are large enough to neglect
the cross-correlation between different bins (we evaluate the accuracy of
this approximation in Sect.~\ref{correlation-redshift}), the integral
(\ref{xib-ij-def}) gives rise to a Kronecker factor $\delta_{i,j}$. Moreover, for 
small angular windows and large enough redshift bins we can use the flat-sky
approximation (\ref{Cij-3}), where the ``conical'' average is approximated by
a ``cylindrical'' average (and spherical harmonics are replaced by plane waves),  
and Limber's approximations (\ref{flat-Limber}), where longitudinal wavenumbers 
are neglected over transverse wavenumbers.
Substituting into Eq.(\ref{Cij-2}) 
we recover the results of \citet{Hu2003},
\beq
C_{i,\alpha;j,\beta}^{\sv} = \delta_{i,j} \int_i \dd\chi \, \cD^5 \, 
\bb_{\alpha} \bb_{\beta} \, \nb_{\alpha} \nb_{\beta} \,
\xicyl ,
\label{Cij-7}
\eeq
where $\xicyl(z)$ was defined in Eq.(\ref{I-thetas-def}).
(We evaluate the accuracy of the approximation (\ref{Cij-7}) in
Sect.~\ref{accuracy-flat-sky-Limber}.)
Thus, while the shot-noise contribution (\ref{Cij-sn}) to the covariance matrix
is diagonal, the sample-variance contribution (\ref{Cij-7})  is only block-diagonal
(for large redshift bins) since within the same redshift bin different mass bins
are correlated.

\begin{figure}[htb]
\begin{center}
\epsfxsize=8 cm \epsfysize=6 cm {\epsfbox{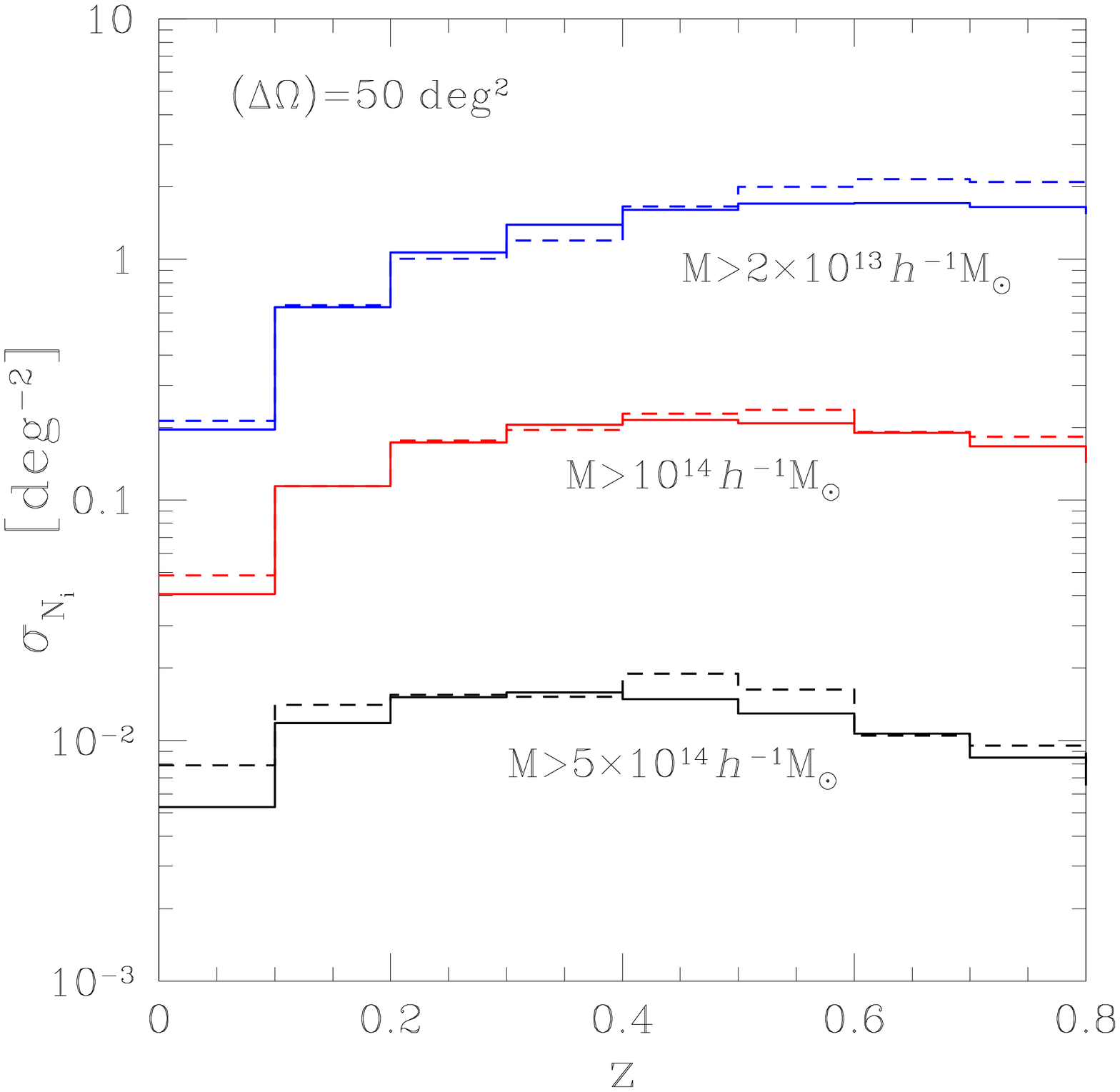}}
\end{center}
\caption{The variance $\sigma_{N_i}$ of the halo angular number densities of
Fig.~\ref{fig_nz_Horizon}, for redshift bins $\Delta z=0.1$ and an angular
window of 50 deg$^2$. We compare our analytical results
(solid lines) with numerical simulations (dashed lines).}
\label{fig_C_nz_Horizon}
\end{figure}

In Fig.~\ref{fig_C_nz_Horizon} we compare with the numerical simulations
our results for the variance $\sigma_{N_i}$ of the halo angular number densities,
where $\sigma_{N_i}$ includes both the shot-noise and sample-variance
contributions (\ref{C-NiNj-sn-sv}),
\beq
\sigma_{N_i} = \sqrt{C_{i,i}^{\sn}+C_{i,i}^{\sv}} .
\label{sigma-Ni-def}
\eeq
We consider an angular window of 50 deg$^2$, which corresponds for
instance to the case of the XXL survey \citep{Pierre2011}.
As for the mean densities of Fig.~\ref{fig_nz_Horizon}, we obtain a reasonable
agreement with the simulations, and we correctly reproduce the dependence
on halo mass and redshift.

\begin{figure}[htb]
\begin{center}
\epsfxsize=8 cm \epsfysize=6 cm {\epsfbox{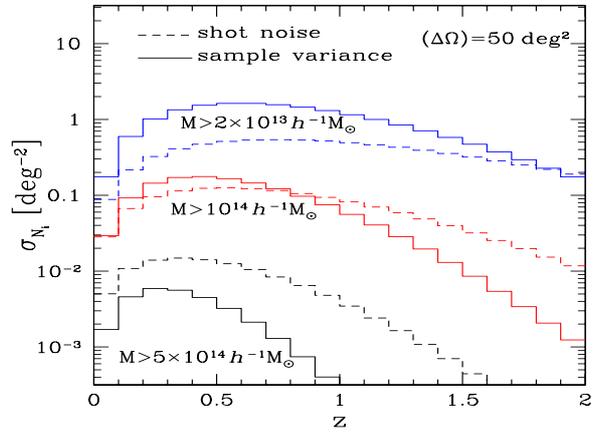}}
\end{center}
\caption{The shot-noise (dashed lines) and sample-variance (solid lines)
errors for the angular number densities shown in Fig.~\ref{fig_nz_Horizon},
associated with a redshift binning of width $\Delta z=0.1$, but up to $z=2$,
and an angular window of 50 deg$^2$.}
\label{fig_C_nz}
\end{figure}

Taking advantage of our analytical model, we compare in Fig.~\ref{fig_C_nz}
the shot-noise and sample-variance contributions to the total error that was
displayed in Fig.~\ref{fig_C_nz_Horizon}, but going up to redshift $z=2$.
Here we define
\beq
\sigma_{N_i}^{\sn} = \sqrt{C_{i,i}^{\sn}} , \;\;\; 
\sigma_{N_i}^{\sv} = \sqrt{C_{i,i}^{\sv}} .
\label{sigma-Ni-sn-sv-def}
\eeq
As expected, we can see that the error of observed number counts
$\hN_i$ is dominated by the shot-noise contribution for rare halos
(high mass or high redshift), where effects associated with the discreteness
of the halo distribution are very important.

\paragraph{Signal-to-noise ratio}
\label{signal-to-noise}
~~\\

\begin{figure}[htb]
\begin{center}
\epsfxsize=8 cm \epsfysize=6 cm {\epsfbox{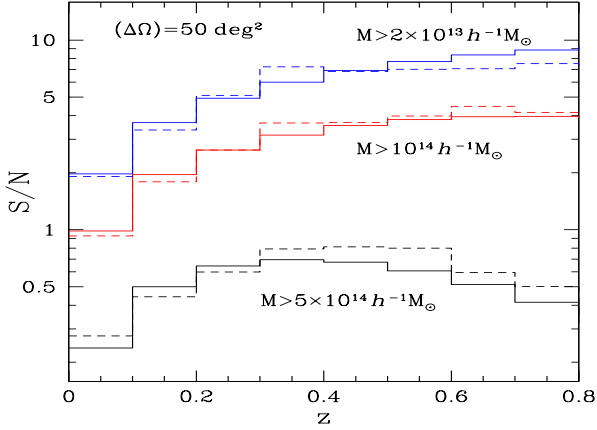}}
\end{center}
\caption{The signal-to-noise ratios of number counts for an angular area
$\Delta\Omega=50$ deg$^2$, as in Figs.~\ref{fig_nz_Horizon} and
\ref{fig_C_nz_Horizon}. We compare our analytical results
(solid lines) with numerical simulations (dashed lines).}
\label{fig_SN_Nz_Horizon}
\end{figure}

From the angular number density $\lag\hN_i\rag$ and its variance $\sigma_{N_i}$
we define the signal-to-noise ratio as
\beq
\frac{S}{N} = \frac{\lag\hN_i\rag}{\sigma_{N_i}} ,
\label{SN-def}
\eeq
which we compute from Eqs.(\ref{Ni-3}) and (\ref{sigma-Ni-def}).
Thus, combining Figs.~\ref{fig_nz_Horizon} and \ref{fig_C_nz_Horizon}, we 
display in Fig.~\ref{fig_SN_Nz_Horizon} this signal-to-noise ratio.
In agreement with these previous figures, we obtain a good match to the numerical
simulations. 
Thus, the analytical results are competitive with the numerical simulations
since they appear to be no less reliable and much faster to compute.

\paragraph{Scalings with survey area and number of subfields}
\label{Scalings-Nz}
~~\\

For practical purposes it is interesting to compare the signal-to-noise ratios
associated with a different number of subfields at fixed total area
$\Delta\Omega$, since this can help for choosing the best observational strategy,
whether one should perform a single wide-field survey or several smaller scale
surveys.

\begin{figure}[htb]
\begin{center}
\epsfxsize=8 cm \epsfysize=6 cm {\epsfbox{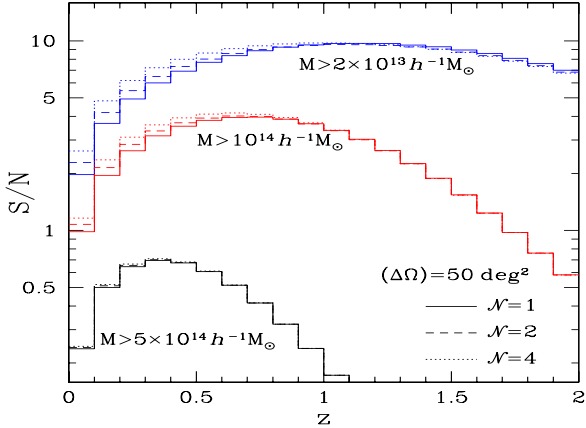}}
\end{center}
\caption{The signal-to-noise ratios of number counts for a total angular area
$\Delta\Omega=50$ deg$^2$, divided over $\cN$ independent subfields.
We show the results obtained for the numbers of subfields
$\cN=1$ (solid lines), $2$ (dashed lines), and $4$ (dotted lines).}
\label{fig_SN_Nz}
\end{figure}

Therefore, let us consider a survey with a total angular window of area
$\Delta\Omega$ that can be split over $\cN$ angular subfields, which we
assume to be independent and to have equal area  $\Delta\Omega/\cN$. 
For instance, the survey may be made of $\cN$ smaller regions that are well
separated on the sky.
Then, the total angular number density $\hN^{\rm tot}_i$
of objects in the redshift bin $[\zim,\zip]$, summed over the $\cN$ 
smaller subfields of index $\alpha$ with angular number densities
$\hN^{(\alpha)}_i$, writes as
\beq
\hN^{\rm tot}_i = \frac{1}{\cN} \sum_{\alpha=1}^{\cN} \hN^{(\alpha)}_i .
\label{Ntot-def}
\eeq
Since all subfields are independent and have the same depth we have,
for any $\alpha$,
\beq
\lag \hN^{\rm tot}_i \rag = \lag\hN^{(\alpha)}_i\rag ,
\label{Ntot-1}
\eeq
which depends neither on $(\Delta\Omega)$ nor $\cN$.
Without mass binning the covariance matrix remains diagonal, see Eq(\ref{Cij-7}),
with a shot-noise contribution
\beq
C^{\rm tot,(s.n.)}_{i,i} = \frac{1}{\cN} \, C^{\rm (\alpha),(s.n.)}_{i,i} = 
\frac{\lag \hN^{\rm tot}_i \rag}{(\Delta\Omega)}
\propto (\Delta\Omega)^{-1} ,
\label{CN-tot-sn}
\eeq
while the sample-variance contribution is
\beq
C^{\rm tot,(s.v.)}_{i,i} = \frac{1}{\cN} \, C^{\rm (\alpha),(s.v.)}_{i,i} \propto 
\xicylalpha/\cN ,
\label{CN-tot-mf}
\eeq
where $\xicylalpha$ is the typical value of the integral
(\ref{I-thetas-def}) within the small angular window $(\Delta\Omega)/\cN$.
Then, the signal-to-noise ratio scales as
\beq
\frac{S}{N} = \frac{\lag \hN^{\rm tot}_i \rag}{\sqrt{C^{\rm tot,}_{i,i}}} \propto 
\sqrt{\frac{(\Delta\Omega)}{1+\frac{(\Delta\Omega)}{\cN} \xicylalpha}} .
\label{SN-Ntot}
\eeq

In the regime where the covariance is dominated by the shot-noise contribution,
the signal-to-noise ratio grows as the square root of the total area and
does not depend on the number of subfields, 
$(S/N) \propto \sqrt{\Delta\Omega}$.

To estimate the scaling in the regime where the covariance is dominated by the
sample-variance contribution, we assume that on the relevant scale $k \sim
1/(\cD\theta_s)$, that is, $k \sim \sqrt{\cN/(\cD^2\Delta\Omega)}$, the
power spectrum behaves as $P(k) \sim k^n$, with $-2<n<1$. (Wavenumbers
where $n<-2$ for CDM cosmologies would correspond to very small angular
windows.)
Then, from Eq.(\ref{I-thetas-def}) we obtain
$\xicylalpha \sim k^{n+2} \sim (\cN/\Delta\Omega)^{(n+2)/2}$,
and the signal-to-noise ratio scales as
$(S/N) \sim \cN^{-n/4} \, (\Delta\Omega)^{(n+2)/4}$.
Thus, in this regime the signal-to-noise ratio still grows with the total
survey area, but there is a weak dependence on the number of subfields,
which may either increase or decrease with $\cN$ depending on the sign of $n$.

For illustration, we show in Fig.~\ref{fig_SN_Nz} the signal-to-noise ratios
of the number counts obtained for a total angular area $\Delta\Omega=50$
deg$^2$, divided over $\cN$ subfields with $\cN=1$, $2$, and $4$.
The case of a single field, $\cN=1$, corresponds to Figs.~\ref{fig_nz_Horizon} and
\ref{fig_C_nz_Horizon}. The curves in Fig.~\ref{fig_SN_Nz} are exact estimates
of the signal-to-noise ratios, obtained from
$\lag \hN^{\rm tot}_i \rag/\sqrt{C^{\rm tot,}_{i,i}}$ and not from the
approximate scaling given in Eq.(\ref{SN-Ntot}).

In agreement with the discussion above and with Fig.~\ref{fig_C_nz_Horizon},
we can check that at high redshift or for high mass, the signal-to-noise ratio
does not depend on $\cN$, since the error is dominated by the shot-noise
contribution, which only depends on the total area as seen in 
Eqs.(\ref{CN-tot-sn}) and (\ref{SN-Ntot}).

At low redshift and low mass, where the error is dominated by the
sample-variance contribution, the signal-to-noise ratio increases slightly with
$\cN$. This can be understood from the fact that the local slope $n$ of the
power spectrum is slightly negative on the scales where the halo correlation
is significant. For instance, within the $\Lambda$CDM cosmology that we
consider in this paper, a window of area $50$ deg$^2$ corresponds at $z=1$
to a radius $\cD\theta_s=164 h^{-1}$Mpc, and the local slope $n(k)$ of the linear
power spectrum, at wavenumber $k \simeq 2\pi/(164 h^{-1}{\rm Mpc})$, is
$n \simeq -0.6$.
This means that for the number counts at low redshift and low masses, it
is slightly advantageous to choose a survey divided over several independent
subfields.

As shown by Figs.~\ref{dndz-horizon-scaling-domega} and
\ref{dndz-horizon-scaling-n06} in App.~\ref{scaling-horizon}, these scalings 
are approximately satisfied by the results obtained from numerical simulations,
for a wide variety of survey area and of number of subfields.
Therefore, the scalings derived from Eq.(\ref{SN-Ntot}) allow a reasonable
estimate of the dependence of the signal-to-noise ratio of number counts
with $(\Delta\Omega)$ and $\cN$.

\subsubsection{Large angular windows}
\label{Large-angular-windows}

\begin{figure}[htb]
\begin{center}
\epsfxsize=8 cm \epsfysize=6 cm {\epsfbox{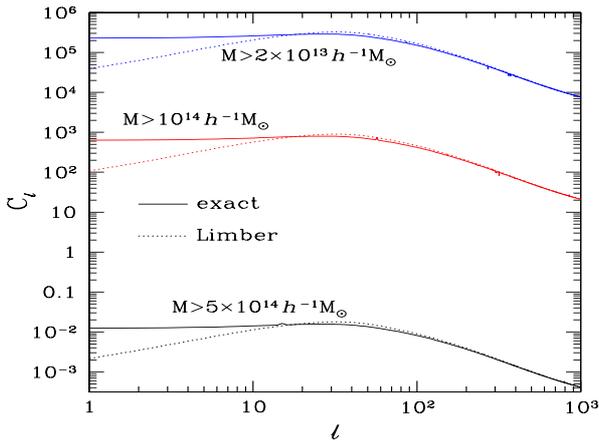}}
\end{center}
\caption{The angular power spectrum of the distribution of halos in the
redshift bin $0.95<z<1.05$. We plot both the exact result (\ref{Cl-noflat})
(solid line) and Limber's approximation (\ref{Cl-flat}) (dotted line).}
\label{fig_Cl_Nz}
\end{figure}

Expression (\ref{Cij-7}) relies on the approximation of small angular windows
for the sample-variance contribution. This allowed us to use
the flat-sky approximation (\ref{Cij-3}), where the observational cone over
the redshift bin $[\zim,\zip]$ is approximated as a cylinder of radius
$\theta_s$ around the central line of sight.
For large angular windows this approximation is no longer valid and we must
decompose over spherical harmonics \citep{Hu2003}, rather than over the plane
waves of Eq.(\ref{Cij-3}).

Rather than using the Eqs.(\ref{Cij-sv-1}) or (\ref{Cij-2}), with a new
expression for $\xiconzj(z)$ that would remain valid for large angles,
it is more convenient to go back to the angular number densities $\hN$,
as in Eq.(\ref{dndz}). 
Thus, for any redshift bin $i$ we expand the observed distribution
$\hN_i(\vOm)$ on the sky over the spherical harmonics,
\beq
\hN_i(\vOm) = \sum_{\ell,m} \hN_i^{(\ell,m)} \, Y_{\ell}^m(\vOm) ,
\label{Ni-lm}
\eeq
and we define the angular power spectrum as
\beq
\lag \hN_i^{(\ell,m)*} \hN_j^{(\ell',m')} \rag_c^{\sv} = \delta_{\ell,\ell'} \,
\delta_{m,m'} \, C_{i,j;\ell} ,
\label{Cl-def}
\eeq
where we only take the sample-variance contribution.
To simplify notations we do not include mass binning, but this can be
added without difficulty, as in Sect.~\ref{small-angles}.
Then, writing the two-point correlation function again under the factored
form (\ref{xij-bb}), introducing the Fourier-space power spectrum as
in Eq.(\ref{xi-Pk}) and expanding the plane-wave exponential factor over spherical
harmonics, a standard calculation gives \citep{Hu2000,Hu2003}
\beqa
C_{i,j;\ell} & = & 4\pi \int \frac{\dd k}{k} \Delta^2_{L0}(k)
\int_i \dd\chi \, \chi^2 \, \bb \, \nb \, D_+ \, j_{\ell}(k\chi) \nonumber \\
&& \times \int_j \dd\chi' \, \chi'^2 \, \bb' \, \nb' \, D'_+ \, j_{\ell}(k\chi')  ,
\label{Cl-noflat}
\eeqa
where $j_{\ell}$ is the spherical Bessel function of order $\ell$.
Here we assumed for simplicity a flat background, which is sufficient for
practical purposes, and we approximated the Fourier-space power spectrum
by the linear power spectrum, $P(k,z) \simeq D_+(z)^2 P_{L0}(k)$, where
$D_+(z)$ is the linear growth rate (normalized to unity at $z=0$).

Limber's approximation can be recovered in the limit of large $\ell$,
for slowly varying $k$-dependent prefactors, by using the property
$\int \dd k \, k^2 j_{\ell}(k\chi) j_{\ell}(k\chi') = \pi/(2\chi^2)\delta_D(\chi-\chi')$,
and the correspondence $k\leftrightarrow (\ell+1/2)/\chi$
\citep{Hu2003,LoVerde2008a}. This yields
\beq
C_{i,j;\ell}^{\rm Limber} = \delta_{i,j} \int \dd \chi \, \chi^5 \, \bb^2 \, \nb^2 \,
\frac{2\pi^2}{(\ell+1/2)^3} \, \Delta^2\!\left(\frac{\ell+1/2}{\chi}\right) .
\label{Cl-flat}
\eeq
Nevertheless, because the structures of Eqs.(\ref{Cl-noflat}) and (\ref{Cij-2}) 
are quite different (the order of the integrations over redshift and wavenumber
is exchanged, and the large-angle expression keeps two integrations over
redshift, while in the small-angle expression one integral over redshift has already
been performed) it is more convenient to treat the small-angle and
large-angle derivations separately.

We plot in Fig.~\ref{fig_Cl_Nz} the angular power spectrum $C_{i,i;\ell}$ obtained
for halos above three mass thresholds in the redshift bin $0.95<z<1.05$.
We can see that Limber's approximation (\ref{Cl-flat}) significantly
underestimates the power at low $\ell$, while it slightly overestimates the power
at high multipoles, $\ell > 20$. It is already rather good at $\ell \sim 20$,
and becomes increasingly accurate at higher $\ell$, although the difference
remains on the order of $10\%$ until $\ell \sim 80$.
These results agree with previous studies of the Limber approximation
\citep{LoVerde2008a,Crocce2011}.
As noticed in \citet{LoVerde2008a}, the latter can be
extended as a series expansion over $(\ell+1/2)^{-1}$, but higher orders 
behave increasingly badly at low $\ell$. Therefore, we do not investigate further
this approach here, since using the exact expression (\ref{Cl-noflat}) is not
more difficult (but slower) to compute and ensures a smooth behavior
over all $\ell$, while the usual Limber approximation (\ref{Cl-flat}) is sufficient
for our purposes on small scales.

\begin{figure}[htb]
\begin{center}
\epsfxsize=8 cm \epsfysize=6 cm {\epsfbox{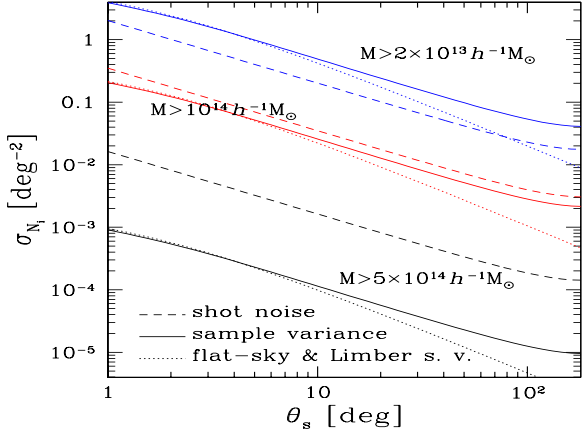}}
\end{center}
\caption{The shot-noise (dashed line) and sample-variance errors
(\ref{sigma-Ni-sn-sv-def}) for the angular number densities in the redshift
bin $0.95<z<1.05$, as a function of the radius $\theta_s$ of the angular window.
The solid line is the exact sample variance, from Eqs.(\ref{Cl-noflat}) and
(\ref{Cii-W2}), while the dotted line is the result (\ref{Cij-7}), which was used
in Fig.~\ref{fig_C_nz} and involves both the flat-sky and Limber's approximations.}
\label{fig_C_nz_noflat}
\end{figure}

Next, the mean angular number densities $\hN_i$ of Eq.(\ref{Ni-1}), smoothed
over the angular window of radius $\theta_s$ and filter $W_2(\vOm)$, read as
\beq
\hN_i = \int\dd\vOm \, \hN_i(\vOm) \, W_2(\vOm) ,
\label{Ni-W2}
\eeq
where $W_2(\vOm)=1/(\Delta\Omega)$ within the angular window
and vanishes outside (but we can choose more general filters).
Then, the sample variance of these number counts writes as
\citep{Hu2003}
\beq
C_{i,j}^{\sv} = \lag \hN_i \hN_j \rag_c^{\sv} = 
\sum_{\ell,m} C_{i,j;\ell} \; | \tW_2^{(\ell,m)}|^2 ,
\label{Cii-W2}
\eeq
where $C_{i,j;\ell}$ is the angular power spectrum (\ref{Cl-noflat}), while
$\tW_2^{(\ell,m)}$ are the angular multipoles of the window $W_2$,
\beq
\tW_2^{(\ell,m)} = \int \dd\vOm \, W_2(\vOm) \, Y_{\ell}^m(\vOm)^* .
\label{tW2-lm-def}
\eeq

For a top-hat window that is symmetric around the azimuthal axis,
we have for $\ell \geq 1$,
\beqa
\tW_2^{(\ell,0)} & = & \frac{2\pi}{(\Delta\Omega)} \int_0^{\theta_s}
\dd \theta \; \sin\theta \; Y_{\ell}^0(\theta) \\
& = & \sqrt{\frac{\pi}{2\ell+1}} \, 
\frac{P_{\ell-1}(\cos\theta_s) -  P_{\ell+1}(\cos\theta_s)}{(\Delta\Omega)}
\label{W2-l0}
\eeqa
and
\beq
\tW_2^{(0,0)} = \frac{1}{2\sqrt{\pi}} , \hspace{0.3cm} \mbox{and}
\hspace{0.3cm} (\Delta\Omega) = 2\pi (1-\cos\theta_s) ,
\label{W2-00}
\eeq
where $P_{\ell}$ are the Legendre polynomials, and $\tW_2^{(\ell,m)} = 0$
for $m\neq 0$.

In the limit of large $\ell$, $Y_{\ell}^0(\theta) \simeq 
\sqrt{\frac{\ell+1/2}{2\pi}} J_0[(\ell+1/2)\theta]$ \citep{Hu2000}, and
we obtain $|\tW_2^{(\ell,0)}|^2 \simeq \frac{k\chi}{2\pi} \tW_2(k\chi\theta_s)^2$,
with $k=(\ell+1/2)/\chi$ and $\tW_2$ the 2D Fourier-space window
(\ref{W-thetas}). This shows that for small angles, where the covariance
is dominated by large $\ell$, the expression (\ref{Cii-W2}) goes to
the flat-sky approximation (\ref{Cij-7}), using
the fact that Limber's approximation (\ref{Cl-flat}) also applies in
this limit (see Fig.~\ref{fig_Cl_Nz}).

We plot in Fig.~\ref{fig_C_nz_noflat} the shot-noise and sample-variance
errors $\sigma_i=\sqrt{C_{i,i}}$, as in Fig.~\ref{fig_C_nz_Horizon} but as a function
of the angular radius $\theta_s$, for the angular number densities in the redshift
bin $0.95<z<1.05$.  
In agreement with Fig.~\ref{fig_Cl_Nz}, we can check that the combination (\ref{Cij-7}) 
of the flat-sky \& Limber's approximations provides a good approximation to the
exact result (\ref{Cii-W2}) on small angles, typically $\theta_s < 10$ deg,
while it underestimates the sample variance on large angles.
In agreement with Fig.~\ref{fig_C_nz}, the shot-noise contribution is dominant
for massive and rare halos, and subdominant for small and numerous halos.
In our case, the transition between the shot-noise and sample-variance
dominated regimes takes place at $M_* \sim 10^{14} h^{-1} M_{\odot}$.

\subsubsection{Accuracy of the "flat-sky + Limber" approximation}
\label{accuracy-flat-sky-Limber}

\begin{figure}[htb]
\begin{center}
\epsfxsize=8 cm \epsfysize=6 cm {\epsfbox{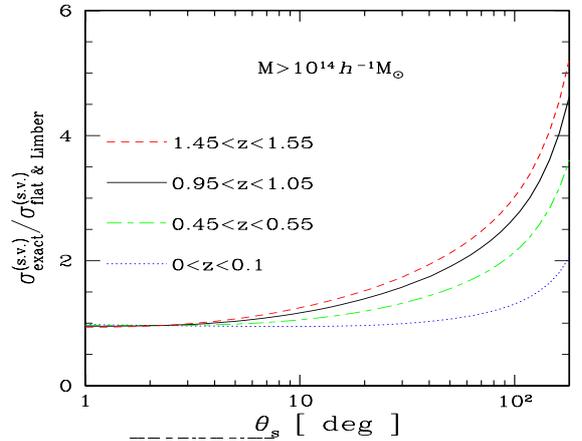}}
\end{center}
\caption{The ratio $\sigma^{\sv}_{\rm exact}/\sigma^{\sv}_{\rm flat \& Limber}$
of the exact sample-variance error (\ref{Cii-W2}) to the approximation (\ref{Cij-7}),
which uses both the flat-sky and Limber's approximations.
We show this ratio as a function of the radius $\theta_s$ of the angular window,
for several redshift bins, for halos above the mass threshold $M>10^{14} 
h^{-1} M_{\odot}$. Higher $z$ corresponds to a higher ratio.}
\label{fig_r_nz_noflat}
\end{figure}

We plot in Fig.~\ref{fig_r_nz_noflat} the ratio of the exact sample-variance error
(\ref{Cii-W2}) to the approximation (\ref{Cij-7}), which used both the flat-sky and Limber's approximations. As in Figs.~\ref{fig_Cl_Nz} and \ref{fig_C_nz_noflat},
we can check that the approximation (\ref{Cij-7}) is reliable for small angular
windows but significantly underestimate the sample-variance error for wide angles,
above $10$ deg.
In the extreme case of full-sky surveys ($\theta_s=180$ deg), it can underestimate
the sample-variance error by a factor from 2 to 5. The effect is actually greater
for higher redshift bins. This may seem somewhat surprising since the ``flat-sky''
approximation (\ref{Cij-3}) is expected to be more accurate at higher redshifts,
where large angles $\theta$ correspond to large distances $\cD\theta$ that are 
weakly correlated and should not significantly contribute (i.e., the CDM power 
spectrum itself yields more weight to pairs separated by small angles).
However, Limber's approximation (\ref{flat-Limber}) goes in the opposite direction
because it relies on the assumption $\Delta\chi \gg \cD \theta_s$ (i.e., longitudinal
wavenumbers are more suppressed by the integration along the line of sight
than transverse wavenumbers, which are only integrated over the smaller
angular distance $\cD \theta_s$). 
For instance, for an Einstein-de Sitter universe,
where $\chi=\cD=2c/H_0[1-(1+z)^{-1/2}]$, this constraint on the angle
$\theta_s$ and the redshift bin width $\Delta z$ writes as
$\theta_s \ll 29 (\Delta z) [(1\!+\!z)^{3/2}\!-\!(1\!+\!z)]^{-1}$ deg. At fixed
$\Delta z$ this upper bound on $\theta_s$ becomes stronger at higher $z$.
Therefore, Fig.~\ref{fig_r_nz_noflat} shows that this second effect, associated
with Limber's approximation, dominates over the first effect, associated with the
flat-sky approximation.

We checked that we obtain very close results for other
mass thresholds (not shown in the figure). For instance, the curves obtained
for the mass threshold $2\times 10^{13}$ and 
$5\times 10^{14}h^{-1} M_{\odot}$ cannot be distinguished from those
plotted in Fig.~\ref{fig_r_nz_noflat}. This is not surprising since within our
bias model (\ref{xij-bb}) the halo correlations are governed by the same
matter density correlation function $\xi(r;z)$.

Figure~\ref{fig_r_nz_noflat} shows that the small-angle approximation (\ref{Cij-7})
that we used in Sect.~\ref{small-angles} was legitimate, since we considered
angular windows of $50$ deg$^2$ or less (i.e. $\theta_s \la 4$ deg).

\subsubsection{Correlation between different redshift bins}
\label{correlation-redshift}

\begin{figure}[htb]
\begin{center}
\epsfxsize=8 cm \epsfysize=6 cm {\epsfbox{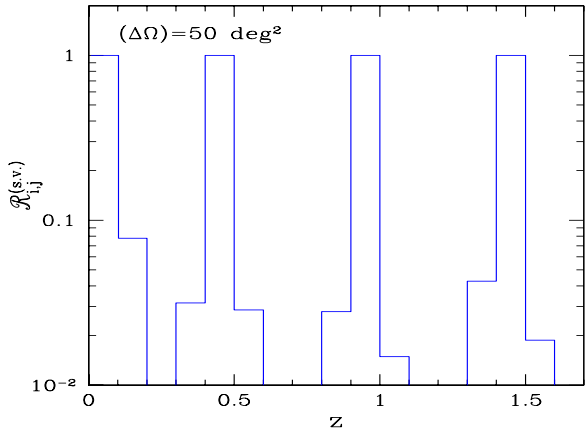}}
\end{center}
\caption{The correlation matrix $\cR^{\sv}_{i,j}$ of Eq.(\ref{correlation-matrix})
between redshift bins of width $\Delta z=0.1$. We show $\cR^{\sv}_{i,j}$ as a
function of $j$, for four values of $i$. In each case, $\cR^{\sv}_{i,j}=1$ at $j=i$.
We consider halos above $10^{14} h^{-1} M_{\odot}$ and an angular window
of area $(\Delta\Omega)=50$ deg$^2$.}
\label{fig_Rn_ij}
\end{figure}

We can use the expression (\ref{Cii-W2}) to compute the correlation between
different redshift bins $i$ and $j$. Thus, we show in Fig.~\ref{fig_Rn_ij} the
correlation matrix $\cR^{\sv}_{i,j}$ (also called normalized covariance matrix) 
defined as
\beq
\cR^{\sv}_{i,j} = \frac{C^{\sv}_{i,j}}{\sqrt{C^{\sv}_{i,i} C^{\sv}_{j,j}}} ,
\label{correlation-matrix}
\eeq
where we only consider the sample-variance contribution. (The shot-noise
contribution (\ref{Cij-sn}) is always diagonal for nonoverlapping redshift bins.)
Thus, $\cR^{\sv}_{i,j}$ is unity along the diagonal and elements $\{i,j\}$ where
$\cR^{\sv}_{i,j}$ is much smaller than one are weakly correlated.
We can check that the decay is always rather fast and correlations between 
neighboring redshift bins, $j=i\pm1$, are already below $10\%$.
This shows that it is appropriate to neglect
cross-correlations between redshift bins of width $\Delta z=0.1$, as we did
in Sect.~\ref{small-angles}.
We also checked that we obtain almost identical results for other angular
windows, such as $(\Delta\Omega)=400$ deg$^2$. (For very large or full-sky
surveys we do not need the approximation of uncorrelated redshift bins
since we use Eq.(\ref{Cii-W2}).)

\section{Real-space two-point correlation function}
\label{Two-point-correlation}

In the previous section we have studied the covariance of the estimators
$\hN_i$, which measure the redshift distribution $\dd n/\dd z$ of the
population of interest (galaxies, clusters, etc.), over a set of finite redshift bins.
This corresponds to one-point statistics. We now study estimators of the
real-space two-point correlation function $\xi(x_{12};M_1,M_2;z)$ of these objects,
which corresponds to two-point statistics, as a function of the comoving
distance $x_{12}$.
In this article we do not investigate redshift-space distortions, which we leave
for future works, and we assume that a real-space map of the population
under study is available or that redshift distortions can be neglected.

Estimators of 3D correlation functions, or power spectra, have already been
studied in many works, mostly in view of their application to galaxy surveys.
However, since we have in mind the application to cluster surveys and,
more generally, to deep surveys of rare objects, we consider
3D correlation functions averaged over a wide redshift bin (in order to
accumulate a large enough number of objects), rather than the usual
local 3D correlation functions at a given redshift.
This means that the quantities that we consider in this section, while being
truly 3D correlations and not 2D angular correlations, nevertheless involve
integrations along the line of sight or, more precisely, the observational cone,
within a finite redshift interval.
This is also why 3D Fourier-space power spectra may not be the most
convenient tool for our purposes, since we do not have homogeneous
and isotropic distributions since the radial direction plays a special  role.

\subsection{Mean correlation}
\label{Mean correlation}

\subsubsection{Peebles \& Hauser estimator}
\label{Simple-estimator}

Following \citet{Peebles1974a}, a simple estimator $\hxi$ for the two-point
correlation function of a point distribution is given by
\beq
\hxi = \frac{DD}{RR} - 1 ,
\label{hxi-DR}
\eeq
where $D$ represents the data field and $R$ an independent Poisson
distribution, both with the same mean density.
More precisely, the estimator $\hxi_i$ introduced in (\ref{hxi-DR}) for the mean
correlation over the radial bin $[\Rim,\Rip]$ corresponds to counting all
pairs ``DD'' in the data field that fall in this pair-separation bin $i$ and
all pairs ``RR'' in the auxiliary Poisson field that fall in the same bin, and to taking
the ratio of these two counts.

Before appropriate rescaling, the mean number density of the actual Poisson
process $R$ is taken as much higher than the observed one, so that the
contribution from fluctuations of the denominator $RR$ to the noise of $\hxi$
can be ignored.
The advantage of form (\ref{hxi-DR}) is that one automatically includes
the geometry of the survey (including boundary effects, cuts, etc.),
because the auxiliary field $R$ is drawn on the same geometry.

In our case, we write the analog $\hxi_i$ of Eq.(\ref{hxi-DR}) for the mean
correlation on scales delimited by $\Rim$ and $\Rip$, integrated over some
redshift range and mass intervals, as
\beqa
1+\hxi_{i;\alpha,\beta} & = & \frac{1}{\QQ_{i;\alpha,\beta}} \int \dd z \, 
\frac{\dd\chi}{\dd z} \, \cD^2 \int\frac{\dd\vOm}{(\Delta\Omega)} 
\int_{\alpha} \frac{\dd M}{M} \nonumber \\
&& \times \,  \int_{\Rim}^{\Rip} \dd\vr' \int_{\beta} \frac{\dd M'}{M'}
\; \frac{\dd\hn}{\dd\ln M} \frac{\dd\hn}{\dd\ln M'} ,
\label{xi-1}
\eeqa
with
\beqa
\QQ_{i;\alpha,\beta} & = & \int\dd z \, \frac{\dd\chi}{\dd z} \, \cD^2
\int\frac{\dd\vOm}{(\Delta\Omega)} \int_{\alpha} \frac{\dd M}{M} 
\int_{\Rim}^{\Rip} \dd\vr' \int_{\beta} \frac{\dd M'}{M'}  \nonumber \\
&& \times \, \frac{\dd n}{\dd\ln M} \frac{\dd n}{\dd\ln M'} .
\label{QQi-def}
\eeqa
Here we denoted $\int_{\Rim}^{\Rip} \dd\vr'$ as the integral over the 3D 
spherical shell of radii $\Rim<\Rip$, and $\int_{\alpha}$ and $\int_{\beta}$
are the integrals over the mass bins $\alpha$ and $\beta$.

The redshift interval $\Delta z$ is not necessarily small, and to increase the
statistics we can choose the whole redshift range of the survey, such as $[0,z_s]$.
If we bin the survey over smaller nonoverlapping redshift intervals, which are
large enough to neglect cross-correlations between different bins 
(see for instance Fig.~\ref{fig_Rn_ij}), we can independently study
each redshift bin.
For simplicity we do not explicitly write the redshift boundaries.

As in Eq.(\ref{hxi-DR}), the counting method that underlies Eq.(\ref{xi-1}) can
be understood as follows \citep{Peebles1974a}.
We span all objects in the ``volume'' $(z,\vOm,\ln M)$, and count all
neighbors at distance $r'$, within the shell $[\Rim,\Rip]$, with a mass
$M'$. We denote with unprimed letters the quantities associated with the
first object, $(z,\vOm,\ln M)$, and with primed letters the quantities associated
with the neighbor of mass $M'$ at distance $r'$.
Thus, with obvious notations, $\dd\hn/\dd\ln M$ and $\dd\hn/\dd\ln M'$
are the observed number densities at the first and second (neighboring) points.
The difference between the quantities $(1+\hxi)$ and $\QQ$ is that
in the latter case we use the mean number densities $\dd n/\dd\ln M$ and
$\dd n/\dd\ln M'$. Therefore, $\QQ$ is not a random quantity so it shows
no noise. In practice, the mean number densities $\dd n/\dd\ln M$ may actually
be measured from the same survey, as described in Sect.~\ref{Number-density}.
However, since these measures do not involve a distance binning over $r'$,
there are many more objects in a redshift bin than
within a small interval $[\Rim,\Rip]$. Then, the one-point quantities
$\lag\hN\rag$ are measured with much greater accuracy than $\lag\hxi\rag$,
so that we can indeed neglect their contribution to the noise of the estimator
$\hxi$. In terms of Eq.(\ref{hxi-DR}) this corresponds to neglecting fluctuations
of ``RR''. (This is achieved in practice by choosing a much higher density
for the field $R$, which is later rescaled.)

In Eq.(\ref{xi-1}) we used a simple average over the shell $[\Rim,\Rip]$,
because we count all pairs with a uniform weight in $\vr'$-space. Through
the change to spherical coordinates $\dd\vr'= \dd r' r'^2\dd\vOm'$, this yields
a geometrical weight $r'^2$ in terms of the radial distance $r'$. 
An alternative would be to add a weight $r'^{-2}$, instead
of the simple 3D top-hat written in Eq.(\ref{xi-1}), to eventually obtain a uniform
weight over the radial distance $r'$.
For simplicity we only consider choice (\ref{xi-1}) in the following, but
such alternative weights could be used with straightforward modifications
in the expressions given below.

Thus, we focus on the behavior of the two-point correlation as a function of
distance $r'$, measured through the binning over the intervals $[\Rim,\Rip]$.
We assume that different bins do not overlap,
$\Rip \leq R_{i+1,-}$, and in practice one usually has $\Rip = R_{i+1,-}$,
to cover a continuous range of scales. On the other hand, these intervals
may depend on redshift, as long as  $\Rip(z) \leq R_{i+1,-}(z)$ at each redshift.

Using Eq.(\ref{Nbz-def}), Eq.(\ref{QQi-def}) also writes as
\beq
\QQ_{i;\alpha,\beta} = \int \dd\chi \, \cD^2 \, \nb_{\alpha} \nb_{\beta} \, \cV_i ,
\label{QQ-1}
\eeq
where the volume $\cV_i$ of the $i$-shell is
\beq
\cV_i(z) = \frac{4\pi}{3} [\Rip(z)^3-\Rim(z)^3] ,
\label{Vi-def}
\eeq
which may depend on $z$. In practice, one would usually choose constant
comoving shells, so that $\cV_i$ does not depend on $z$.
To obtain Eq.(\ref{QQ-1}) we used that $\dd n/\dd\ln M$ and
$\dd n/\dd\ln M'$ have no scale dependence (because they correspond to a
uniform distribution of objects) and we neglected edge
effects. (These finite-size effects are discussed and evaluated in
App.~\ref{Finite-size}.)

Because of the finite distance $r'$ between the two objects $M$ and $M'$
in Eq.(\ref{xi-1}), there is no shot-noise contribution to the average
of the quadratic term $(\dd\hn/\dd\ln M)\times(\dd\hn/\dd\ln M')$.
Within the framework presented for Eqs.(\ref{hNi})-(\ref{hn-alpha}),
the integration in Eq.(\ref{xi-1}) does not contain common small (infinitesimal)
cells, because of the finite-size distance $r'>\Rim$. Therefore, the
average of the statistical estimator (\ref{xi-1}) reads as
\beqa
1\!+\!\lag\hxi_{i;\alpha,\beta}\rag & \! = \! & \frac{1}{\QQ_{i;\alpha,\beta}}
\int\!\! \dd\chi \, \cD^2 \int\!\! \frac{\dd\vOm}{(\Delta\Omega)}
\int_{\alpha}\frac{\dd M}{M} \int_i \dd\vr' \int_{\beta} \frac{\dd M'}{M'} \nonumber \\
&& \times \, \frac{\dd n}{\dd\ln M} \frac{\dd n}{\dd\ln M'}
\, [1+\xih(r';M,M';z) ] ,
\label{xi-2}
\eeqa
where $\xih(r';M,M';z)$ is the two-point correlation function of the objects,
as in Eq.(\ref{Ni-Nj-2}).
Here we denoted $\int_i \dd\vr'$ as the integral $\int_{\Rim}^{\Rip}\dd\vr'$ over
the 3D spherical shell $i$. 
Comparing with Eq.(\ref{QQ-1}) we clearly see that
$\hxi_i$ is an unbiased estimator of the two-point correlation function $\xih$,
averaged over the shell $[\Rim,\Rip]$ (with a geometrical weight $r'^2$),
whence the name ``$\hxi$''.

As in Eq.(\ref{QQ-1}), in Eq.(\ref{xi-2}) and in the following we neglect finite-size
effects, which arise because the integration over $\vr'$ should be restricted to
the observational
cone of the survey. This leads to a smaller available volume than the spherical
shell $[\Rim,\Rip]$ close to the survey boundaries. This does not affect the mean
value of the estimator $\hxi_i$, because this effect cancels out between the
numerator of Eq.(\ref{xi-2}) and the denominator $\QQ_i$. However, it will have
a small effect on our estimate of the covariance matrix.
As described in App.~\ref{Finite-size}, at $z=1$ for a circular survey area
$\Delta\Omega=50$deg$^2$, and for a radial bin at $r=30 h^{-1}$Mpc,
by geometrical counting we overestimate the number of pairs by $10\%$
and the signal-to-noise ratio by $5\%$.

As in Sect.~\ref{covariance-counts}, in order to make progress we assume that
the two-point correlation can be factored in as in Eq.(\ref{xij-bb}), so that
Eq.(\ref{xi-2}) reads as
\beq
\lag\hxi_{i;\alpha,\beta}\rag = \frac{1}{Q_{i;\alpha,\beta}} \int\dd\chi \, \cD^2 \, 
\bb_{\alpha} \bb_{\beta} \, \nb_{\alpha} \nb_{\beta} \, \cV_i \, \overline{\xir_{i'}}(z) ,
\label{xi-3}
\eeq
with
\beq
\overline{\xir_{i'}}(z) = \int_i\frac{\dd\vr'}{\cV_i} \, \xi(r';z) .
\label{xi-i-i'-def}
\eeq
We have introduced the superscript ``$(r)$'' to recall that Eq.(\ref{xi-i-i'-def})
is the radial average of $\xi$, over the 3D spherical shell associated with the
radial bin $i$, to distinguish it from the angular averages that we encounter
in Sect.~\ref{Angular-correlation} below.
The prime in the subscript ``$i'$'' also recalls that we integrate
over a neighboring point $\vr'$, with respect to a given point $(\chi,\cD\vOm)$
of the observational cone, to distinguish it from the integration over an
unrelated point within the observational cone as in the ``cylindrical'' average
(\ref{xib-ij-def}).
We give in Eq.(\ref{I3-def}) in App.~\ref{Computation-of-the-mean-of-the-estimator}
the Fourier-space expression of $\overline{\xir_{i'}}(z)$, which is more convenient
for numerical computations.

\subsubsection{Landy \& Szalay estimator}
\label{Landy-Szalay-estimator}

As shown in \citet{Landy1993}, a better estimator than (\ref{hxi-DR}) is given by
\beq
\hxiLS = \frac{DD-2DR+RR}{RR} ,
\label{hxi-LS-DR}
\eeq
which involves the product $DR$ between the data and the auxiliary field.
Within our framework, where the mean quantity $Q$ plays the role of $R$,
this second estimator reads as
\beqa
\hxiLS_{i;\alpha,\beta} & = & \frac{1}{\QQ_{i;\alpha,\beta}} \int \dd z \, 
\frac{\dd\chi}{\dd z} \, \cD^2 \int\frac{\dd\vOm}{(\Delta\Omega)} 
\int_{\alpha}\frac{\dd M}{M} \int_i \dd\vr' \int_{\beta} \frac{\dd M'}{M'}  
\nonumber \\
&& \times \, \frac{\dd\hn}{\dd\ln M} \frac{\dd\hn}{\dd\ln M'} - 2 \, 
\frac{1}{\QQ_{i;\alpha,\beta}} \int \dd z \, \frac{\dd\chi}{\dd z} \, \cD^2
\int\frac{\dd\vOm}{(\Delta\Omega)} \nonumber \\
&& \times \, \int_{\alpha} \frac{\dd M}{M} \int_i \dd\vr' 
\int_{\beta}\frac{\dd M'}{M'} \,  \frac{\dd\hn}{\dd\ln M}
\frac{\dd n}{\dd\ln M'}  + 1 .
\label{xi-LS-1}
\eeqa
The difference between the terms associated with $DD$ and $DR$ is that
in the former we have a product of two observed number densities,
$(\dd\hn/\dd\ln M)\times(\dd\hn/\dd\ln M')$, while in the latter we have
a crossproduct between the observed and the mean number densities,
$(\dd\hn/\dd\ln M)\times(\dd n/\dd\ln M')$.

As checked in App.~\ref{Computation-Landy-Szalay}, the mean of this second
estimator $\hxiLS_{i;\alpha,\beta}$ is equal to the mean of the estimator
$\hxi_{i;\alpha,\beta}$ studied in Sect.~\ref{Simple-estimator},
\beq
\lag\hxiLS_{i;\alpha,\beta}\rag = \lag\hxi_{i;\alpha,\beta}\rag  .
\label{xi-LS-2}
\eeq
To simplify the notations, in the following we do not consider binning over
mass (i.e., we independently consider the correlation functions of halos above 
some mass thresholds), so that Eq.(\ref{xi-3}) readily simplifies as
\beq
\lag\hxiLS_i\rag = \lag\hxi_i\rag = \frac{1}{Q_i} \int\dd\chi \, \cD^2 \, 
\bb^2 \, \nb^2 \, \cV_i \, \overline{\xir_{i'}}(z) ,
\label{xi-4}
\eeq
and a similar simplification holds for $Q_i$. If needed, it is not difficult to
include a mass binning in the expressions given in the following.

\subsubsection{Comparison with simulations}
\label{Comparison}

\begin{figure}
\begin{center}
\epsfxsize=8.5 cm \epsfysize=6.5 cm {\epsfbox{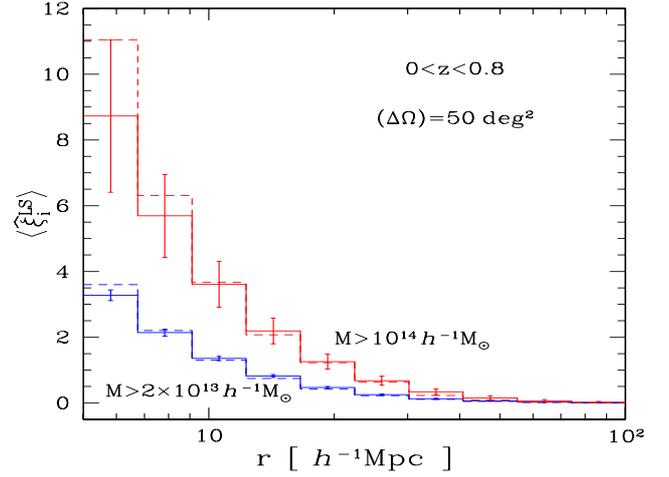}}
\end{center}
\caption{The mean halo correlation, $\lag\hxiLS_i\rag$, over ten
comoving distance bins within $5<r<100 h^{-1}$Mpc, equally spaced in $\log(r)$.
We integrate over halos within the redshift interval $0<z<0.8$ and
we compare our analytical results (solid lines) with numerical simulations
(dashed lines).}
\label{fig_XiR_Horizon}
\end{figure}

We compare in Fig.~\ref{fig_XiR_Horizon} the mean correlation (\ref{xi-4})
with results from numerical simulations (which use the Landy \& Szalay
estimator) for halos above the thresholds 
$M>2\times 10^{13}$ and $10^{14} h^{-1} M_{\odot}$, within the redshift
range $0<z<0.8$.
The error bars are the $3-\sigma$ statistical errors obtained from the
covariance matrices derived in Sect.~\ref{high-order-terms} for 34 fields
of 50 deg$^2$ as used in the simulations.
We obtain reasonable agreement
with the simulations, although we appear to underestimate the
halo correlation of the most massive halos at small radius, $r<7 h^{-1}$Mpc.
This may be due to a scale-dependent halo bias or to a small discrepancy in
the definition of the halo mass, which depends on the halo-finder algorithm
\citep{2011MNRAS.tmp..819K}.

\subsection{Covariance matrices for the halo correlation}
\label{Covariance-matrix-for-the-halo-correlation}

We now consider the covariance of the estimators $\hxi_i$ and $\hxiLS_i$.
As described in App.~\ref{Computation-of-the-covariance-of-the-estimator}, the
covariance of the Peebles \& Hauser estimator is given by
\beq
C_{i,j} = \lag\hxi_i\hxi_j\rag - \lag\hxi_i\rag \lag\hxi_j\rag = 
C_{i,j}^{(2)} +  C_{i,j}^{(3)} + C_{i,j}^{(4)} ,
\label{xii-xij-2}
\eeq
with (see also \citet{Landy1993} for a computation of low-order terms)
\beqa
C_{i,j}^{(2)} & = & \delta_{i,j} \, \frac{2}{(\Delta\Omega) \QQ_i^2} \int \dd\chi_i \, 
\cD_i^2 \frac{\dd\vOm_i}{(\Delta\Omega)} \frac{\dd M_i}{M_i} \int_i \dd\vr_{i'} 
\frac{\dd M_{i'}}{M_{i'}} \nonumber \\
&& \times \, \frac{\dd n}{\dd\ln M_i} \frac{\dd n}{\dd\ln M_{i'}}
\left[ 1+\xih_{i,i'} \right] ,
\label{C2-def}
\eeqa
\beqa
C_{i,j}^{(3)} & = & \frac{4}{(\Delta\Omega) \QQ_i\QQ_j} \int \dd\chi_i \, \cD_i^2
\frac{\dd\vOm_i}{(\Delta\Omega)} \frac{\dd M_i}{M_i} \int_i \dd\vr_{i'} 
\frac{\dd M_{i'}}{M_{i'}} \nonumber \\
&& \times \int_j \dd\vr_{j'} \frac{\dd M_{j'}}{M_{j'}} \; \frac{\dd n}{\dd\ln M_i}
\frac{\dd n}{\dd\ln M_{i'}} \frac{\dd n}{\dd\ln M_{j'}} \nonumber \\
&& \times \left[ 1+ \xih_{i,i'} + \xih_{i,j'} + \xih_{i',j'} + \zetah_{i,i',j'} \right] ,
\label{C3-def}
\eeqa
\beqa
C_{i,j}^{(4)} & \!\!\! = \! & \frac{1}{\QQ_i\QQ_j} \int\!\! \dd\chi_i \cD_i^2
\frac{\dd\vOm_i}{(\Delta\Omega)} \frac{\dd M_i}{M_i} \int_i \! \dd\vr_{i'}
\frac{\dd M_{i'}}{M_{i'}}  \frac{\dd n}{\dd\ln M_i} \frac{\dd n}{\dd\ln M_{i'}}
\nonumber \\
&& \hspace{-0.6cm} \times \int \dd\chi_j \, \cD_j^2 \frac{\dd\vOm_j}{(\Delta\Omega)}
\frac{\dd M_j}{M_j} \int_j \dd\vr_{j'} \frac{\dd M_{j'}}{M_{j'}}
\frac{\dd n}{\dd\ln M_j} \frac{\dd n}{\dd\ln M_{j'}} \nonumber \\
&& \hspace{-0.6cm} \times \left[ 4 \xih_{i;j} + 2 \zetah_{i;j,j'} + 2 \zetah_{i,i';j}
+ 2 \xih_{i;j'} \xih_{i';j} + \etah_{i,i';j,j'} \right] ,
\label{C4-def}
\eeqa
where $\xih$, $\zetah$, and $\etah$, are the two-point, three-point, and
four-point correlation functions of the objects.
To make the expressions compact but easy to understand, we introduced the
following notation in Eqs.(\ref{C2-def})-(\ref{C4-def}).
Variables associated with the object at the center of the $\cV_i$-shell are noted
by the label $i$ (e.g., $\chi_i,M_i,..$) and those associated with the object within 
the $\cV_i$-shell are noted by the label $i'$ (e.g., $\vr_{i'},M_{i'},..$).
This corresponds to the primed and unprimed variables in Eqs.(\ref{xi-1}) and
(\ref{xi-2}), and we may speak of objects $i, i', j$, and $j'$. 
Then, in the indices of the correlation functions, we separate with a semicolon,
as in $\xih_{i;j}$ of Eq.(\ref{C4-def}), objects $i$ and $j$ that are located at
unrelated positions
$(\chi_i,\vOm_i)$ and $(\chi_j,\vOm_j)$ in the observational cone, whereas we
separate with a comma, as in $\xih_{i,i'}$ of Eq.(\ref{C2-def}), objects that are located
at a fixed distance $r'$. (More precisely, the distance $r'$ is restricted to a radial
bin $\cV$.)

The label $C^{(n)}$ refers to quantities that involve $n$ distinct objects.
Thus, the contributions $C^{(2)}$
and $C^{(3)}$ arise from shot-noise effects (as is apparent through the prefactors
$1/(\Delta\Omega)$), associated with the discreteness
of the number density distribution, and they would vanish for continuous
distributions. However, they also involve the two-point
and three-point correlations, and as such they couple discreteness effects
with the underlying large-scale correlations of the population.
In case of zero large-scale correlations, they remain nonzero because of the
unit factors in the brackets and become purely shot-noise contributions,
arising solely from discreteness effects.

More precisely, contribution (\ref{C2-def}) arises from the coupled identification
$i=j$ and $i'=j'$ (or $i=j'$ and $i'=j$), whereas contribution (\ref{C3-def})
arises from the single identification $i=j$ (or either one of $i=j'$, $i'=j$, $i'=j'$).
Thus, in Eq.(\ref{C3-def}) the object $i$ is at the center of both shells $\cV_i$ and
$\cV_j$.

Contribution $C^{(4)}$ is a pure sample-variance contribution and
does not depend on the discreteness of the number density distribution
(hence there is no $1/(\Delta\Omega)$ prefactor).

As shown in App.~\ref{Computation-Landy-Szalay}, the covariance matrix
of the Landy \& Szalay estimator reads as 
(see also \citet{Szapudi2001a,Bernardeau2002})
\beq
C^{\rm LS}_{i,j} = C^{\rm LS (2)}_{i,j} + C^{\rm LS (3)}_{i,j} +  C^{\rm LS (4)}_{i,j} ,
\label{Cij-LS-2}
\eeq
where the first term is equal to Eq.(\ref{C2-def}),
\beq
C_{i,j}^{\rm LS (2)} = C_{i,j}^{(2)} ,
\label{C2-LS-def}
\eeq
and
\beqa
C_{i,j}^{\rm LS (3)} & = & \frac{4}{(\Delta\Omega)\QQ_i\QQ_j} \int \dd\chi_i \, 
\cD_i^2 \frac{\dd\vOm_i}{(\Delta\Omega)} \frac{\dd M_i}{M_i} \int_i
\dd\vr_{i'} \frac{\dd M_{i'}}{M_{i'}} \nonumber \\
&& \times \int_j \dd\vr_{j'} \frac{\dd M_{j'}}{M_{j'}} \; \frac{\dd n}{\dd\ln M_i}
\frac{\dd n}{\dd\ln M_{i'}} \frac{\dd n}{\dd\ln M_{j'}} \nonumber \\
&& \times \left[ \xih_{i',j'} + \zetah_{i,i',j'} \right] ,
\label{C3-LS-def}
\eeqa
\beqa
\!C_{i,j}^{\rm LS (4)} & \!\!\!\! = \!\! & \frac{1}{\QQ_i\QQ_j}\!  \int \!\! 
\dd\chi_i \cD_i^2 \frac{\dd\vOm_i}{(\Delta\Omega)} \frac{\dd M_i}{M_i} \!
\int_i \! \dd\vr_{i'} \frac{\dd M_{i'}}{M_{i'}} \frac{\dd n}{\dd\ln M_i}
\frac{\dd n}{\dd\ln M_{i'}} \nonumber \\
&& \hspace{-0.6cm} \times \int \dd\chi_j \, \cD_j^2 
\frac{\dd\vOm_j}{(\Delta\Omega)} \frac{\dd M_j}{M_j} \int_j \dd\vr_{j'}
\frac{\dd M_{j'}}{M_{j'}} \frac{\dd n}{\dd\ln M_j} \frac{\dd n}{\dd\ln M_{j'}} 
\nonumber \\
&& \hspace{-0.6cm} \times \left[ 2 \xih_{i;j'} \xih_{i';j} + \etah_{i,i';j,j'} \right] .
\label{C4-LS-def}
\eeqa
By comparison with Eqs.(\ref{C3-def})-(\ref{C4-def}) we can see that many
terms have been canceled \citep{Landy1993,Szapudi1998}.
This confirms that the estimator (\ref{xi-LS-1}) is
more efficient than (\ref {xi-1}), since its covariance will be smaller.

\subsubsection{Low-order terms}
\label{xi-Low-order terms}

In this section we assume that the radial bins $[\Rim,\Rip]$ are restricted to large
enough scales to neglect three and four-point correlation functions, as well as
products such as $\xih_{i;j'} \xih_{i';j}$.
We compute these high-order terms in Sect.~\ref{high-order-terms} and
Figs.~\ref{fig_CiiR_NG} and \ref{fig_Rij_CXiR_NG}
show the range where they can be neglected. Along the diagonal, for halos above
$10^{14} h^{-1} M_{\odot}$ this corresponds to the full range $5 <r< 100 h^{-1}$ Mpc. For lower mass halos, $M>2\times 10^{13} h^{-1} M_{\odot}$,
all scales receive significant contributions from high-order terms, but the low-order
terms contribute to about $50\%$ for $r<15 h^{-1}$ Mpc.
This is sufficient for our purposes in this section, which are to compare the
Peebles \& Hauser and the Landy \& Szalay estimators, the shot-noise and 
sample-variance effects, and the scalings with survey area and number of subfields.
Accurate computation of the covariance matrix requires taking all terms into account,
which we do in Sect.~\ref{high-order-terms}.

Thus, in this section we only keep the contributions that
are constant or linear over the two-point correlation function $\xih$ of the
objects, and we again assume that the two-point correlation function can be
factored as in Eq.(\ref{xij-bb}). Then, as shown in
App.~\ref{Computation-of-the-covariance-of-the-estimator}, for the Peebles \&
Hauser estimator we obtain from Eqs.(\ref{C2-def})-(\ref{C4-def}), at this order, 
\beqa
C_{i,j} & = & \delta_{i,j} \frac{2}{(\Delta\Omega)\QQ_i} (1+\lag\hxi_i\rag)
+ \frac{4}{(\Delta\Omega)\QQ_i\QQ_j} \nonumber \\
&& \times \int \dd\chi \, \cD^2 \,
\nb^3 \, \cV_i \cV_j \left[ 1 + \bb^2 \, \left(\overline{\xir_{i'}}  \!+\! 
\overline{\xir_{j'}} \!+\! \overline{\xir_{i',j'}} \right) \right] \nonumber \\
&& + \frac{4}{\QQ_i\QQ_j} \int \dd\chi \, \cD^5
\, \bb^2 \nb^4 \, \cV_i \cV_j \, \xicyl ,
\label{Cij-tot}
\eeqa
where we introduced
\beq
\overline{\xir_{i',j'}}(z) = \int_i\frac{\dd\vr_{i'}}{\cV_i} \int_j\frac{\dd\vr_{j'}}{\cV_j} 
\; \xi(|\vr_{i'}-\vr_{j'}|;z) .
\label{I3-ij-xi-def}
\eeq
Following the notation explained earlier, below Eq.(\ref{C4-def}), the comma and
the primes in $\overline{\xir_{i',j'}}$ mean that this is a ``spherical average'',
more precisely the average over the two spherical shells $\cV_i$ and $\cV_j$,
in contrast to $\xiconzj$ in Eq.(\ref{xib-ij-def}), which was a ``conical''
average within the observational cone.
There are two indices, $i'$ and $j'$, because we integrate over the two shells
$\cV_i$ and $\cV_j$, whereas in $\overline{\xir_{i'}}$ of Eq.(\ref{xi-i-i'-def})
there was only one index $i'$ because we integrated over a single shell $\cV_i$.
The Fourier-space expression of Eq.(\ref{I3-ij-xi-def}), which can be convenient
for numerical computations, is given in Eq.(\ref{I3-ij-def}) in 
App.~\ref{Computation-of-the-covariance-of-the-estimator}.

Again, to obtain Eq.(\ref{Cij-tot}) we neglected finite-size effects, that is, we
did not take the fact into account that close to the boundaries of the survey
part of the shell $\cV_i$ is not observed. As explained in App.~\ref{Finite-size},
this only leads to an overestimate of $5\%$ of the signal-to-noise ratio,
for a radial bin of $30 h^{-1}$Mpc in a circular survey window of $50$deg$^2$.
This error decreases for wider surveys or smaller radial bins.

For the Landy \& Szalay estimator, at the same order the covariance matrix reads
from Eqs.(\ref{C2-LS-def})-(\ref{C4-LS-def}) as
\beqa
C_{i,j}^{\rm LS} & = & \delta_{i,j} \frac{2}{(\Delta\Omega)\QQ_i} (1+\lag\hxi_i\rag) 
+ \frac{4}{(\Delta\Omega)\QQ_i\QQ_j} \nonumber \\
&& \times \int \dd\chi \, \cD^2 \, \bb^2 \, \nb^3 \, \cV_i \cV_j \,
\overline{\xir_{i',j'}} .
\label{Cij-LS-tot}
\eeqa
Again, as compared with Eq.(\ref{Cij-tot}) several terms have been canceled.
Moreover, at this order only shot-noise terms, whether coupled to large-scale 
correlations or not, contribute to the Landy \& Szalay covariance (\ref{Cij-LS-tot}), 
as can be seen
from the prefactors $1/(\Delta\Omega)$. In contrast, at the same order in the
Peebles \& Hauser covariance (\ref{Cij-tot}), we have two more shot-noise terms
(coupled to the large-scale correlations through the means $\overline{\xir_{i'}}$
and $\overline{\xir_{j'}}$) and one additional sample-variance-only contribution
(i.e., the last term, without the prefactor $1/(\Delta\Omega)$).

\paragraph{Comparison of Peebles \& Hauser and Landy \& Szalay covariance matrices}
\label{Comparison-of-matrices}
~~\\

\begin{figure}
\begin{center}
\epsfxsize=8.5 cm \epsfysize=6.5 cm {\epsfbox{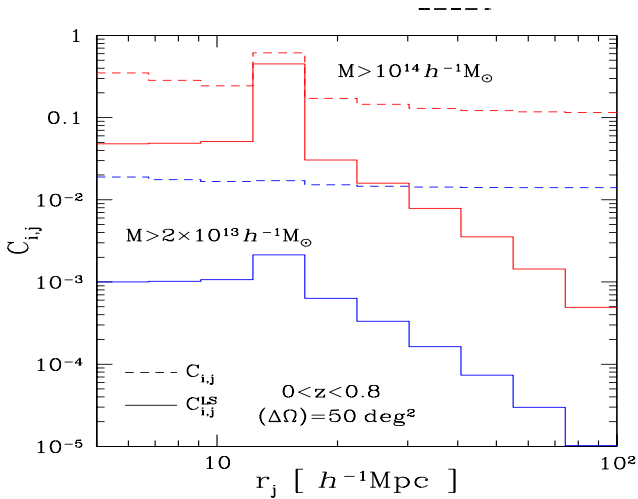}}
\end{center}
\caption{The covariance matrices $C_{i,j}^{\rm LS}$ (solid line) and 
$C_{i,j}$ (dashed line) of the estimators $\hxiLS_i$ and $\hxi_i$,
for $i=4$ associated with the distance bin $12.3<r<16.6h^{-1}$Mpc, as
a function of $j$. We show the results obtained for halos in the redshift range
$0<z<0.8$ with an angular window of $50$ deg$^2$.
Here we only consider the low-order terms given by Eqs.(\ref{Cij-tot}) and
(\ref{Cij-LS-tot}).}
\label{fig_Rij_CXiR_PH_LS}
\end{figure}

We show in Fig.~\ref{fig_Rij_CXiR_PH_LS} one row of the covariance matrices
$C_{i,j}$ and $C_{i,j}^{\rm LS}$, as a function of $j$ at fixed $i$.
We consider halos in the redshift range $0<z<0.8$, for a window of $50$ deg$^2$.
The covariance is larger for the case of higher mass
threshold. In agreement with Eqs.(\ref{Cij-tot}) and (\ref{Cij-LS-tot})
and with standard results \citep{Kerscher2000},
the covariance of the Landy \& Szalay estimator
(\ref{hxi-LS-DR}) is smaller than for the Peebles \& Hauser estimator
(\ref{hxi-DR}), especially
for the lower mass threshold (the higher mass threshold case being more
dominated by the common shot-noise contribution (\ref{C2-LS-def})).

As shown by Fig.~\ref{fig_Rij_CXiR_PH_LS}, another advantage of the Landy \&
Szalay estimator is that its covariance matrix is much more diagonal than for
the Peebles \& Hauser estimator.
This can be checked by comparing the left and middle panels of
Fig.~\ref{fig_Rij_CXi}, where we show the correlation matrices $\cR_{i,j}$
defined as in Eq.(\ref{correlation-matrix}), but where we include all shot-noise
and sample-variance contributions of Eqs.(\ref{Cij-tot}) and (\ref{Cij-LS-tot}).

\paragraph{Comparison of sample-variance and shot-noise effects}
\label{Comparison-of-sv-and-sn}
~~\\

\begin{figure}
\begin{center}
\epsfxsize=8.5 cm \epsfysize=6.5 cm {\epsfbox{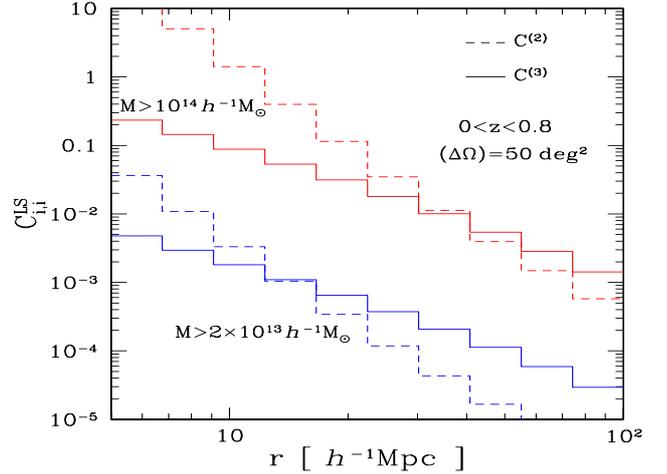}}
\end{center}
\caption{The contributions $C^{(2)}$ and $C^{(3)}$ to the covariance
of the Landy \& Szalay estimator, along the diagonal $i=j$.
As in Fig.~\ref{fig_Rij_CXiR_PH_LS}, we only consider the low-order terms, given by
Eq.(\ref{Cij-LS-tot}).}
\label{fig_CiiR_sn_sv}
\end{figure}

We compare in Fig.~\ref{fig_CiiR_sn_sv} the contributions $C^{(2)}$
(first term in Eq.(\ref{Cij-LS-tot})) and $C^{(3)}$ (second term in
Eq.(\ref{Cij-LS-tot})), again keeping only these low-order terms.
We consider the same survey properties as in Fig.~\ref{fig_Rij_CXiR_PH_LS}
but plot these contributions along the diagonal, $i=j$.
Let us recall that both contributions $C^{(2)}$ and $C^{(3)}$ are
shot-noise contributions (i.e., they arise from the discreteness of the
halo distribution). However, they also involve the underlying large-scale
correlations, as apparent through the factors $\xi$. In particular, $C^{(3)}$,
which arises from a single pair identification, vanishes if there are no large-scale
correlations, whereas $C^{(2)}$, which arises from two pair identifications,
remains nonzero if $\xi=0$ (the term associated with the factor 1 is thus
a ``pure shot-noise'' contribution).
Therefore, by comparing $C^{(2)}$ and $C^{(3)}$ we can assess the 
relative importance of shot-noise and sample-variance effects,
$C^{(2)}$ involving an extra degree of shot noise (one more 
pair identification).
As expected, $C^{(2)}$ is dominant for small distance bins, which
correspond to small volumes, $\cV_i \propto r^3$, and
contain few halos. It also remains dominant up to larger scales in the case
of more massive halos, which are rarer.
Since the contribution $C^{(2)}$ is diagonal, as shown by the Kronecker
prefactor in Eq.(\ref{Cij-tot}), it implies that covariance matrices are more
strongly diagonal for high-mass halos, as can be checked in Fig.~\ref{fig_Rij_CXi}
where we show the correlation matrices $\cR_{i,j}$ of small (upper row)
and large (lower row) halos.

\paragraph{Scalings with survey area and number of subfields}
\label{Scalings-xi}
~~\\

As in Sect.~\ref{Scalings-Nz}, we consider the dependence of the signal-to-noise
ratio on the total survey area $\Delta\Omega$ and on the number $\cN$ of
subfields. Thus, we define the estimator $\hxi_i^{\rm LS,tot}$ as the mean of
the estimators $\hxi_i^{\rm LS,(\alpha)}$ of Eq.(\ref{xi-LS-1}) of the subfields,
\beq
\hxi_i^{\rm LS,tot} = \frac{1}{\cN} \sum_{\alpha=1}^{\cN}
\hxi_i^{\rm LS,(\alpha)} .
\label{xi-LS-tot-def}
\eeq
Of course, the expectation value is independent of $(\Delta\Omega)$ and
$\cN$,
\beq
\lag \hxi_i^{\rm LS,tot} \rag = \lag \hxi_i^{\rm LS,(\alpha)} \rag \;
\mbox{is independent of} \; (\Delta\Omega) \; \mbox{and} \; \cN .
\label{xi-LS-tot-1}
\eeq
From Eq.(\ref{QQ-1}) we can check that $Q_i^{(\alpha)}$ does not depend
on $(\Delta\Omega)$ nor $\cN$, so that for each subfield $\alpha$, of area
$(\Delta\Omega)/\cN$, the covariance (\ref{Cij-LS-tot}) scales as
\beq
C_{i,j}^{\rm LS,(\alpha)} \propto \frac{\cN}{(\Delta\Omega)} .
\label{Cij-LS-alpha}
\eeq
Both terms in Eq.(\ref{Cij-LS-tot}) scale in the same fashion, so that
the structure of the covariance matrix does not change with $(\Delta\Omega)$ nor
$\cN$ (i.e. it does not become more or less diagonal), if we neglect
boundary effects. Then, the covariance matrix of the averaged estimator
(\ref{xi-LS-tot-def}) scales as
\beq
C_{i,j}^{\rm LS,tot} = \frac{1}{\cN} C_{i,j}^{\rm LS,(\alpha)} \propto
\frac{1}{(\Delta\Omega)} ,
\label{Cij-LS-tot-alpha}
\eeq
so that the signal-to-noise ratio scales as
\beq
\frac{S}{N}=\frac{\lag \hxi_i^{\rm LS,tot} \rag}{\sqrt{C_{i,j}^{\rm LS,tot}}} \propto
\sqrt{(\Delta\Omega)} .
\label{SN-xi-tot}
\eeq
Therefore, a single wide-field survey and a combination of several independent
smaller surveys, with the same total area, show the same efficiency.
This is because both terms in Eq.(\ref{Cij-LS-tot}) scale in
the same way with the survey geometry, as $1/(\Delta\Omega)$, because
the sample-variance effects involved in these mixed contributions
arise from the correlation between objects separated by a
distance $r<\Rip+\Rjp$, independently of the angular size of the survey.
This is different from the sample-variance contribution (\ref{Cij-7}) to the
covariance of the number counts, which explicitly depends
on the large-scale correlation over the survey angular size $\theta_s$,
see Eq.(\ref{I-thetas-def}), because it arises from the correlation between 
objects located at any position in the survey cone.
Of course, result (\ref{SN-xi-tot}) only applies to small length scales, 
$\Rip+\Rjp \ll \cD\theta_s$, where it is legitimate to neglect finite-size effects.
For long wavelengths a wider survey is clearly more efficient, and the only
possible choice for scales that are close to the larger survey diameter.

\subsubsection{High-order terms for the covariance of $\hxiLS$}
\label{high-order-terms}

We now estimate the high-order terms for the covariance $C_{i,j}^{\rm LS}$
of the Landy \& Szalay estimator $\hxiLS_i$ that we neglected in
Eq.(\ref{Cij-LS-tot}), where we only kept terms of order zero or one over the
two-point correlation function.
To evaluate the contributions associated with the factors $\zetah_{i,i',j'}$ in
Eq.(\ref{C3-LS-def}) and $\etah_{i,i';j,j'}$ in Eq.(\ref{C4-LS-def}),
we use the model for the three- and four-point halo correlation functions
described in Sect.~\ref{three-point}.
Then, as shown in App.~\ref{Computation-high-order-terms}, the contribution
associated with the product $\xih_{i;j'} \xih_{i';j}$ in Eq.(\ref{C4-LS-def}) is given by
\beq
C_{i,j}^{\rm LS (\xi\xi)} = \frac{2}{\QQ_i\QQ_j}
\int \dd\chi \, \cD^5 \, \bb^4 \, \nb^4 \, \cV_i \cV_j \, \overline{\xir_{i;j'} \xir_{i';j}} ,
\label{CLS-xixi-1}
\eeq
the term $\zetah_{i,i',j'}$ of Eq.(\ref{C3-LS-def}) yields
\beqa
C_{i,j}^{\rm LS (\zeta)} & = & \frac{4}{(\Delta\Omega)\QQ_i\QQ_j} \int \dd\chi \,
\cD^2 \, \bb^3 \, \nb^3 \, \cV_i \cV_j \, \frac{S_3}{3}  \nonumber \\
&& \times \, \left[ \overline{\xir_{i'}} \times \overline{\xir_{j'}} 
+ \overline{\xir_{i',i} \xir_{i',j'}} +\overline{\xir_{j',i} \xir_{j',i'}} \right] ,
\label{CLS-zeta-1}
\eeqa
and the term $\etah_{i,i';j,j'}$ of Eq.(\ref{C4-LS-def}) gives
\beqa
C_{i,j}^{\rm LS (\eta)} & = & \frac{2}{\QQ_i\QQ_j} \int \dd\chi \,
\cD^5 \, \bb^4 \, \nb^4 \, \cV_i \cV_j \, \frac{S_4}{16} \left[ \overline{\xir_{i'}}
\times \overline{\xi_{i;j} \xir_{i;j'}} \right . \nonumber \\
&& + \overline{\xir_{j'}} \times \overline{\xi_{i;j} \xir_{j;i'}} + 2 \,
\overline{\xir_{i'}} \times \overline{\xir_{j'}} \times \xicyl \nonumber \\
&& \left. + 2 \, \overline{\xir_{j';i}\xi_{i;j}\xir_{j;i'}} 
+ \overline{\xir_{j';i}\xir_{i,i'}\xir_{i';j}} + \overline{\xir_{i';j}\xir_{j,j'}\xir_{j';i}} 
 \right] 
\label{CLS-eta-1}
\eeqa 
where the various factors are given in App.~\ref{Computation-high-order-terms},
and we used for Eqs.(\ref{CLS-zeta-1})-(\ref{CLS-eta-1}) the
``hierarchical clustering ansatz'', described in Figs.~\ref{fig-zeta} and
\ref{fig-eta} and given by Eqs.(\ref{zeta-def}) and (\ref{eta-def}). 

The terms $C_{i,j}^{\rm LS (\xi\xi)}$ and $C_{i,j}^{\rm LS (\eta)}$ are ``pure
sample-variance'' contributions. Thus, there is no prefactor $1/(\Delta\Omega)$
and they involve large-scale correlations among four halos, $i,i',j,j'$.
The term $C_{i,j}^{\rm LS (\zeta)}$ is a coupled shot-noise and sample-variance
contribution, as shown by the prefactor $1/(\Delta\Omega)$ and the fact
that it involves large-scale correlations among three halos, $i,i',j'$.
(The discreteness of the halo distribution has led to the identification
$i=j$, i.e. a shot-noise effect, which leaves three distinct halos.)

\begin{figure}
\begin{center}
\epsfxsize=8.5 cm \epsfysize=6.5 cm {\epsfbox{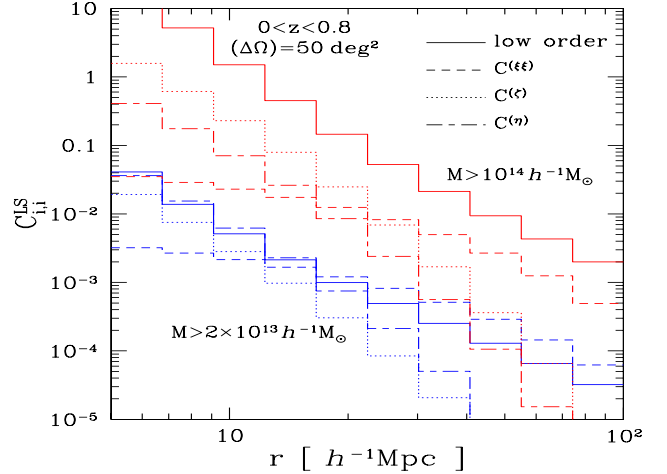}}
\end{center}
\caption{The low- and high-order contributions to the covariance matrix
$C_{i,j}^{\rm LS}$ along its diagonal. We again consider halos in the redshift
range $0<z<0.8$, with an angular window of $50$ deg$^2$, above two mass
thresholds.}
\label{fig_CiiR_NG}
\end{figure}

\begin{figure}
\begin{center}
\epsfxsize=8.5 cm \epsfysize=6.5 cm {\epsfbox{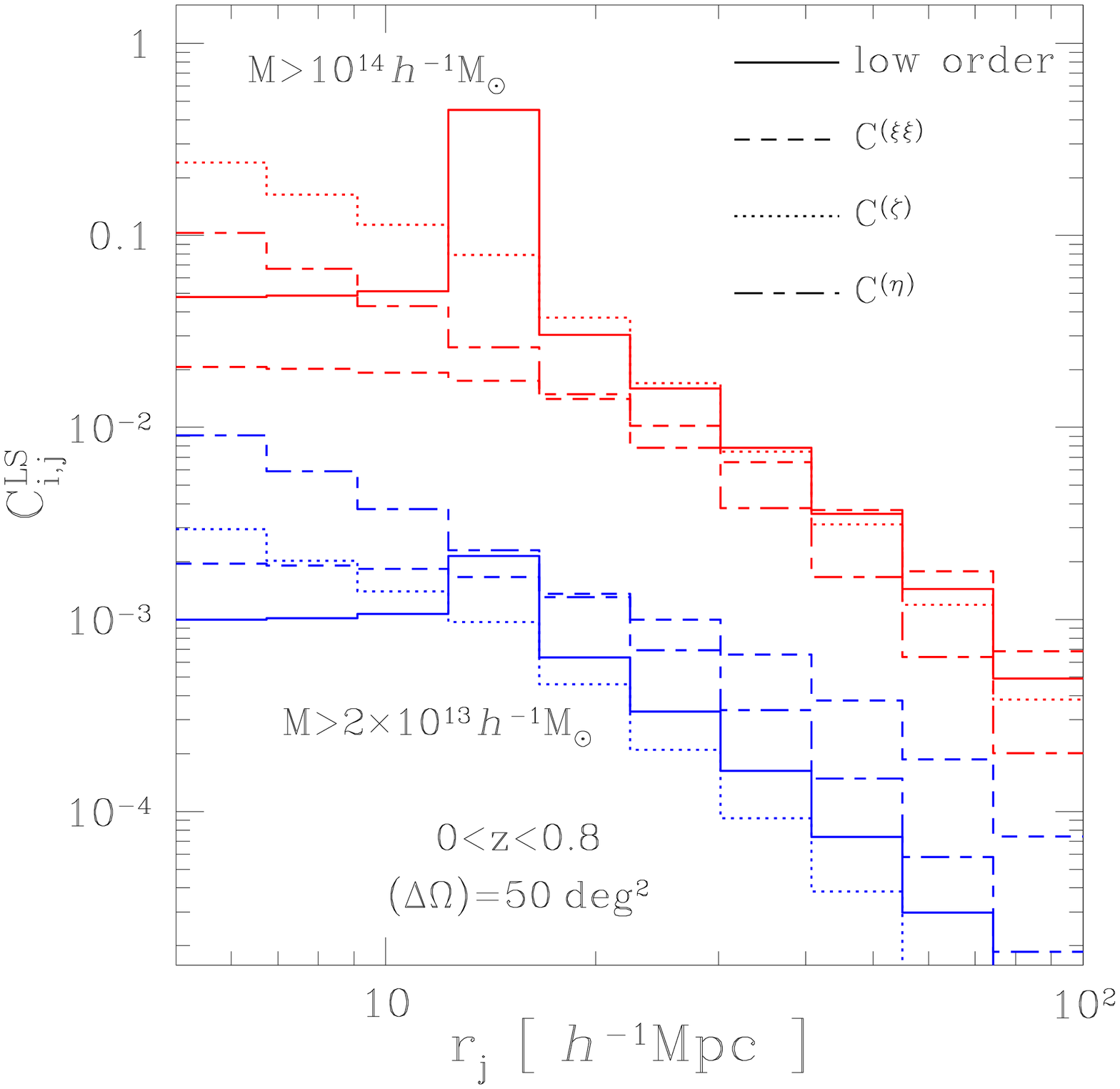}}
\end{center}
\caption{The low- and high-order contributions to the covariance matrix
$C_{i,j}^{\rm LS}$, as in Fig.~\ref{fig_CiiR_NG}, but along one row.
This corresponds to the fixed bin $i=4$, associated with the distance bin
$12.3<r<16.6h^{-1}$Mpc, as a function of $j$.}
\label{fig_Rij_CXiR_NG}
\end{figure}

We compare in Figs.~\ref{fig_CiiR_NG} and \ref{fig_Rij_CXiR_NG} the low-order
contributions (\ref{Cij-LS-tot}) with these high-order contributions
(\ref{CLS-xixi-1})-(\ref{CLS-eta-1}).
We can see that the latter can be non-negligible on these scales,
$5<r<100 h^{-1}$Mpc.
Along the diagonal, $i=j$, shown in Fig.~\ref{fig_CiiR_NG},
they are always significantly smaller than the low-order contribution (which
includes both sample-variance and shot-noise effects) for massive halos,
$M>10^{14}h^{-1} M_{\odot}$, but are close to it or larger for
$M>2\times 10^{13}h^{-1} M_{\odot}$. On large scales the main high-order
contribution is the term (\ref{CLS-xixi-1}), associated with a product $\xi\xi$,
while the terms (\ref{CLS-zeta-1}) and (\ref{CLS-eta-1}),
associated with the three- and four-point correlation functions, dominate
on small scales.
Indeed, the former does not increase much on small scales, whereas the latter
are very sensitive to the smoothing scales $R_i$ and $R_j$ and show a steep
growth on small scales, even though formally $\zeta$ is also of order $\xi\xi$
within the model (\ref{zeta-def}). This is because the term
(\ref{CLS-xixi-1}) involves the product of two correlations between two
distinct lines of sight, as seen in Eq.(\ref{C4-LS-def}), so that each $\xi$ is
averaged along the radial direction, while the term (\ref{CLS-zeta-1}),
which arises from one shot-noise contraction that has removed one line-of-sight
integration, involves the product of two correlations between a central point
and two points at distances $R_i$ and $R_j$, as seen in Eq.(\ref{C3-LS-def}).

As seen in Fig.~\ref{fig_Rij_CXiR_NG}, at fixed $i$  the relative importance of these
high-order contributions to $C_{i,j}^{\rm LS}$ increases as the bin $j$ 
shifts to smaller scales. Again, we can see that among these contributions
the ``$\xi\xi$'' term (\ref{CLS-xixi-1}) dominates on large scales and saturates
on small scales, while the ``$\zeta$'' and ``$\eta$'' terms (\ref{CLS-zeta-1})
and (\ref{CLS-eta-1}) dominate on small scales and strongly depend on the
smoothing scales.

That high-order contributions can become dominant as one of the
bins $i$ and $j$ shifts to small scales agrees with expectations, as one probes
deeper into the nonlinear regime where three- and four-point correlation
functions become important, and with some previous studies 
\citep{Meiksin1999,Scoccimarro1999}.
This implies that the covariance matrix is less diagonal once we take
these contributions into account, and it decreases the number of effectively 
independent modes. This can be checked in Fig.~\ref{fig_Rij_CXi}, where we show the
correlation matrices $\cR_{i,j}^{\rm LS}$ without (middle panels) and with
(right panels) these high-order contributions, for the mass thresholds
$M>2\times 10^{14}h^{-1} M_{\odot}$ and $M>10^{14}h^{-1} M_{\odot}$.
Therefore, for survey characteristics such as those of
Figs.~\ref{fig_CiiR_NG}-\ref{fig_Rij_CXiR_NG}, it is necessary to include
high-order contributions to the covariance matrix of two-point estimators
for moderate-mass halos that are not dominated by shot-noise effects.

\subsubsection{Comparison with numerical simulations}
\label{comparison-Cij-xi}

\begin{figure}
\begin{center}
\epsfxsize=8.5 cm \epsfysize=6.5 cm {\epsfbox{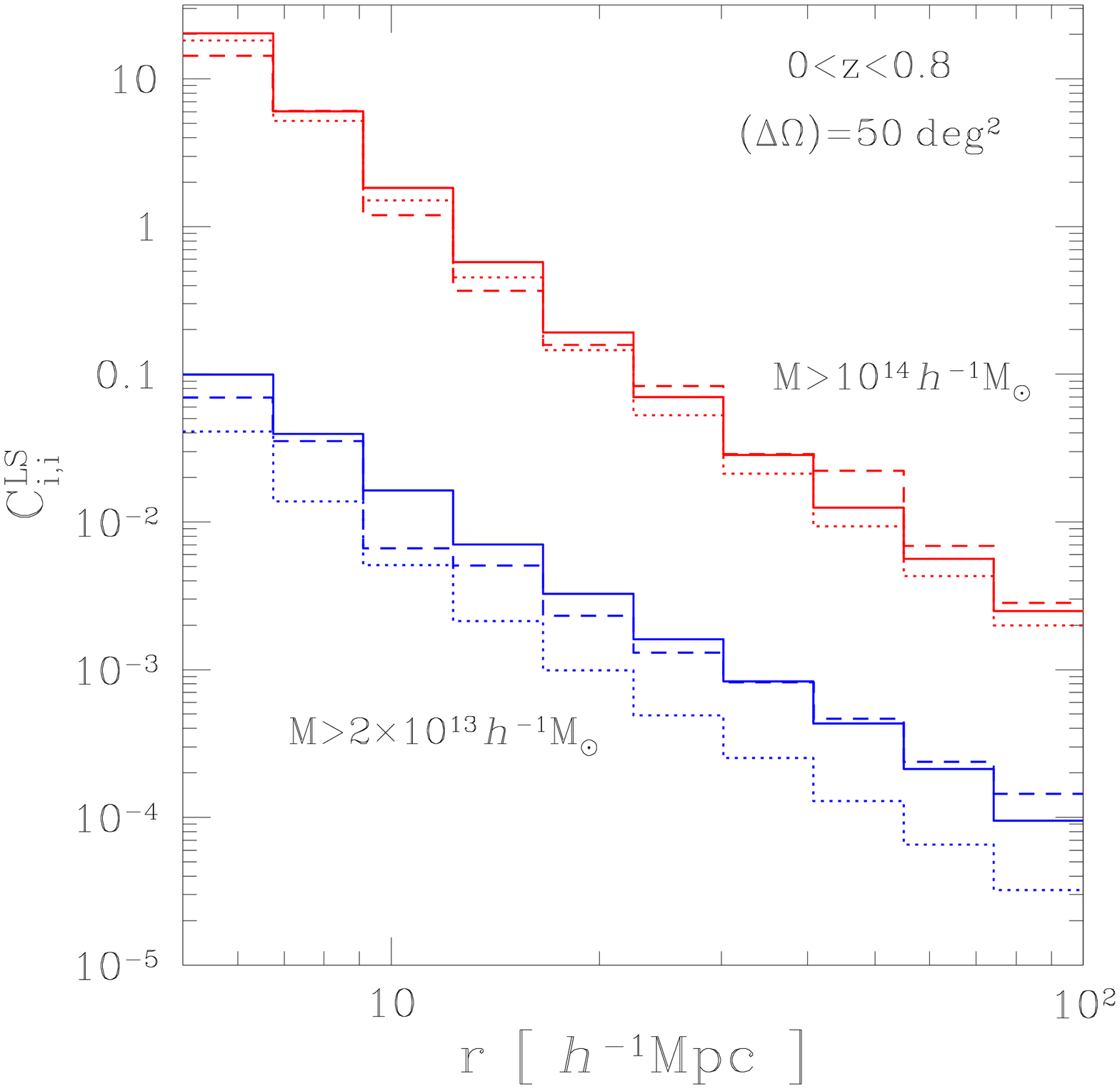}}
\end{center}
\caption{The covariance matrix $C_{i,j}^{\rm LS}$ of the Landy \& Szalay 
estimator, along the diagonal $i=j$. We show our analytical results including
all contributions (solid lines) or only low-order terms (dotted lines), and results 
from numerical simulations (dashed lines).}
\label{fig_CiiR_Horizon}
\end{figure}

\begin{figure}
\begin{center}
\epsfxsize=8.5 cm \epsfysize=6.5 cm {\epsfbox{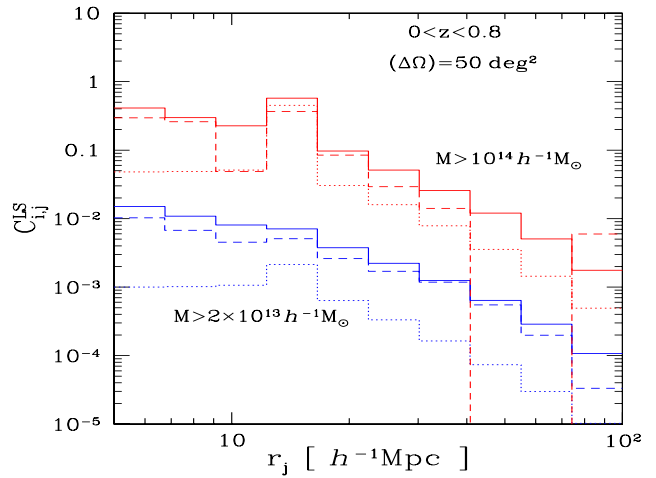}}
\end{center}
\caption{The covariance matrix $C_{i,j}^{\rm LS}$, as in
Fig.~\ref{fig_CiiR_Horizon}, but along one row.
This corresponds to the fixed bin $i=4$, associated with the distance bin
$12.3<r<16.6h^{-1}$Mpc, as a function of $j$.}
\label{fig_Rij_CXiR_Horizon}
\end{figure}

We display in Fig.~\ref{fig_CiiR_Horizon} the covariance matrix
$C_{i,j}^{\rm LS}$ of the Landy \& Szalay estimator, along its diagonal. 
We show our results obtained when we include the high-order contributions
of Sect.~\ref{high-order-terms}, see Eqs.(\ref{CLS-xixi-1})-(\ref{CLS-eta-1}),
and when we only take the low-order terms of Eq.(\ref{Cij-LS-tot}) into account.
We obtain a good match
to the numerical simulations, especially on the largest scales, which are also
more reliable. In particular, we recover the strong dependence on radius and
halo mass. We can see that, for moderate-mass halos,
$M>2\times 10^{13} h^{-1} M_{\odot}$, the high-order contributions
are not negligible (because the low-order shot-noise contribution is relatively
smaller).

We show the same covariance matrix along its fourth row in
Fig.~\ref{fig_Rij_CXiR_Horizon}. The results from the numerical simulations are
somewhat noisy, especially for the rare massive halos at low radii. However,
where they are reliable they show reasonably good agreement with
our analytical results. In agreement with Sect.~\ref{high-order-terms}, 
it is clear that, even more than along the diagonal, the high-order contributions
of Eqs.(\ref{CLS-xixi-1})-(\ref{CLS-eta-1}) cannot be neglected in order to
obtain a good estimate of the off-diagonal terms of the covariance
matrix (see also Fig.~\ref{fig_Rij_CXi}).

\subsubsection{Correlation matrices}
\label{correlation-matrices-R}

\begin{figure*}
\begin{center}
\epsfxsize=5.7 cm \epsfysize=5.7 cm {\epsfbox{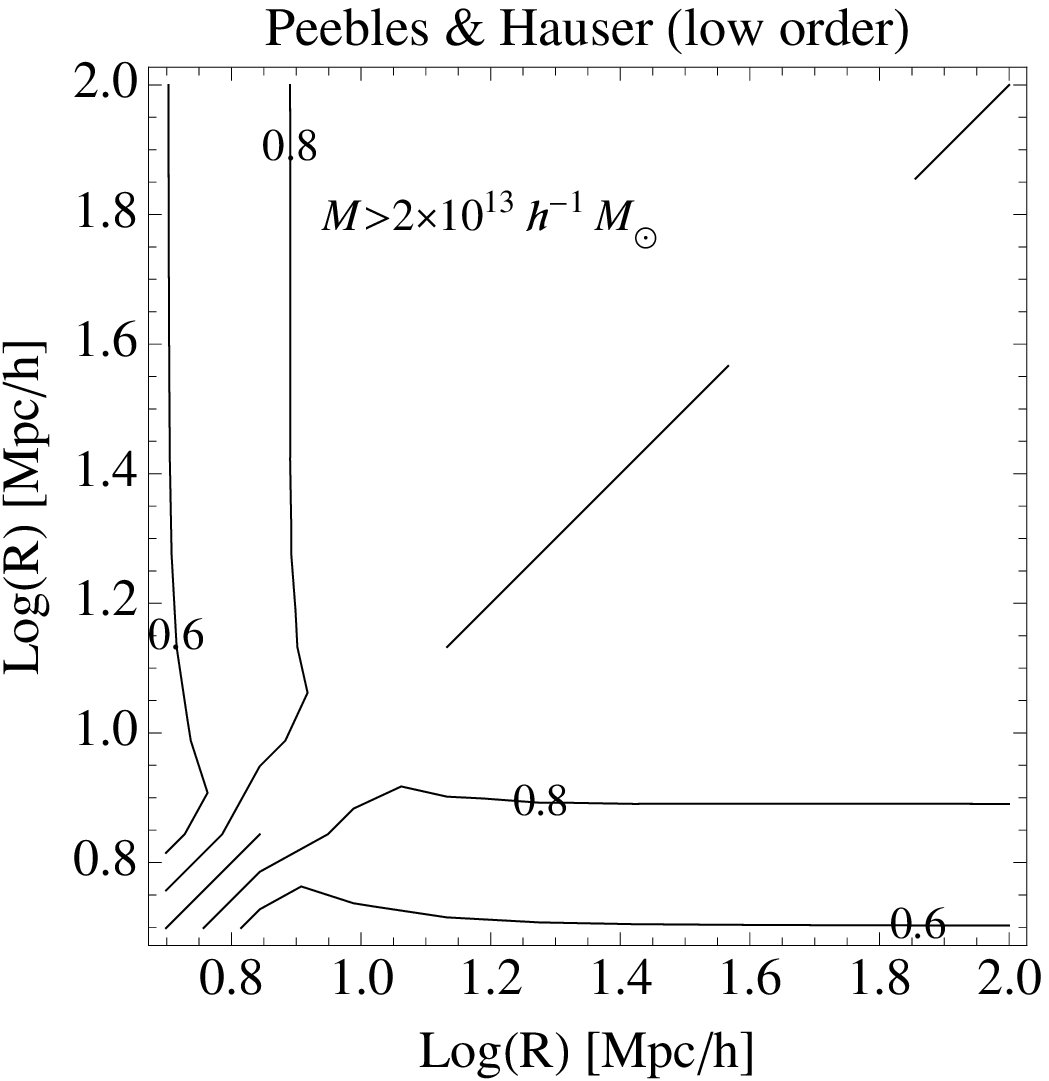}}
\epsfxsize=5.7 cm \epsfysize=5.7 cm {\epsfbox{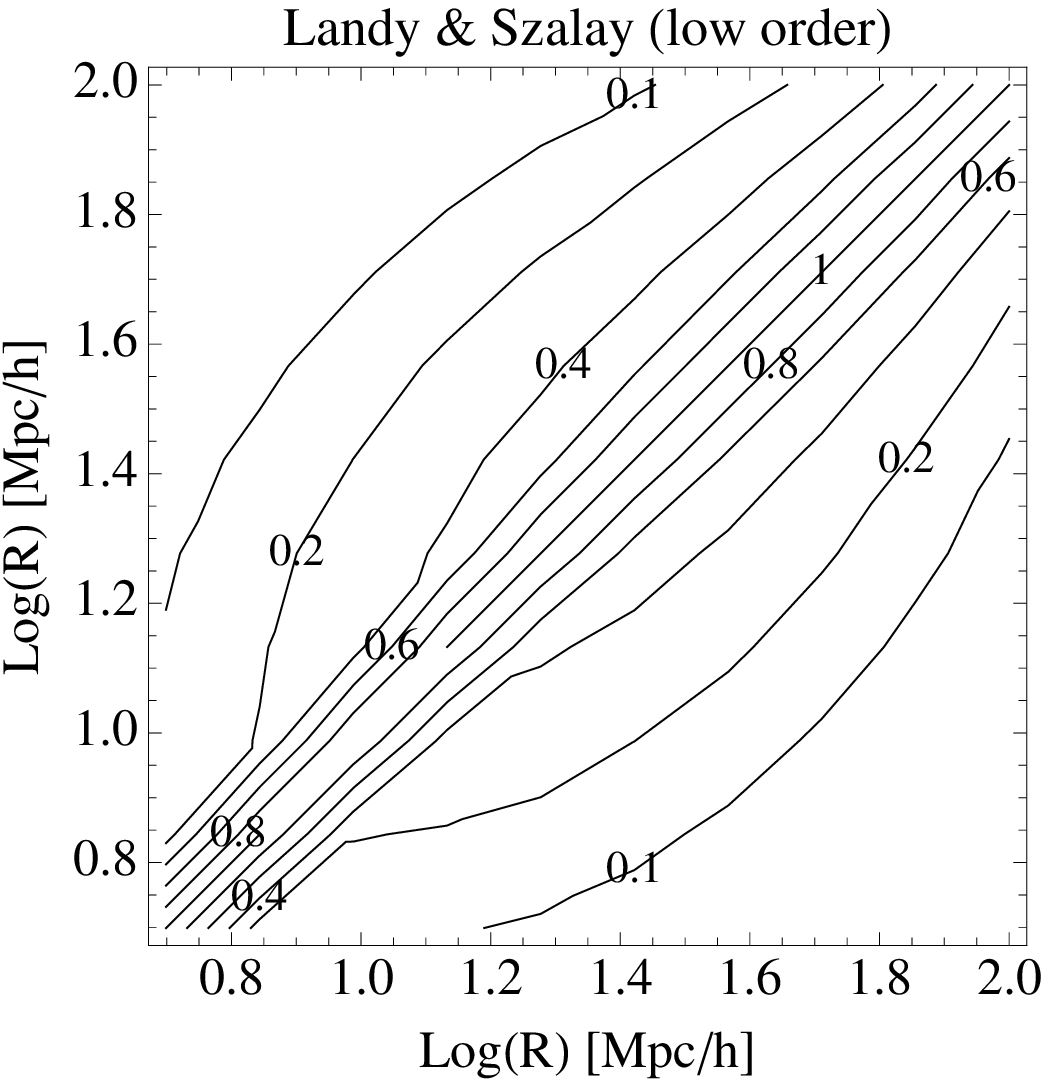}}
\epsfxsize=5.7 cm \epsfysize=5.7 cm {\epsfbox{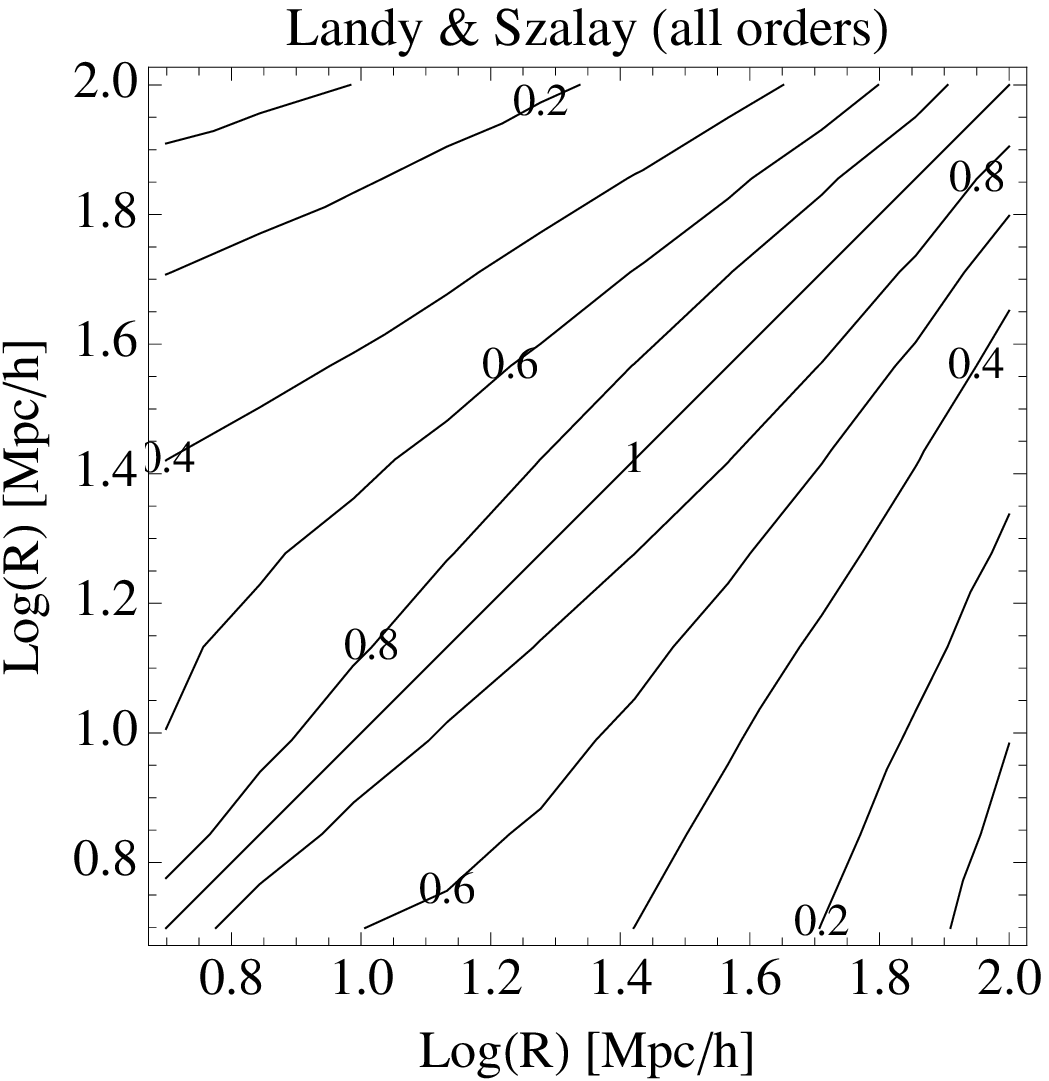}}\\
\epsfxsize=5.7 cm \epsfysize=5.7 cm {\epsfbox{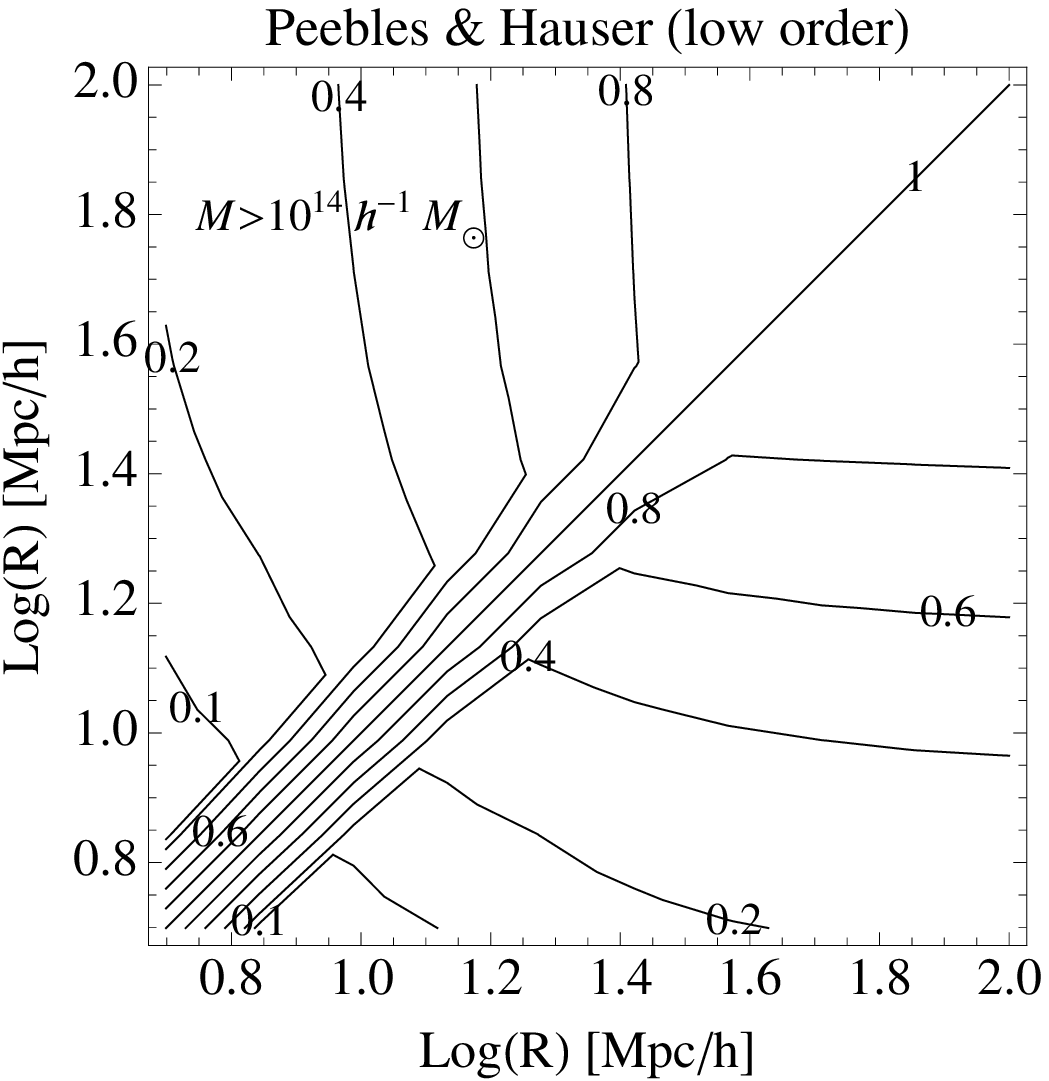}}
\epsfxsize=5.7 cm \epsfysize=5.7 cm {\epsfbox{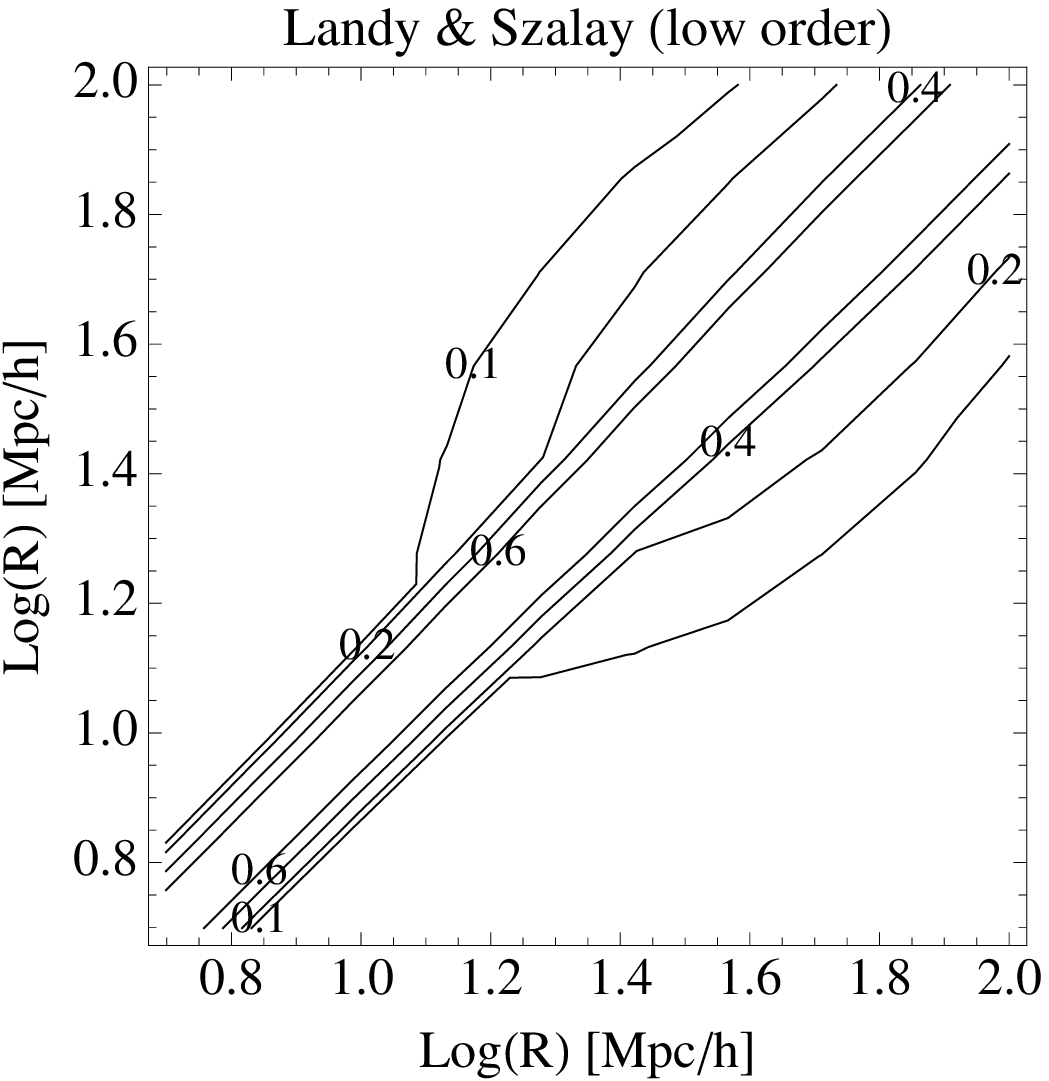}}
\epsfxsize=5.7 cm \epsfysize=5.7 cm {\epsfbox{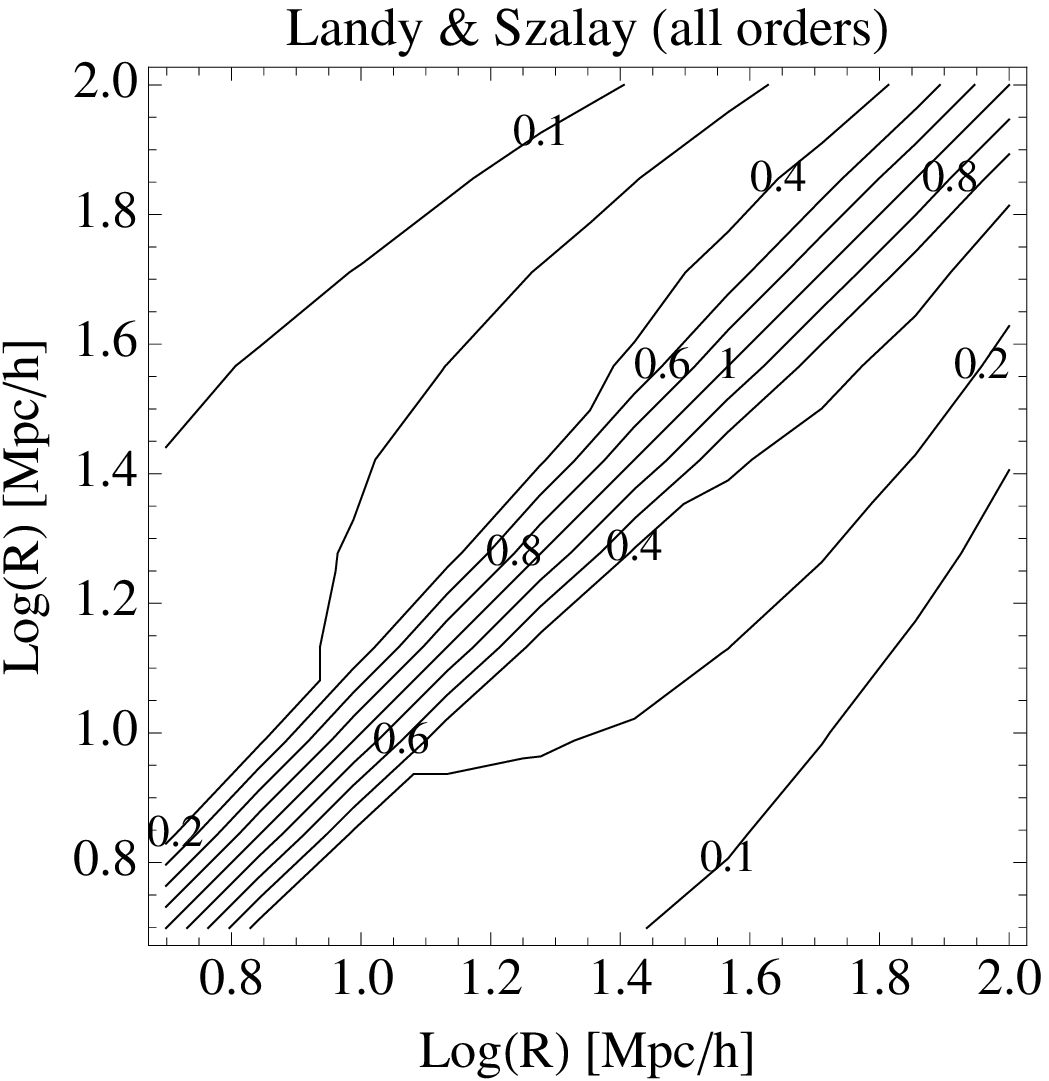}}
\end{center}
\caption{Contour plots for the correlation matrix $\cR_{i,j}$, defined
as in Eq.(\ref{correlation-matrix}) but for the full covariance matrix $C_{ij}$
of the halo correlation. 
There are ten distance bins, over $5<r<100 h^{-1}$Mpc, equally spaced in
$\log(r)$, as in previous figures.
We consider halos in the redshift range $0<z<0.8$,
within an angular window of $50$ deg$^2$, above the mass thresholds
$M>2\times 10^{13}h^{-1} M_{\odot}$ in the {\it upper row}, and
$M>10^{14}h^{-1} M_{\odot}$ in the {\it lower row}.
{\it Left panels:} Low-order contributions (\ref{Cij-tot}) for the Peebles \&
Hauser estimator.
{\it Middle panels:} Low-order contributions (\ref{Cij-LS-tot}) for the
Landy \& Szalay estimator.
{\it Right panels:} Full correlation matrix, including the high-order
contributions of Eqs.(\ref{CLS-xixi-1})-(\ref{CLS-eta-1}), for the
Landy \& Szalay estimator.}
\label{fig_Rij_CXi}
\end{figure*}

We show in Fig.~\ref{fig_Rij_CXi} the correlation matrices, defined as in
Eq.(\ref{correlation-matrix}) but for the full covariance matrix $C_{ij}$
of the halo correlation. 
(Although $\cR_{i,j}$ is a discrete $10\times 10$ matrix, it is still possible to 
draw a contour plot by interpolation. This gives clear figures that are easier to 
read than a density plot where each cell is colored with a level of gray that 
depends on the entry $\cR_{i,j}$.)

The left and middle panels of Fig.~\ref{fig_Rij_CXi} clearly show the strong 
improvement associated with the use of the Landy \& Szalay estimator in place of 
the Peebles \& Hauser estimator. 
In agreement with Fig.~\ref{fig_CiiR_sn_sv} and the discussion in
Sect.~\ref{Comparison-of-sv-and-sn}, the correlation matrix is more diagonal
for massive halos, where the diagonal shot-noise contribution $C^{(2)}$
of Eq.(\ref{C2-def}) is more important. Indeed, shot-noise effects become
dominant for rare objects.
For the same reason, high-order contributions to the covariance matrix,
which are due to sample-variance effects, are more important for low-mass
halos, as shown by the comparison between the middle and right panels.
The slope of the contour lines in the right panels, especially in the low-mass
case, shows that high-order terms are more important on small scales and also
increase the correlation between small and large scales while making the
matrix less diagonal.

Thus, for low-mass halos there are rather strong correlations between all scales
in the range $5<r<100 h^{-1}$Mpc, and to obtain accurate estimates of error
bars on cosmological parameters it is necessary to take off-diagonal
entries and high-order contributions to the covariance matrix into account.

\section{Angular correlation function}
\label{Angular-correlation}

In the previous section we considered the real-space 3D correlation
function, which requires knowledge of the radial position of the halos
(or more generally of the objects of interest).
If this information is not available (e.g., redshift estimates are too noisy
or distance measures are highly contaminated by redshift-space distortions),
it is still possible to derive some constraints on cosmology from the
angular distribution of the objects on the sky 
\citep{Peebles1980,Eisenstein2001,Maller2005}.
Therefore, in this section we apply the formalism developed in
Sect.\ref{Two-point-correlation} to the angular two-point correlation function
$w(\theta)$.

\subsection{Mean correlation}
\label{Mean-correlation-w}

\subsubsection{Peebles \& Hauser estimator}
\label{Simple-estimator-w}

As in Sect.~\ref{Large-angular-windows}, we write the observed number density 
of objects on the sky as $\hN(\vOm)$, but we omit the index $i$ of Eq.(\ref{Ni-W2})
since we consider a single redshift bin. As in Sect.~\ref{Two-point-correlation},
the width $\Delta z$ is not necessarily small and may cover the whole
redshift range of the survey.
Then, using notations that are similar to Eq.(\ref{xi-1}), we can write the
Peebles \& Hauser estimator $\hw_i$ as
\beq
1+\hw_i = \frac{1}{\Qw_i} \int \frac{\dd\vOm}{(\Delta\Omega)} \, \hN(\vOm)
\int_{\thetaim}^{\thetaip} \dd\vtheta' \hN(\vOm') ,
\label{wi-1}
\eeq
with
\beq
\Qw_i = \int \frac{\dd\vOm}{(\Delta\Omega)} \, \Nb(\vOm)
\int_{\thetaim}^{\thetaip} \dd\vtheta' \, \Nb(\vOm') ,
\label{Qwi-def}
\eeq
where $\Nb(\vOm)$ is the mean angular number density on the direction
$\vOm$, given by
\beq
\Nb = \int\! \dd\chi \, \cD^2 \, \nb(z) = \int\! \dd\chi \, \cD^2 \!
\int\! \frac{\dd M}{M} \, \frac{\dd n}{\dd\ln M}(M,z) .
\label{Nb-Om}
\eeq
Here we used Eq.(\ref{Nbz-def}), and we assumed that the sky coverage is the same
over the survey window $(\Delta\Omega)$, so that $\Nb(\vOm)$ is actually
a constant that does not depend on $\vOm$ (but the formalism is readily extended
to the more general case where we add a filter that depends on $\vOm$).

Here and in the following, the index $i$ of Eqs.(\ref{wi-1})-(\ref{Qwi-def}) refers
to the angular bin $[\thetaim,\thetaip]$, over which we estimate the angular
correlation $w(\theta)$.
Again, we denote with unprimed letters the quantities associated with the
first object, such as its position $\vOm$ on the sky, and with primed letters
the quantities associated with the neighbor at distance $\theta'$, such as its
position $\vOm'$. We use the flat-sky and Limber's approximations, which are
typically valid for angular radii below $10$ deg, as seen in 
Fig.~\ref{fig_r_nz_noflat}.

The quantity $\Qw_i$ introduced in Eq.(\ref{Qwi-def}) can be written as
\beq
\Qw_i = \cA_i \, \Nb^2 , \hspace{0.5cm}  ,
\label{Qw-1}
\eeq
using that $\Nb$ defined in Eq.(\ref{Nb-Om}) does not depend on $\vOm$
in our case, and $\cA_i$ is the area of the $i$-ring,
\beq
\cA_i = \pi ( \thetaip^2 - \thetaim^2) .
\label{Ai-def}
\eeq
Then, we proceed as in Sect.~\ref{Two-point-correlation}. Substituting
the observed 3D number density $\dd \hn/\dd\ln M$ as in Eq.(\ref{Ni-1}),
introducing the halo two-point correlation $\xih$ when we take the
average as in Eq.(\ref{xi-2}), and using the factorization (\ref{xij-bb}), 
we obtain
\beq
\lag\hw_i\rag = \frac{1}{\Nb^2} \int\dd\chi \, \cD^5 \, \bb^2 \, \nb^2 \, 
\overline{\xith_{i'}}(z) ,
\label{wi-2}
\eeq
with
\beq
\overline{\xith_{i'}}(z) = \int_i \, \frac{\dd\vtheta'}{\cA_i}
\int \frac{\dd\chi'}{\cD} \, \xi(r';z) .
\label{w-i-i'-def}
\eeq
The superscript ``$(\theta)$'' recalls that Eq.(\ref{w-i-i'-def}) is an average
over the angular ring $\cA_i$, instead of the 3D spherical shell $\cV_i$ 
of Eq.(\ref{xi-i-i'-def}). The prime in the subscript ``$i'$'' also recalls that we
integrate over a neighboring point $\vtheta'$, with respect to a given point
$(\chi,\cD\vOm)$ of the observational cone. However, because the two points
are only close in the 2D angular space (i.e., in the $i$-ring), we also integrate
over the longitudinal coordinate $\chi'$ along the full line of sight in
Eq.(\ref{w-i-i'-def}).

Explicit expressions for $\overline{\xith_{i'}}(z)$ are given in
App.~\ref{Computation-of-the-mean-of-the-estimator-w}.
In contrast to the number counts studied in Sect.~\ref{Number-density},
where, for large angles above a few degrees, it is necessary to go beyond
Limber's approximation, as found in Figs.~\ref{fig_C_nz_noflat} and
\ref{fig_r_nz_noflat}, for our study
of the angular correlation function Limber's approximation is sufficient because
we consider much smaller angular scales of a few arcmin.

\subsubsection{Landy \& Szalay  estimator}
\label{Landy-Szalay-w}

As in Sect.~\ref{Landy-Szalay-estimator}, the measure of the angular correlation
can be made more accurate by using the Landy \& Szalay  estimator instead
of the Peebles \& Hauser estimator (\ref{wi-1}) \citep{Landy1993,Szapudi1998}.
As in Eq.(\ref{xi-LS-1}), this reads as
\beqa
\hwLS_i & = & \frac{1}{\Qw_i} \int \frac{\dd\vOm}{(\Delta\Omega)} \,
\hN(\vOm) \int \dd\vtheta' \, \hN(\vOm') \nonumber \\ 
&& - \frac{2}{\Qw_i} \int \frac{\dd\vOm}{(\Delta\Omega)} \, \hN(\vOm)
\int \dd\vtheta' \, \Nb(\vOm') + 1 ,
\label{wi-LS-1}
\eeqa
and we can check that its mean is again equal to the average (\ref{wi-2}).

\subsubsection{Comparison with simulations}
\label{Comparison-w}

\begin{figure}
\begin{center}
\epsfxsize=8.5 cm \epsfysize=6.5 cm {\epsfbox{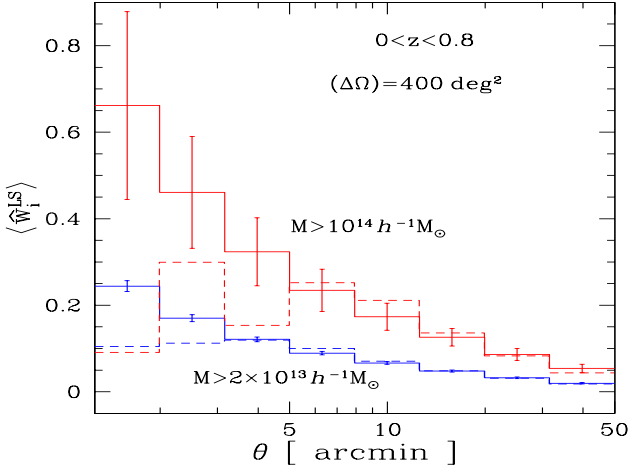}}
\end{center}
\caption{The mean angular correlation, $\lag\hwLS_i\rag$, over eight angular
bins within $1.25<\theta<50$ arcmin, equally spaced in $\log(\theta)$.
We compare our analytical results (solid lines) with numerical simulations
(dashed lines).}
\label{fig_wi_Horizon}
\end{figure}

We compare in Fig.~\ref{fig_wi_Horizon} the mean correlation (\ref{wi-2})
with results from numerical simulations.
The error bars are the $3-\sigma$ statistical errors obtained from the
covariance matrices derived in Sect.~\ref{High-order-w} for 41 fields
of 400 deg$^2$ as used in the simulations.
Above $5$ arcmin we obtain a good
match between our results and the numerical simulations.
This could be expected from Sect.~\ref{Two-point-correlation}
since the angular correlation is a projection of the 3D correlation.
On lower angular scales the discrepancy may be due to the finite size
of the clusters. This implies that $\xih=-1$ at distances below the sum
of the two cluster radii (exclusion effect), but we have not included this
effect in our bias model (\ref{xij-bb}). Since a typical cluster at $z=0.5$
(with a size of $1 h^{-1}$ Mpc) corresponds to an angle of
$\sim 2.5$ arcmin and projection effects are rare (since clusters are rare
objects with a surface density $\sim 10$ deg$^{-2}$), this exclusion effect
indeed occurs at $\theta \la 5$ arcmin and appears at slightly larger angles for 
more massive halos. This explains the behavior found on these scales in
Fig.~\ref{fig_wi_Horizon}.

\subsection{Covariance matrices for the halo angular correlation}
\label{Covariance-w}

The covariance matrices of the estimators $\hw_i$ and $\hwLS_i$ can be
computed following the procedure used in Sect.~\ref{Two-point-correlation}
for the 3D correlation. Denoting again the covariance matrices as $C_{i,j}$
and $C^{\rm LS}_{i,j}$, they decompose as in Eq.(\ref{xii-xij-2}),
\beq
C_{i,j} = C_{i,j}^{(2)} +  C_{i,j}^{(3)} + C_{i,j}^{(4)} ,
\label{Cij-w-1}
\eeq
where $C_{i,j}^{(4)}$ is a pure sample-variance contribution, whereas
$C_{i,j}^{(2)}$ and $C_{i,j}^{(3)}$ are shot-noise contributions that arise
when either one pair or two pairs of objects are identified.
Again, the contributions $C_{i,j}^{(2)}$ and $C_{i,j}^{(3)}$ also involve the
two-point and three-point correlations; i.e., they contain terms that couple
discreteness effects with large-scale density correlations.

For the Peebles \& Hauser estimator (\ref{wi-1}) we obtain, as in
Eqs.(\ref{C2-def})-(\ref{C4-def}),
\beqa
C_{i,j}^{(2)} & = & \delta_{i,j} \, \frac{2}{(\Delta\Omega)\Qw_i^2} \int \!\!
\frac{\dd\vOm_i}{(\Delta\Omega)} \dd\chi_i \,
\cD_i^2 \frac{\dd M_i}{M_i} \int \!\! \dd\vtheta_{i'}\dd\chi_{i'}\cD_{i'}^2
\frac{\dd M_{i'}}{M_{i'}} \nonumber \\
&& \times \frac{\dd n}{\dd\ln M_i} \frac{\dd n}{\dd\ln M_i'}
\left[ 1+\xih_{i,i'} \right] ,
\label{C2-w-def}
\eeqa
\beqa
C_{i,j}^{(3)} & = & \frac{4}{(\Delta\Omega)\Qw_i\Qw_j} \int \!\!
\frac{\dd\vOm_i}{(\Delta\Omega)} \dd\chi_i \, \cD_i^2
\frac{\dd M_i}{M_i} \int \!\! \dd\vtheta_{i'}\dd\chi_{i'}\cD_{i'}^2
\frac{\dd M_{i'}}{M_{i'}}  \nonumber \\
&& \times \int \dd\vtheta_{j'}\dd\chi_{j'}\cD_{j'}^2
\frac{\dd M_{j'}}{M_{j'}} \frac{\dd n}{\dd\ln M_i} \frac{\dd n}{\dd\ln M_{i'}}
\frac{\dd n}{\dd\ln M_{j'}}  \nonumber \\
&& \times \left[ 1+ \xih_{i,i'} + \xih_{i,j'} + \xih_{i',j'} + \zetah_{i,i',j'} \right] ,
\label{C3-w-def}
\eeqa
\beqa
C_{i,j}^{(4)} & = & \frac{1}{\Qw_i\Qw_j} \int \!\! \frac{\dd\vOm_i}{(\Delta\Omega)}
\dd\chi_i \, \cD_i^2 \frac{\dd M_i}{M_i} \dd\vtheta_{i'}\dd\chi_{i'}\cD_{i'}^2 
\frac{\dd M_{i'}}{M_{i'}} \frac{\dd n}{\dd\ln M_i} \nonumber \\
&& \hspace{-0.7cm} \times \frac{\dd n}{\dd\ln M_{i'}}  \int \!\! 
\frac{\dd\vOm_j}{(\Delta\Omega)} \dd\chi_j
\, \cD_j^2 \frac{\dd M_j}{M_j} \dd\vtheta_{j'}\dd\chi_{j'}\cD_{j'}^2 
\frac{\dd M_{j'}}{M_{j'}} \frac{\dd n}{\dd\ln M_j} \nonumber \\
&& \hspace{-0.7cm} \times \frac{\dd n}{\dd\ln M_j'}  
\left[ 4 \xih_{i;j} \!+\! 2 \zetah_{i;j,j'} \!+\! 2 \zetah_{i,i';j}
\!+\! 2 \xih_{i;j'} \xih_{i';j} \!+\! \etah_{i,i';j,j'} \right] . \nonumber \\
&& 
\label{C4-w-def}
\eeqa

For the Landy \& Szalay estimator (\ref{wi-LS-1}) only a few of these terms
remain, as in Eqs.(\ref{C2-LS-def})-(\ref{C4-LS-def}), and we obtain
\beq
C_{i,j}^{\rm LS (2)} = C_{i,j}^{(2)} ,
\label{C2-LS-w-def}
\eeq
\beqa
C_{i,j}^{\rm LS (3)} & \!\!  = \!\! & \frac{4}{(\Delta\Omega)\Qw_i\Qw_j} \int\!\! 
\frac{\dd\vOm_i}{(\Delta\Omega)} \dd\chi_i \, \cD_i^2 \frac{\dd M_i}{M_i}
\int \!\! \dd\vtheta_{i'}\dd\chi_{i'}\cD_{i'}^2 \frac{\dd M_{i'}}{M_{i'}} \nonumber \\
&& \times \int \dd\vtheta_{j'}\dd\chi_{j'}\cD_{j'}^2
\frac{\dd M_{j'}}{M_{j'}} \frac{\dd n}{\dd\ln M_i} \frac{\dd n}{\dd\ln M_{i'}}
\frac{\dd n}{\dd\ln M_{j'}}  \nonumber \\
&& \times \left[ \xih_{i',j'} + \zetah_{i,i',j'} \right] ,
\label{C3-LS-w-def}
\eeqa
\beqa
C_{i,j}^{\rm LS (4)} & = & \frac{1}{\Qw_i\Qw_j} \int \!\! 
\frac{\dd\vOm_i}{(\Delta\Omega)} \dd\chi_i \, \cD_i^2
\frac{\dd M_i}{M_i} \dd\vtheta_i'\dd\chi_i'\cD_i'^2 \frac{\dd M_i'}{M_i'} 
\frac{\dd n}{\dd\ln M_i} \nonumber \\
&& \hspace{-0.7cm} \times \frac{\dd n}{\dd\ln M_i'}  
\int \!\! \frac{\dd\vOm_j}{(\Delta\Omega)} \dd\chi_j \, \cD_j^2
\frac{\dd M_j}{M_j} \dd\vtheta_j'\dd\chi_j'\cD_j'^2 \frac{\dd M_j'}{M_j'}
\frac{\dd n}{\dd\ln M_j} \nonumber \\
&& \hspace{-0.7cm} \times \frac{\dd n}{\dd\ln M_j'}  
\left[ 2 \xih_{i;j'} \xih_{i';j} + \etah_{i,i';j,j'} \right] ,
\label{C4-LS-w-def}
\eeqa
see also \citet{Szapudi2001a}, and \citet{Bernstein1994} who considers
(up to order $\nb^{-2}$ over the inverse of the mean density)
the additional terms associated with fluctuations of the denominator in the 
estimator (\ref{hxi-LS-DR}), when the latter is normalized to the number counts
in the same field.

\subsubsection{Low-order terms}
\label{Low-order-w}

Keeping only the contributions that are constant or linear over the two-point
halo correlation $\xih$, as in Eqs.(\ref{Cij-tot}) and (\ref{Cij-LS-tot}), we obtain
\beqa
C_{i,j} & = & \delta_{i,j} \, \frac{2}{(\Delta\Omega)\Qw_i} 
\left[ 1+\lag\hw_i\rag\right] + \frac{4}{(\Delta\Omega)\Nb} 
+ \frac{4}{(\Delta\Omega)\Nb^3} \nonumber \\
&& \times \int\dd\chi \, \cD^5 \, \bb^2 \, \nb^2
\, \left[ \overline{\xith_{i'}} + \overline{\xith_{j'}} + \overline{\xith_{i',j'}} \right]
\nonumber \\
&& + \frac{4}{\Nb^2} \int\dd\chi \, \cD^5 \, \bb^2 \, \nb^2 \, \xicyl ,
\label{Cij-w-tot}
\eeqa
and
\beq
C_{i,j} ^{\rm LS} = \delta_{i,j} \, \frac{2 [ 1+\lag\hw_i\rag]}{(\Delta\Omega)\Qw_i} 
+ \frac{4}{(\Delta\Omega)\Nb^3} \int\dd\chi \, \cD^5 \, \bb^2 \, \nb^2 \,
\overline{\xith_{i',j'}} 
\label{Cij-LS-w-tot}
\eeq
where, in a fashion similar to Eq.(\ref{I3-ij-xi-def}), we introduced the average
\beq
\overline{\xith_{i',j'}}(z) = \int_i \frac{\dd\vtheta_{i'}}{\cA_i}
\int_j \frac{\dd\vtheta_{j'}}{\cA_j} \int \frac{\dd\chi_{j'}}{\cD} \; 
\xi(|\vx_{i'}-\vx_{j'}|;z) .
\label{I2ij-def-xi}
\eeq
The Fourier-space expression of Eq.(\ref{I2ij-def-xi}) is given in Eq.(\ref{I2ij-def}).

\paragraph{Comparison of Peebles \& Hauser and Landy \& Szalay covariance matrices}
\label{Comparison-of-matrices-w}
~~\\

\begin{figure}
\begin{center}
\epsfxsize=8.5 cm \epsfysize=6.5 cm {\epsfbox{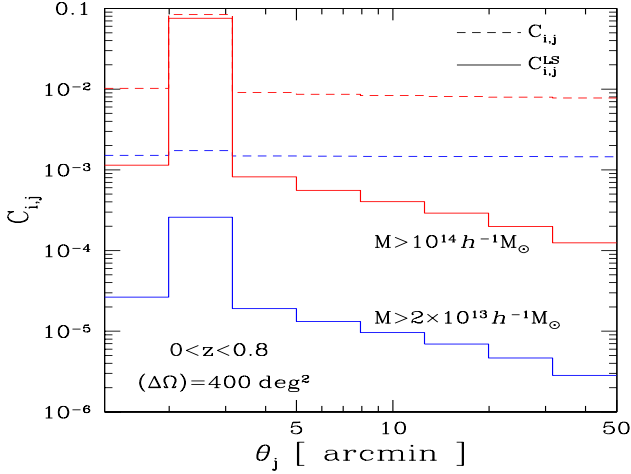}}
\end{center}
\caption{The covariance matrices $C_{i,j}^{\rm LS}$ (solid line) and 
$C_{i,j}$ (dashed line) of the estimators $\hwLS_i$ and $\hw_i$,
for $i=2$ associated with the angular bin $2<\theta<3.2$ arcmin, as
a function of $j$. We show the results obtained for halos in the redshift range
$0<z<0.8$, with an angular window of $400$ deg$^2$, above the mass thresholds
$M_*=2\times 10^{13}$ and $10^{14}h^{-1} M_{\odot}$, from bottom to top.
Here we only consider the low-order contributions, given by
Eqs.(\ref{Cij-w-tot}) and (\ref{Cij-LS-w-tot}).}
\label{fig_Rij_CXiw_PH_LS}
\end{figure}

We compare in Fig.~\ref{fig_Rij_CXiw_PH_LS} the covariances matrices
(\ref{Cij-w-tot}) and (\ref{Cij-LS-w-tot}) as a function of $j$ at fixed $i$.
As in Fig.~\ref{fig_wi_Horizon}, we consider halos in the redshift range
$0<z<0.8$ in a survey of area $400$ deg$^2$.
As was the case for the 3D real-space correlation $\xi$ shown in
Fig.~\ref{fig_Rij_CXiR_PH_LS}, and in agreement with previous works
\citep{Kerscher2000}, the covariance is much smaller and more diagonal for
the Landy \& Szalay estimator (\ref{wi-LS-1}) than for the Peebles \& Hauser 
estimator (\ref{wi-1}).
This can also be clearly seen from the comparison of the left and middle
panels of Fig.~\ref{fig_Rij_Cw}, where we show the correlation matrices
associated with Eqs.(\ref{Cij-w-tot}) and (\ref{Cij-LS-w-tot}).

\paragraph{Comparison of sample-variance and shot-noise effects}
\label{Comparison-of-sv-and-sn-w}
~~\\

\begin{figure}
\begin{center}
\epsfxsize=8.5 cm \epsfysize=6.5 cm {\epsfbox{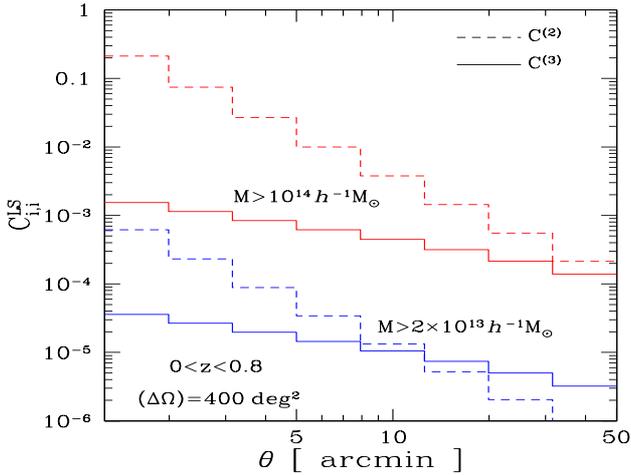}}
\end{center}
\caption{The contributions $C^{(2)}$ and $C^{(3)}$ to the covariance
of the Landy \& Szalay estimator, along the diagonal $i=j$.
As in Fig.~\ref{fig_Rij_CXiw_PH_LS}, we only consider the low-order terms,
given by Eq.(\ref{Cij-LS-w-tot}).}
\label{fig_Ciiw_sn_sv}
\end{figure}

Next, we compare in Fig.~\ref{fig_Ciiw_sn_sv} the contributions
$C^{(2)}$ (first term in Eq.(\ref{Cij-LS-w-tot})) and $C^{(3)}$ (second term
in Eq.(\ref{Cij-LS-w-tot})), again keeping only these low-order terms of
the covariance of the Landy \& Szalay estimator.
As compared with $C^{(3)}$, $C^{(2)}$ involves an extra degree
of shot noise (one more pair identification).
Taking only these low-order terms into account, the covariance is dominated by 
$C^{(2)}$ (whence shot-noise effects are dominant) below $10$ arcmin
for halos above $M_*=2\times 10^{13}h^{-1} M_{\odot}$, and below
$50$ arcmin for halos above $M_*=10^{14}h^{-1} M_{\odot}$.
As for the 3D correlation, shot-noise effects dominate up to larger scales
for more massive and rare halos, and this also implies that their covariance
matrix is more strongly diagonal.

\subsubsection{High-order terms}
\label{High-order-w}

\begin{figure}
\begin{center}
\epsfxsize=8.5 cm \epsfysize=6.5 cm {\epsfbox{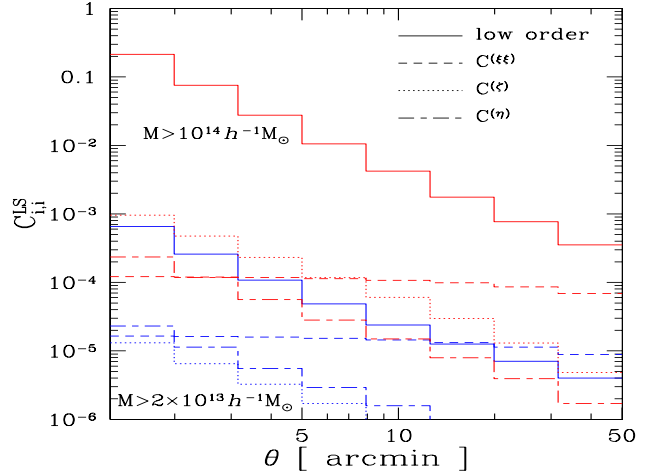}}
\end{center}
\caption{The low- and high-order contributions to the covariance matrix
$C_{i,j}^{\rm LS}$ along its diagonal. We again consider halos in the redshift
range $0<z<0.8$, with an angular window of $400$ deg$^2$, above two mass
thresholds.}
\label{fig_Ciiw_NG}
\end{figure}

\begin{figure}
\begin{center}
\epsfxsize=8.5 cm \epsfysize=6.5 cm {\epsfbox{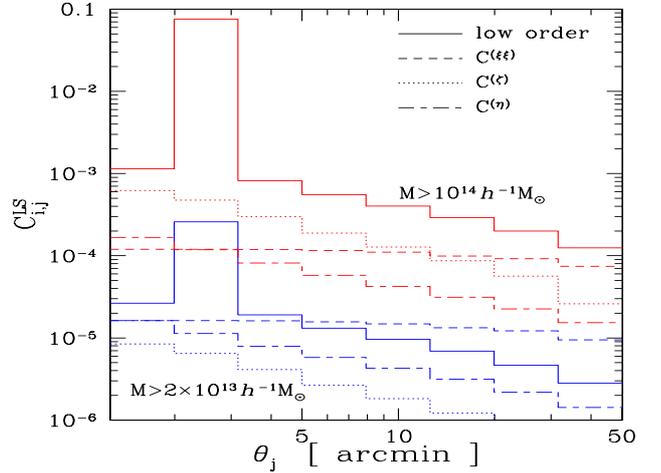}}
\end{center}
\caption{The low- and high-order contributions to the covariance matrix
$C_{i,j}^{\rm LS}$, as in Fig.~\ref{fig_Ciiw_NG}, but along one row.
This corresponds to the fixed bin $i=2$, associated with the angular bin
$2<\theta<3.2$ arcmin, as a function of $j$.}
\label{fig_Rij_CXiw_NG}
\end{figure}

At small angular separations, the high-order terms in Eqs.(\ref{C3-LS-w-def})
and (\ref{C4-LS-w-def}), associated with the product $\xi\xi$ and the three- and
four-point correlations $\zeta$ and $\eta$, are not negligible.
As in Sect.~\ref{high-order-terms} and in \citet{Bernstein1994},
to estimate these high-order correlations
we use the ``hierarchical clustering ansatz'' shown in Figs.~\ref{fig-zeta}
and \ref{fig-eta} and given by Eqs.(\ref{zeta-def})-(\ref{S4-def}).
As described in App.~\ref{Computation-high-order-terms-w}, we follow the
procedure that we have already used in
App.~\ref{Computation-high-order-terms} to compute
the high-order terms associated with the 3D correlation $\xi$. Then,
the contribution associated with the product $\xih_{i;j'} \xih_{i';j}$ in
Eq.(\ref{C4-LS-w-def}) writes as
\beq
C_{i,j}^{\rm LS (\xi\xi)} = \frac{2\pi^2}{\Nb^4} \int_0^2\dd y \, y \, A^{(2)}(y)
B_i^{(2)}(y\theta_s) B_j^{(2)}(y\theta_s) ,
\label{CLS-w-xixi-1}
\eeq
the term $\zetah_{i,i',j'}$ of Eq.(\ref{C3-LS-w-def}) yields
\beqa
C_{i,j}^{\rm LS (\zeta)} & = & \frac{4}{(\Delta\Omega)\Nb^4} \int \dd\chi \,
\cD^8 \, \bb^3 \, \nb^3 \, \frac{S_3}{3} \left[ \overline{\xith_{i'}} \times
\overline{\xith_{j'}} \right. \nonumber \\
&& \left. + \overline{\xith_{i',i}\xith_{i',j'}} + \overline{\xith_{j',j}\xith_{j',i'}} \right] ,
\label{CLS-w-zeta-1}
\eeqa
and the term $\etah_{i,i';j,j'}$ of Eq.(\ref{C4-LS-w-def}) gives
\beqa
C_{i,j}^{\rm LS (\eta)} & = & \frac{2}{\Nb^4} \int \dd\chi \, \cD^{11} \, 
\bb^4 \, \nb^4 \, \frac{S_4}{16} \left[ \overline{\xith_{i'}} \times
\overline{\xi_{i;j}\xith_{i;j'}} \right. \nonumber \\
&& + \overline{\xith_{j'}} \times
\overline{\xi_{i;j}\xith_{j;i'}} + 2  \, \overline{\xith_{i'}} \times \overline{\xith_{j'}}
\times \xicyl \nonumber \\
&& \left. + 2 \, \overline{\xith_{j',i}\xi_{i;j}\xith_{j;i'}}
+ \overline{\xith_{j';i}\xith_{i,i'}\xith_{i';j}}
+ \overline{\xith_{i';j}\xith_{j,j'}\xith_{j';i}} \right]
\label{CLS-w-eta-1}
\eeqa
where the various factors are given in
App.~\ref{Computation-high-order-terms-w}.

We compare in Figs.~\ref{fig_Ciiw_NG} and \ref{fig_Rij_CXiw_NG} these high-order
contributions (\ref{CLS-w-xixi-1})-(\ref{CLS-w-eta-1}) with the low-order
contribution (\ref{Cij-LS-w-tot}), for the covariance matrix of the Landy \& Szalay
estimator. We recover the qualitative behavior encountered in
Figs.~\ref{fig_CiiR_NG} and \ref{fig_Rij_CXiR_NG} for the estimator of the 3D
correlation function. The ``$\zeta$" and ``$\eta$'' terms (\ref{CLS-w-zeta-1})
and (\ref{CLS-w-eta-1}) show a strong dependence on the smoothing scales,
while the ``$\xi\xi$'' term (\ref{CLS-w-xixi-1}) shows a very weak dependence.
Again, this is because the contribution (\ref{CLS-w-xixi-1}) involves
the product of two correlations between two distinct lines of sight, so that
each $\xi$ is averaged over the angular window $\theta_s$ of the survey, as seen
in Eq.(\ref{C4-LS-w-def}), whereas the contribution (\ref{CLS-w-zeta-1}) involves
the product of two correlations between a central point and two points at
angular distances $\theta_i$ and $\theta_j$, as seen in Eq.(\ref{C3-LS-w-def}).

For the case of massive halos, $M>10^{14}h^{-1} M_{\odot}$, 
the high-order terms are negligible along the diagonal, which is dominated by the
shot-noise term, and only give a modest
contribution to off-diagonal entries. Then, the covariance matrix remains strongly
diagonal (for the angular bins studied here).

For the case of low-mass halos, $M>2\times 10^{13}h^{-1} M_{\odot}$, the
high-order contribution (\ref{CLS-w-xixi-1}) to the diagonal is no longer
negligible for
$\theta>5$ arcmin, while the two other contributions (\ref{CLS-w-zeta-1}) and
(\ref{CLS-w-eta-1}), which involve the three- and four-point correlation
functions, are always subdominant on these scales. This is a convenient
property since the modelization of high-order many-body correlations 
is increasingly difficult. However, this was not the case for the 3D correlation
$\xi$, as seen in Figs.~\ref{fig_CiiR_NG} and \ref{fig_Rij_CXiR_NG}, except on large
scales. For off-diagonal entries, the high-order contribution (\ref{CLS-w-xixi-1})
can become dominant for widely separated angular scales, while on small scales,
$\theta\sim 1$ arcmin, all contributions are of the same order of magnitude.

\subsubsection{Comparison with numerical simulations}
\label{comparison-Cij-w}

\begin{figure}
\begin{center}
\epsfxsize=8.5 cm \epsfysize=6.5 cm {\epsfbox{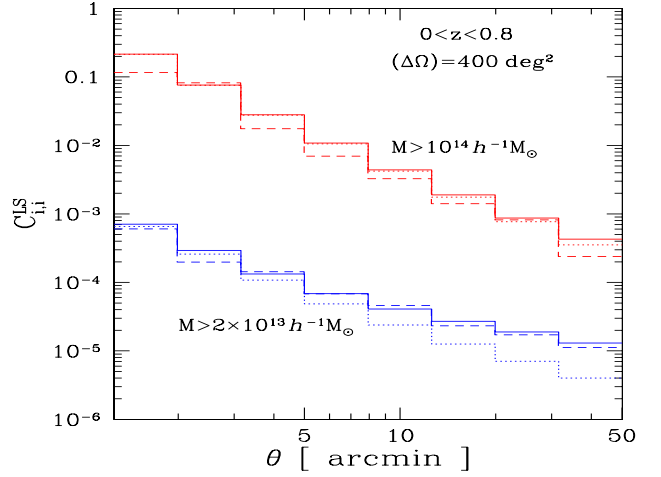}}
\end{center}
\caption{The covariance matrix $C_{i,j}^{\rm LS}$ along its diagonal. We show
our analytical results including all contributions (solid lines) or only low-order
terms (dotted lines), and results from numerical simulations (dashed lines).}
\label{fig_Cwii_Horizon}
\end{figure}

\begin{figure}
\begin{center}
\epsfxsize=8.5 cm \epsfysize=6.5 cm {\epsfbox{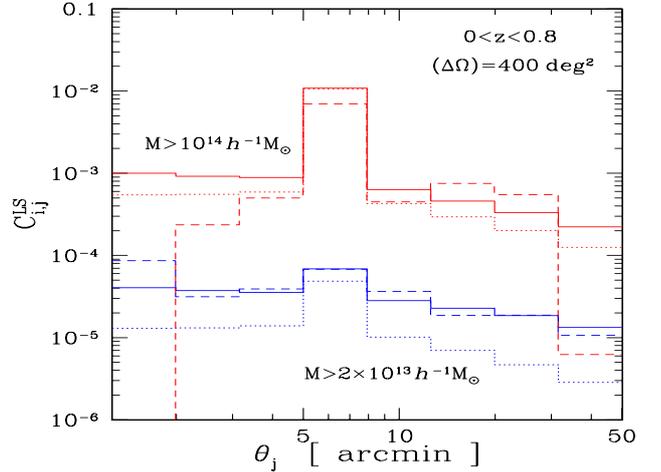}}
\end{center}
\caption{The covariance matrix $C_{i,j}^{\rm LS}$, as in
Fig.~\ref{fig_Cwii_Horizon}, but along one row. This corresponds to the
fixed bin $i=4$, associated with the angular bin $5<\theta<8$ arcmin, as a
function of $j$.}
\label{fig_Rij_Cw_Horizon}
\end{figure}

As for the 3D correlation, we show the covariance matrix $C_{i,j}^{\rm LS}$,
along its diagonal and along one row, in Figs.~\ref{fig_Cwii_Horizon} and
\ref{fig_Rij_Cw_Horizon}.
Again we obtain a reasonable agreement with the numerical simulations.
For moderate-mass halos, the high-order contributions are again necessary
to obtain a good match on large scales for diagonal entries and on most scales
for off-diagonal entries. 
The off-diagonal terms of the covariance matrix obtained from the numerical
simulations are rather noisy, and our analytical results are competitive
in obtaining reliable estimates.

\subsubsection{Correlation matrices}
\label{correlation-matrices-w}

\begin{figure*}
\begin{center}
\epsfxsize=5.7 cm \epsfysize=5.7 cm {\epsfbox{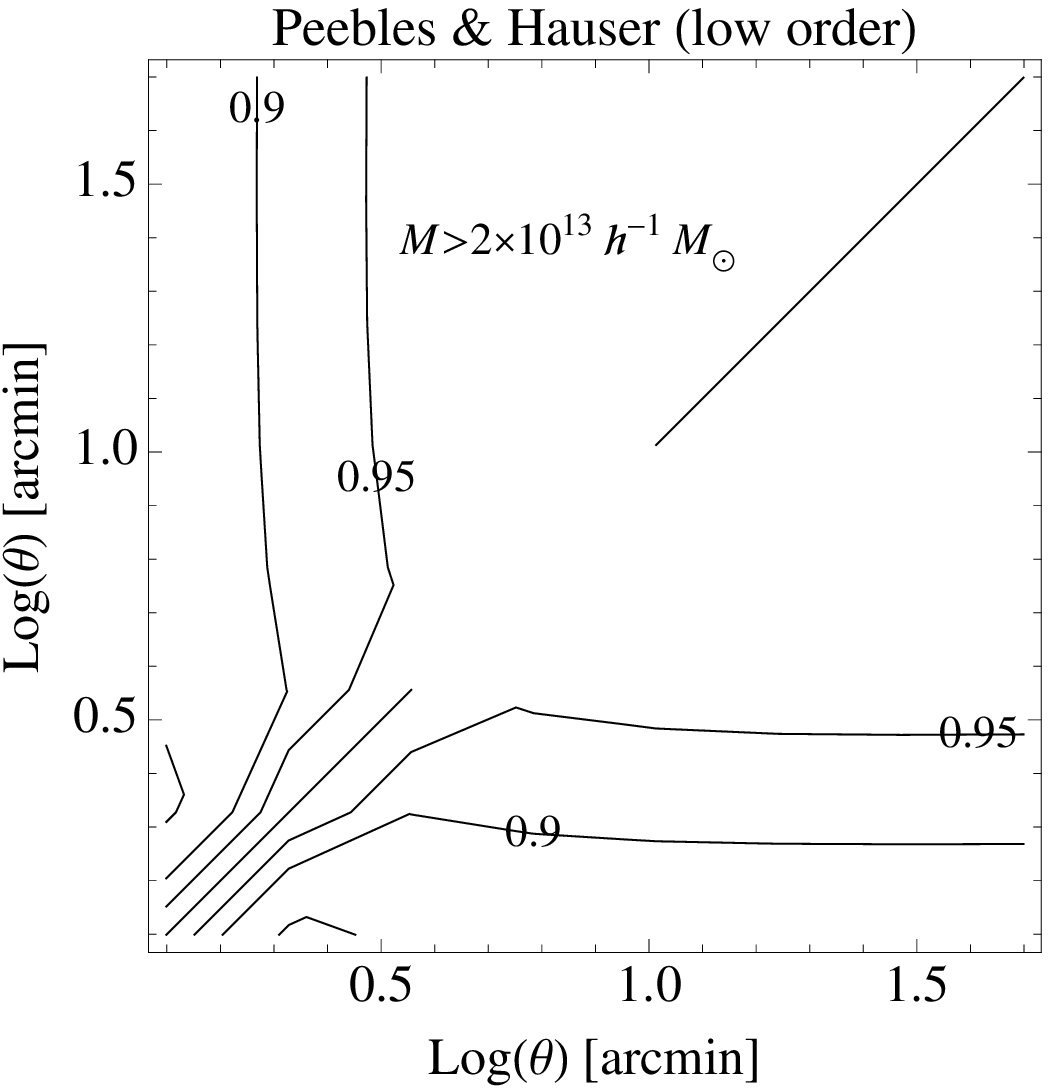}}
\epsfxsize=5.7 cm \epsfysize=5.7 cm {\epsfbox{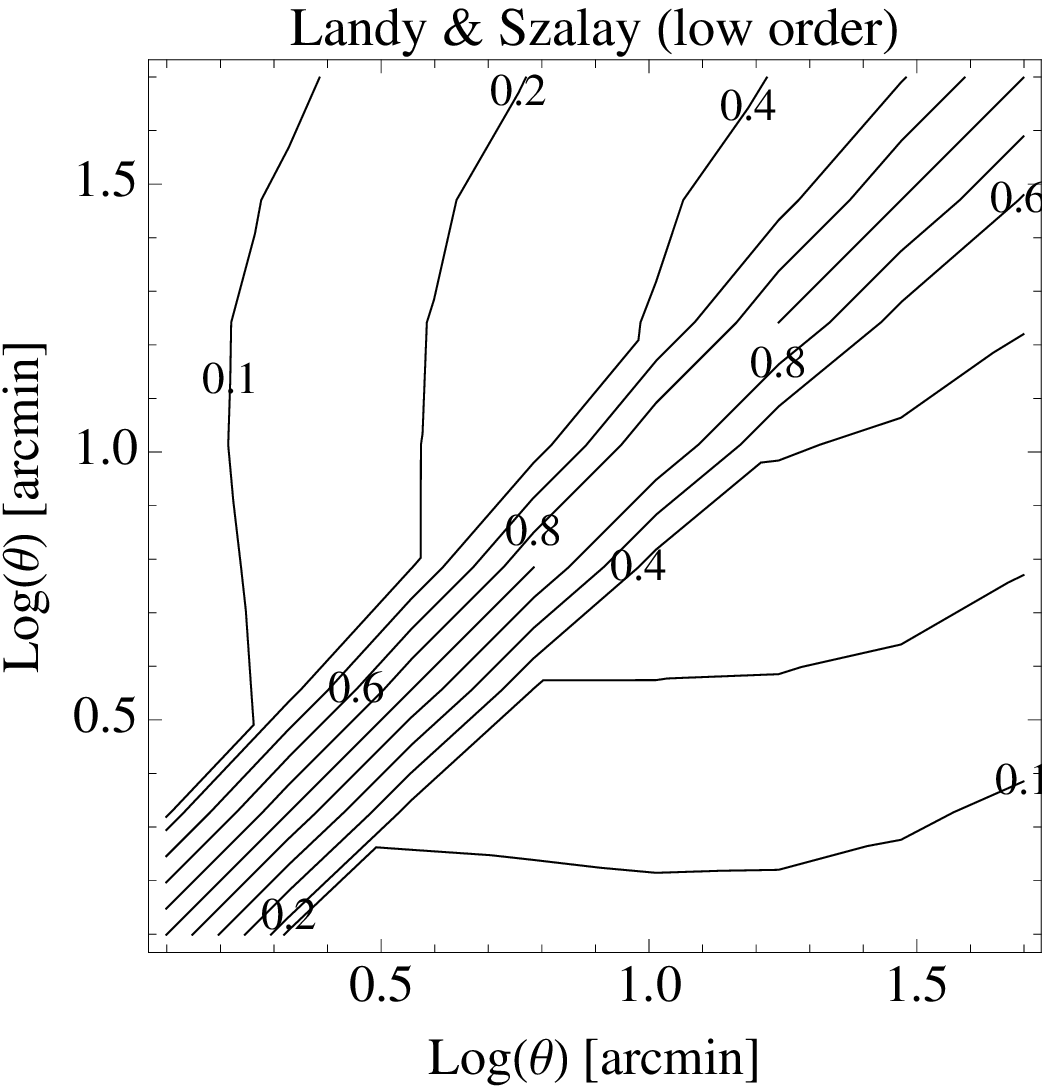}}
\epsfxsize=5.7 cm \epsfysize=5.7 cm {\epsfbox{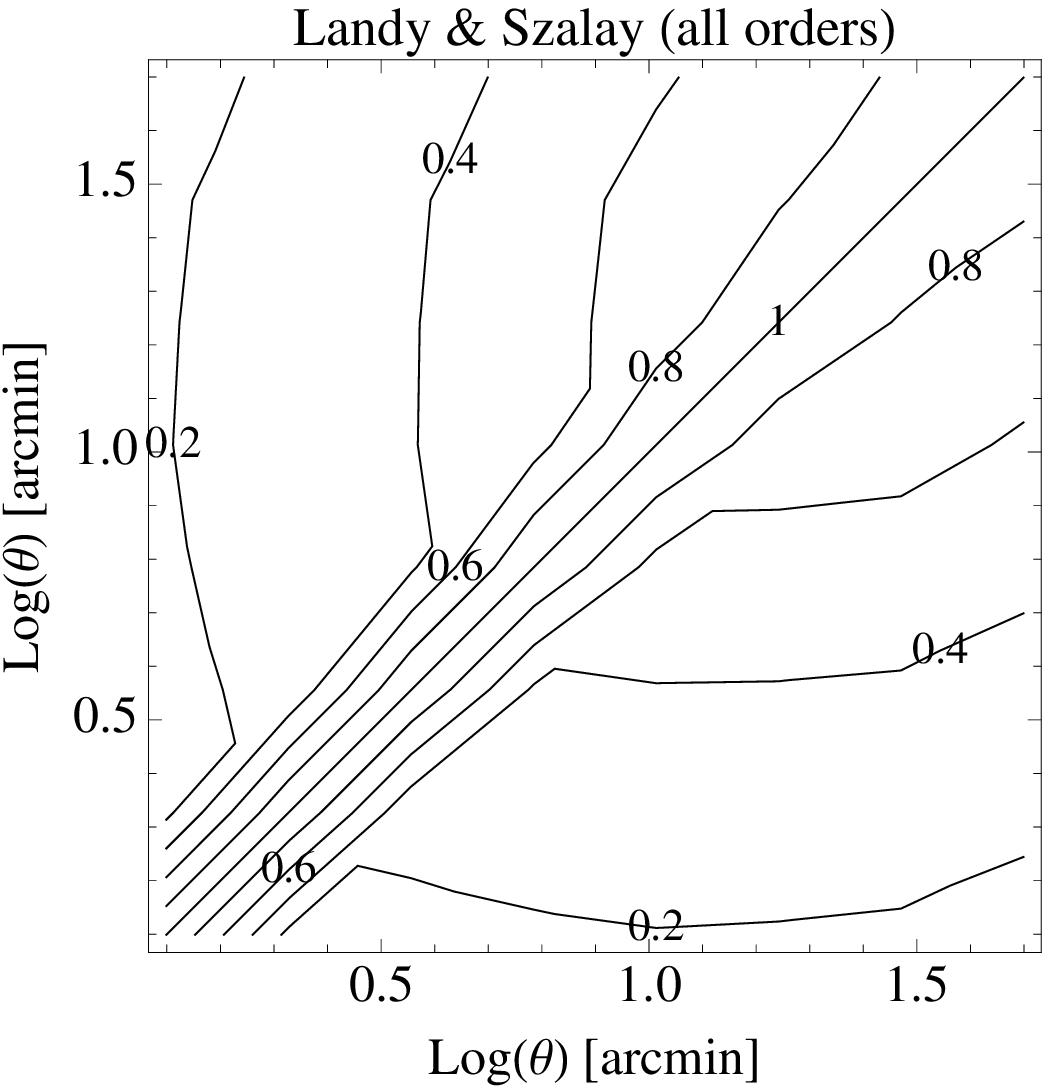}}\\
\epsfxsize=5.7 cm \epsfysize=5.7 cm {\epsfbox{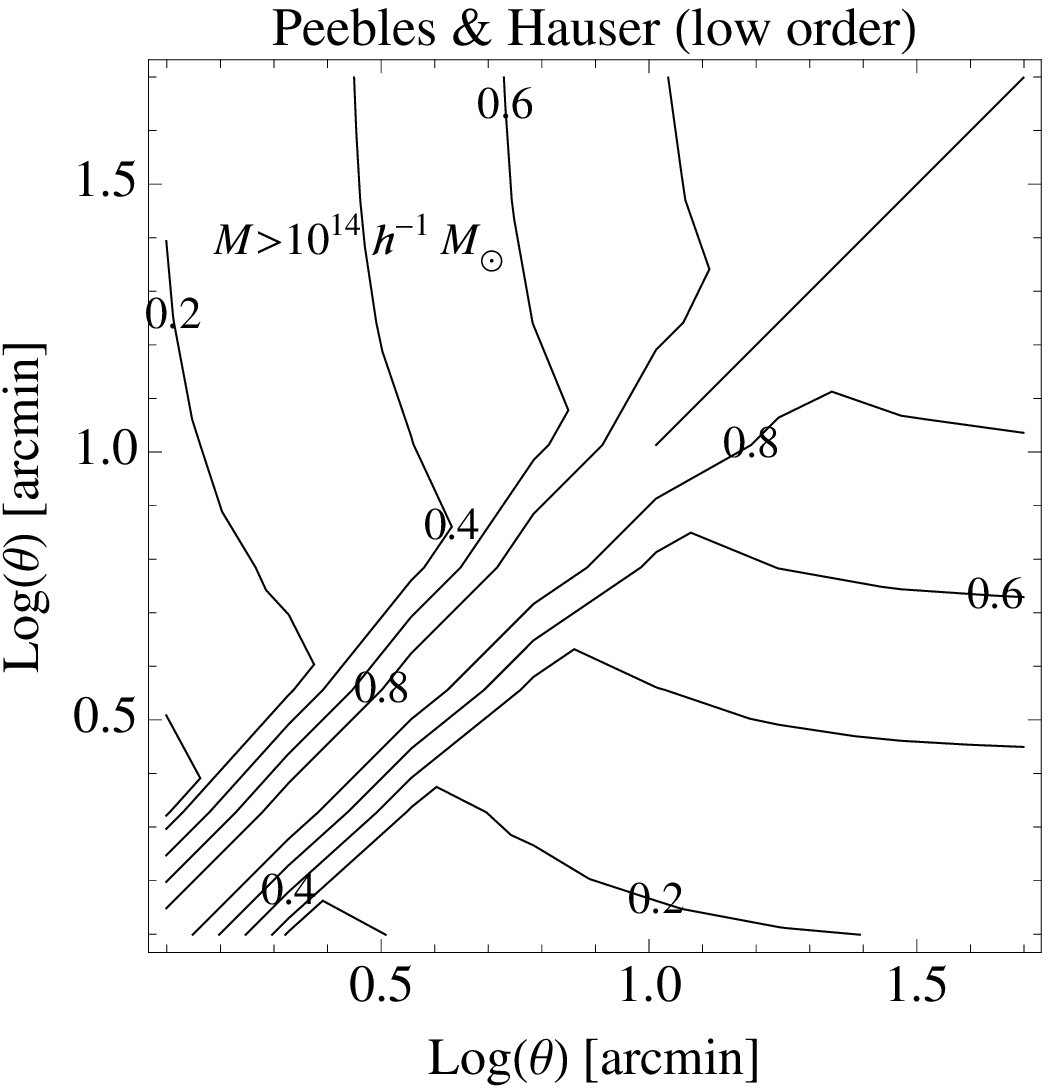}}
\epsfxsize=5.7 cm \epsfysize=5.7 cm {\epsfbox{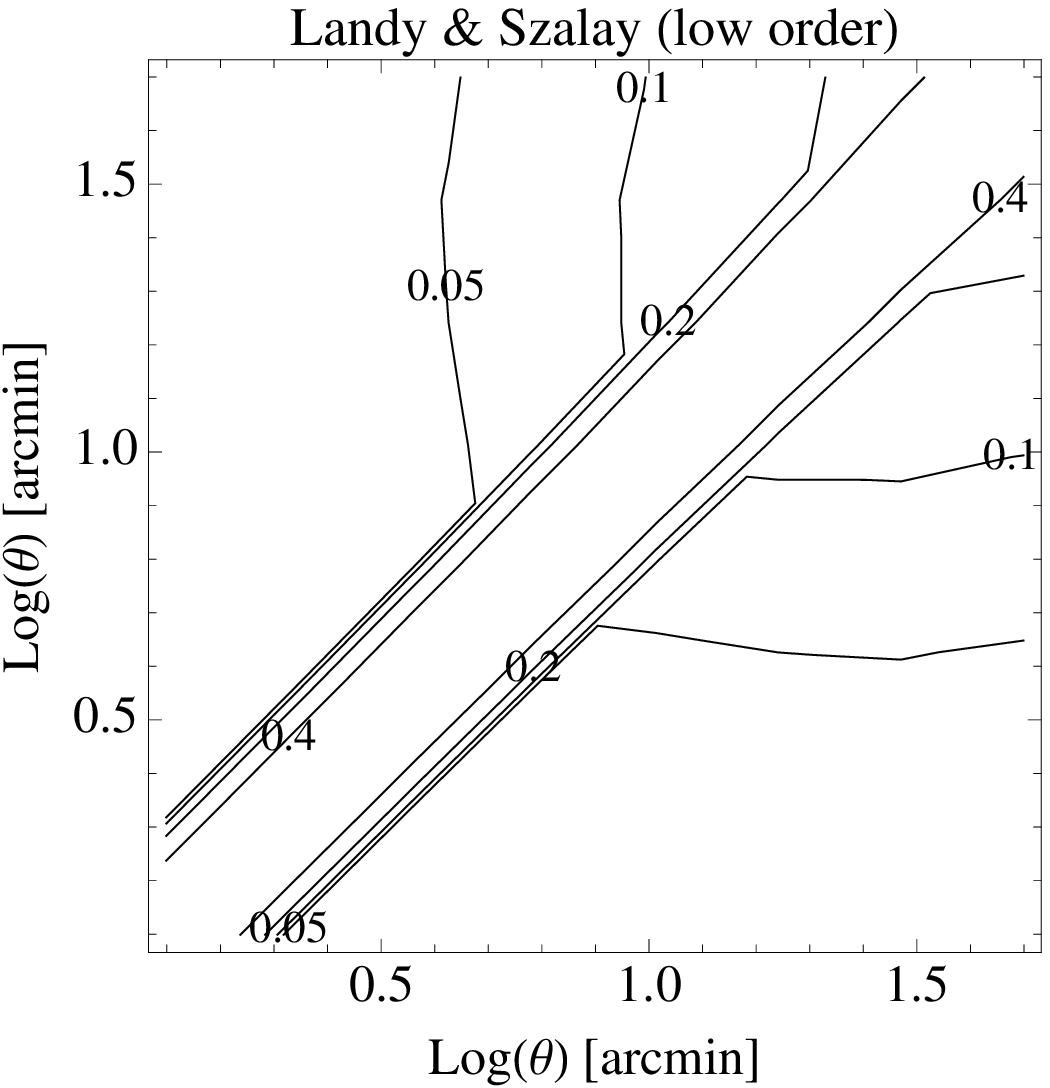}}
\epsfxsize=5.7 cm \epsfysize=5.7 cm {\epsfbox{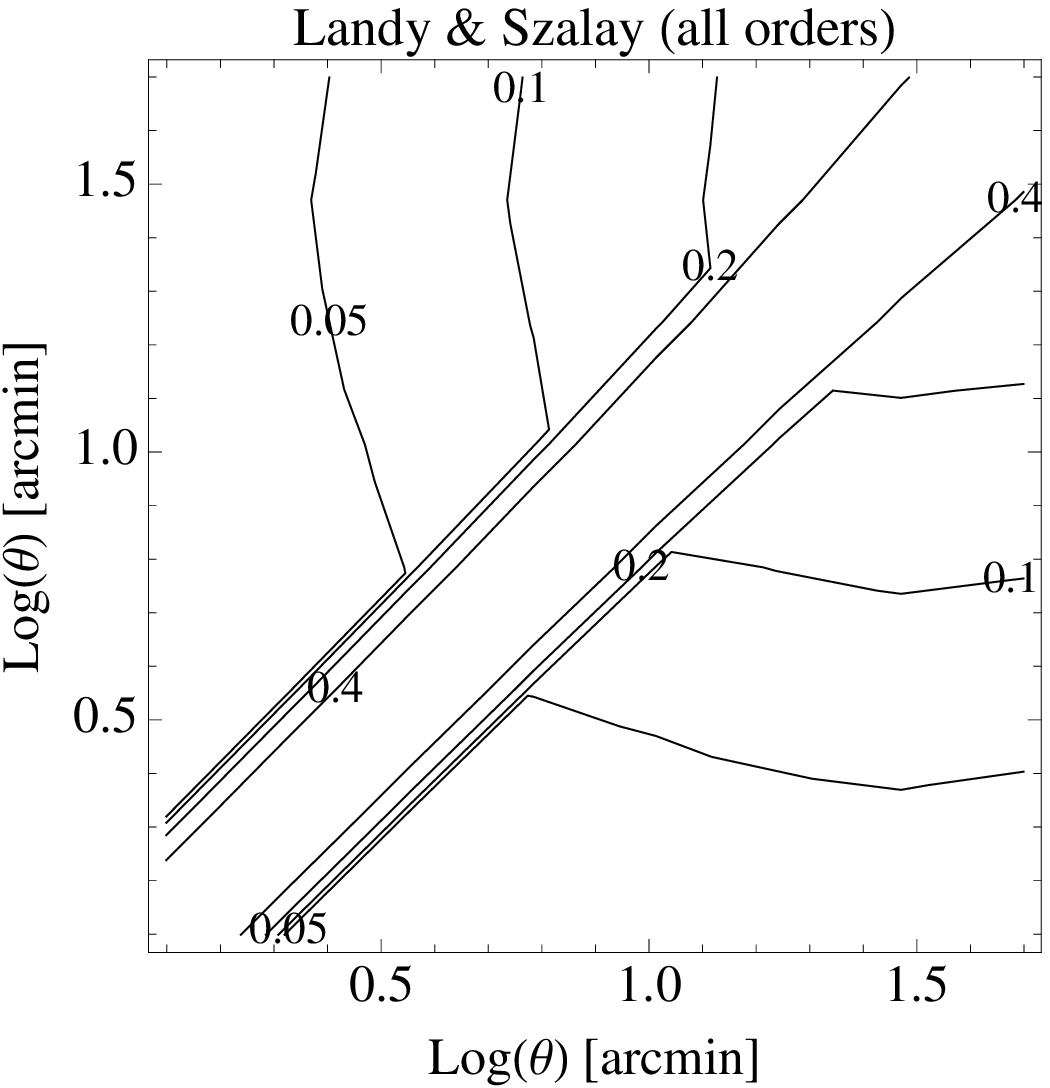}}
\end{center}
\caption{Contour plots for the correlation matrix $\cR_{i,j}$, defined
as in Eq.(\ref{correlation-matrix}) but for the full covariance matrix $C_{ij}$
of the halo angular correlation. 
There are eight angular bins, over $1.25<r<50$arcmin, equally spaced in
$\log(\theta)$, as in previous figures.
We consider halos in the redshift range $0<z<0.8$, with an angular window
of $400$ deg$^2$, above the mass thresholds 
$M>2\times 10^{13}h^{-1} M_{\odot}$ in the {\it upper row}, and
$M>10^{14}h^{-1} M_{\odot}$ in the {\it lower row}.
{\it Left panels:} Low-order contributions (\ref{Cij-w-tot}) for the Peebles \&
Hauser estimator.
{\it Middle panels:} Low-order contributions (\ref{Cij-LS-w-tot}) for the
Landy \& Szalay estimator.
{\it Right panels:} Full correlation matrix, including the high-order
contributions of Eqs.(\ref{CLS-w-xixi-1})-(\ref{CLS-w-eta-1}), for the
Landy \& Szalay estimator.}
\label{fig_Rij_Cw}
\end{figure*}

We show in Fig.~\ref{fig_Rij_Cw} the correlation matrices $\cR_{i,j}$, defined
as in Eq.(\ref{correlation-matrix}), but for the full covariance matrices
$C_{i,j}$ of the estimators of the halo angular correlation.
As for the 3D correlation, we can check that, keeping only low-order terms,
the correlation matrix of the Landy \& Szalay estimator (\ref{wi-LS-1}) is
much more diagonal than for the Peebles \& Hauser estimator (\ref{wi-1}).
Taking high-order contributions into account makes the matrix slightly
less diagonal, but it still remains significantly diagonal,
in agreement with Fig.~\ref{fig_Rij_CXiw_NG}.
As in the 3D case, the correlation matrix is much more diagonal for massive
halos, where shot-noise effects are more important.
The full angular correlation matrix is more diagonal than
its 3D counterpart shown in the right hand panels of Fig.~\ref{fig_Rij_CXi},
and the correlations between small and large angular scales are not as strong
as the correlations between the small and large radii found in
Fig.~\ref{fig_Rij_CXi}.
In particular, off-diagonal entries and high-order contributions play
a less important role (although for low-mass halos it is still useful to take them
into account).

\section{Applications to real survey cases}
\label{Applications}

In this section, we compare the statistical significance of the number counts and
of the 3D correlation function for future large cosmological cluster surveys
(we give in App.~\ref{selection} the selection functions that we use for some of
these surveys).
Here we must note that, while redshift-space distortions only have a low
impact on angular number densities (number counts and angular correlations)
for wide redshift bins, they can more strongly affect 3D clustering.
In principle, redshift distortions could be corrected to recover a real-space map
if the velocity field is known, and when also applying a finger-of-god compression, 
but this would require a rather complete spectroscopic follow up, so it is not very
practical.
Therefore, observations instead provide redshift-space 3D correlations.
Then, the results discussed in this section for 3D correlations should be
seen as a first step toward more accurate computations.

Nevertheless, a simple estimate shows that these redshift-space distortions should
not strongly affect our results. Indeed, we find in the numerical simulations
that at $z=0.5$ for instance clusters have peculiar velocities $v$ on the order of
$300$ km/s along each axis. The redshift-space coordinate $s_{\parallel}$ along the
line of sight is given by $s_{\parallel}=x_{\parallel}+v_{\parallel}/(aH)$.
This yields a typical error $\Delta x_{\parallel}$
for the cluster comoving coordinate on the order of $3.6 h^{-1}$ Mpc.
This is not much larger than the typical size of the clusters, which ranges from
$1$ to $2 h^{-1}$ Mpc.
Then, for distance bins that are larger than $20 h^{-1}$ Mpc we can expect
redshift distortions to affect our results on the covariance matrices by about
$20\%$. The net effect should actually be smaller because the 3D estimators
also include information on clustering along the transverse directions, which
are not contaminated by the cluster peculiar velocities.
We leave an explicit computation of these redshift-space distortions to future works.

\subsection{Surveys of limited areas}
\label{limited-areas}

\begin{figure}[htb]
\begin{center}
\epsfxsize=8.5 cm \epsfysize=6.5 cm {\epsfbox{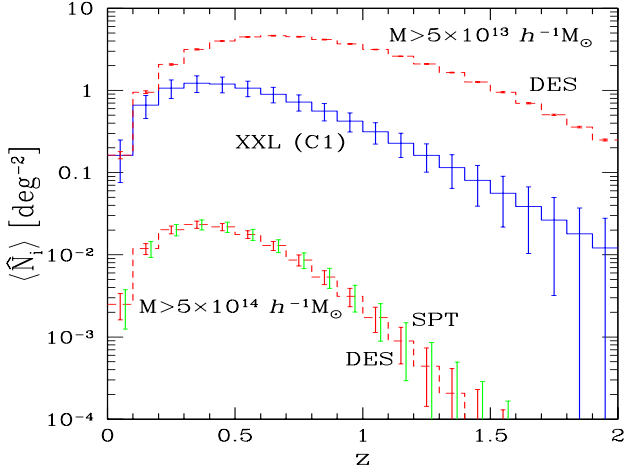}}
\end{center}
\caption{The mean angular number densities of X-ray clusters per square degree,
within redshift bins of width $\Delta z= 0.1$, for the XXL, DES, and SPT surveys.
Error bars contain both the shot-noise and sample-variance contributions,
from Eqs.(\ref{Cij-sn}) and (\ref{Cij-7}). 
For DES we consider the mass thresholds $M>5\times 10^{13}h^{-1} M_{\odot}$
and $M>5\times 10^{14}h^{-1} M_{\odot}$ (smaller error bars), and for SPT
the mass threshold $M>5\times 10^{14}h^{-1} M_{\odot}$ (larger error bars
shifted to the right).}
\label{fig_lnz_XXLall}
\end{figure}

We first consider several surveys of clusters of galaxies on limited angular
windows.

- The XXL survey \citep{Pierre2011} is an XMM Very Large Programme 
specifically designed to constrain the equation of state of the dark energy 
by using clusters of galaxies. It consists of two $5\times 5$ \degs\ areas and 
probes massive clusters out to a redshift of $\sim 2$. The well-characterized 
cluster selection function relies on the fact that clusters of galaxies are the 
only extended extragalactic sources, so that the selection operates in a 
two-dimensional  parameter space (equivalent to flux and spatial extent), allowing 
for different degrees of contamination by misclassified point sources.
We show the mass detection probabilities as a function of redshift in the left hand
panel of Fig.~\ref{fig_Fs_XXL_Planck_Erosita}, for the C1 selection. The 
space density of this population is $\sim 6 {\rm deg}^{-2}$.
This complex selection function $F(M,z)$, which differs from a simple mass
or X-ray flux threshold (see also \citet{Pacaud2006,Pacaud2007}), is readily
included in our formalism through a redefinition of the halo mass function,
$n(M,z) \rightarrow F(M,z) n(M,z)$.

- The Dark Energy Survey (DES) is an optical imaging survey to cover $5,000$
\degs\ with the Blanco 4-meter telescope at the Cerro Tololo Inter-American
Observatory\footnote{https://www.darkenergysurvey.org/index.shtml}.
We consider the expected mass threshold
$M>5\times 10^{13}h^{-1} M_{\odot}$, as well as the subset of massive 
clusters $M>5\times 10^{14}h^{-1} M_{\odot}$, since a binning over 
mass should help in deriving tighter constraints on cosmology.

- The South Pole Telescope (SPT) operates at millimeter
wavelengths\footnote{http://pole.uchicago.edu/}. It  will cover some
$2,500$ \degs\ at three frequencies, aiming at detecting clusters of galaxies 
from the Sunyaev-Zel'dovich (S-Z) effect. A preliminary survey of
$178$ \degs\ at 150 GHz reveals some 20 clusters down to a depth of
18 $\mu K$. Extensive simulations allow the determination of the mass 
completeness level, above a given significance for these secondary CMB anisotropies 
\citep{Vanderlinde2010}. This gives a mass threshold on the order of
$5\times 10^{14}h^{-1} M_{\odot}$.

\begin{figure}
\begin{center}
\epsfxsize=8.5 cm \epsfysize=6.5 cm {\epsfbox{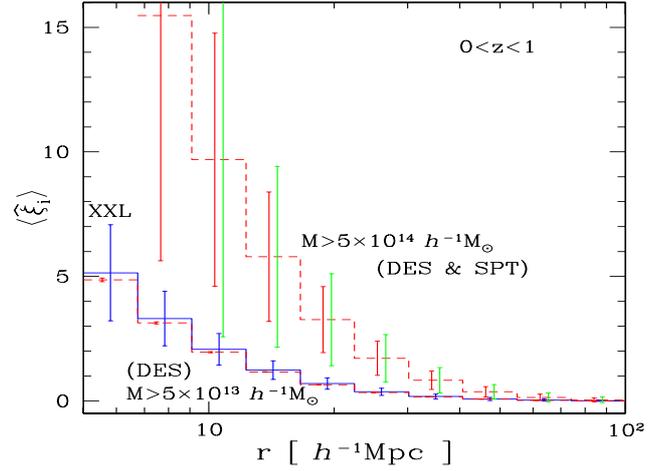}}
\end{center}
\caption{The mean correlation, $\lag\hxi_i\rag$ from
Eq.(\ref{xi-4}), over ten comoving distance bins within $5<r<100 h^{-1}$Mpc,
equally spaced in $\log(r)$. We integrate over halos within the redshift
interval $0<z<1$, for the XXL, DES, and SPT surveys, as in
Fig.~\ref{fig_lnz_XXLall} (again the error bars for SPT are slightly larger and shifted
to the right with respect to those of DES, for $M>5\times 10^{14}h^{-1} M_{\odot}$).
The error bars show the diagonal part of the covariance,
$\sqrt{C_{i,i}^{\rm LS}}$, for the Landy \& Szalay estimator, from
Eqs.(\ref{Cij-LS-tot}) and (\ref{CLS-xixi-1})-(\ref{CLS-eta-1}).}
\label{fig_XiR_z0to1_XXLall}
\end{figure}

\begin{figure}[htb]
\begin{center}
\epsfxsize=8.5 cm \epsfysize=6.5 cm {\epsfbox{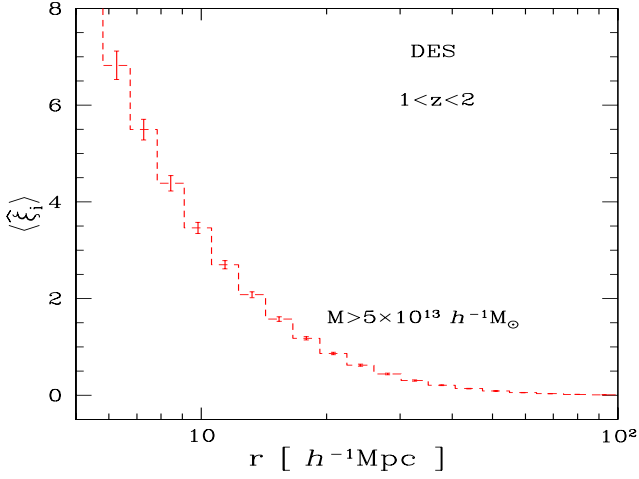}}
\end{center}
\caption{The mean correlation, $\lag\hxi_i\rag$, for the clusters detected by DES
over the redshift interval $1<z<2$. Here we consider $20$ distance bins within $5<r<100 h^{-1}$Mpc, equally spaced in $\log(r)$ (i.e. twice as many as in
Fig.~\ref{fig_XiR_z0to1_XXLall}).}
\label{fig_XiR_z1to2_DES}
\end{figure}

We show in Figs.~\ref{fig_lnz_XXLall}, \ref{fig_XiR_z0to1_XXLall}, and 
\ref{fig_XiR_z1to2_DES} the angular number densities and 3D correlations
expected for these various surveys. The error bars include all shot-noise and
sample-variance contributions (including high-order terms).
For the higher redshift interval, $1<z<2$, we only show the correlation of clusters
above $5\times 10^{13} h^{-1} M_{\odot}$ for DES, because in other cases
the error bars are too large to allow an accurate measure.
On the other hand, to take advantage of the good expected accuracy of this case
we consider in Fig.~\ref{fig_XiR_z1to2_DES} distance bins that are half the size
of those of Fig.~\ref{fig_XiR_z0to1_XXLall}.

As expected, the DES provides the best measures of cluster number
counts and correlations, hence the tightest constraints on cosmology,
thanks to its wide size, which provides a large number of objects.
However, the much smaller XXL survey already provides a meaningful measure
of both the abundance and the correlation of clusters, and appears to be
a promising tool. The SPT survey allows a useful measure of the number counts
as a function of redshift, but its rather high mass threshold leads to a relatively
small number of objects, hence large error bars for the 3D correlations,
even though a positive signal should still be within reach.
Assuming its expected mass threshold of
$M>5\times 10^{13}h^{-1} M_{\odot}$ remains valid over $1<z<2$, the DES
is the only survey among these three that allows an accurate measure of the
cluster correlation at high redshift, which should help to further constrain the
cosmology.

\subsection{All-sky surveys}
\label{All-sky}

\begin{figure}[htb]
\begin{center}
\epsfxsize=8.5 cm \epsfysize=6.5 cm {\epsfbox{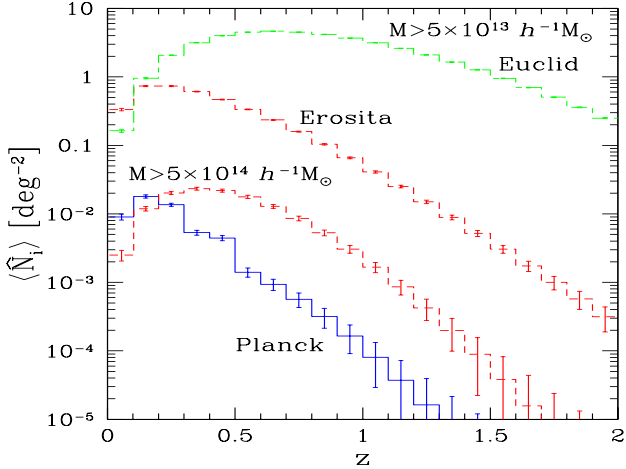}}
\end{center}
\caption{The mean angular number densities of clusters within redshift bins of width
$\Delta z= 0.1$. From top to bottom, we show a) halos above
$5\times 10^{13}h^{-1} M_{\odot}$ in Euclid, b) halos detected by Erosita
with the selection function of the right panel in Fig.~\ref{fig_Fs_XXL_Planck_Erosita},
c), halos above $5\times 10^{14}h^{-1} M_{\odot}$ in either Erosita or Euclid,
and d) halos detected by Planck with the selection function of the middle panel
in Fig.~\ref{fig_Fs_XXL_Planck_Erosita}.}
\label{fig_lnz_Planckall}
\end{figure}

\begin{figure}
\begin{center}
\epsfxsize=8.5 cm \epsfysize=6.5 cm {\epsfbox{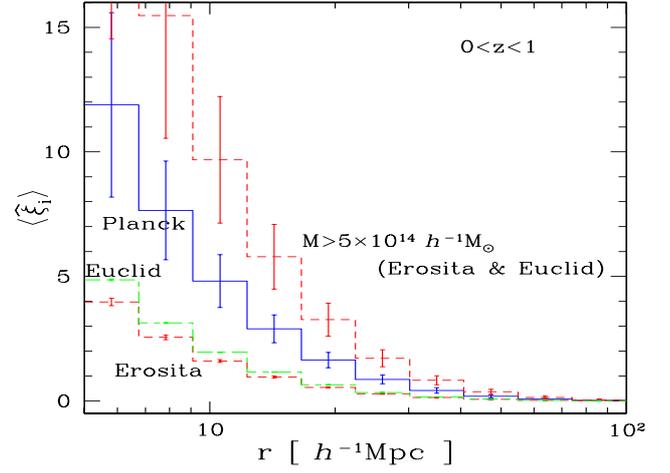}}
\end{center}
\caption{The mean correlation, $\lag\hxi_i\rag$, integrated over $0<z<1$, as in 
Fig.~\ref{fig_XiR_z0to1_XXLall}.
From top to bottom, we show a) halos above $5\times 10^{14}h^{-1} M_{\odot}$ in 
either Erosita or Euclid, b) halos detected by Planck with the selection function of the 
middle panel in Fig.~\ref{fig_Fs_XXL_Planck_Erosita}, c) halos above
$5\times 10^{13}h^{-1} M_{\odot}$ in Euclid, and d) halos detected by Erosita
with the selection function of the right panel in Fig.~\ref{fig_Fs_XXL_Planck_Erosita}.}
\label{fig_XiR_z0to1_Planckall}
\end{figure}

\begin{figure}
\begin{center}
\epsfxsize=8.5 cm \epsfysize=6.5 cm {\epsfbox{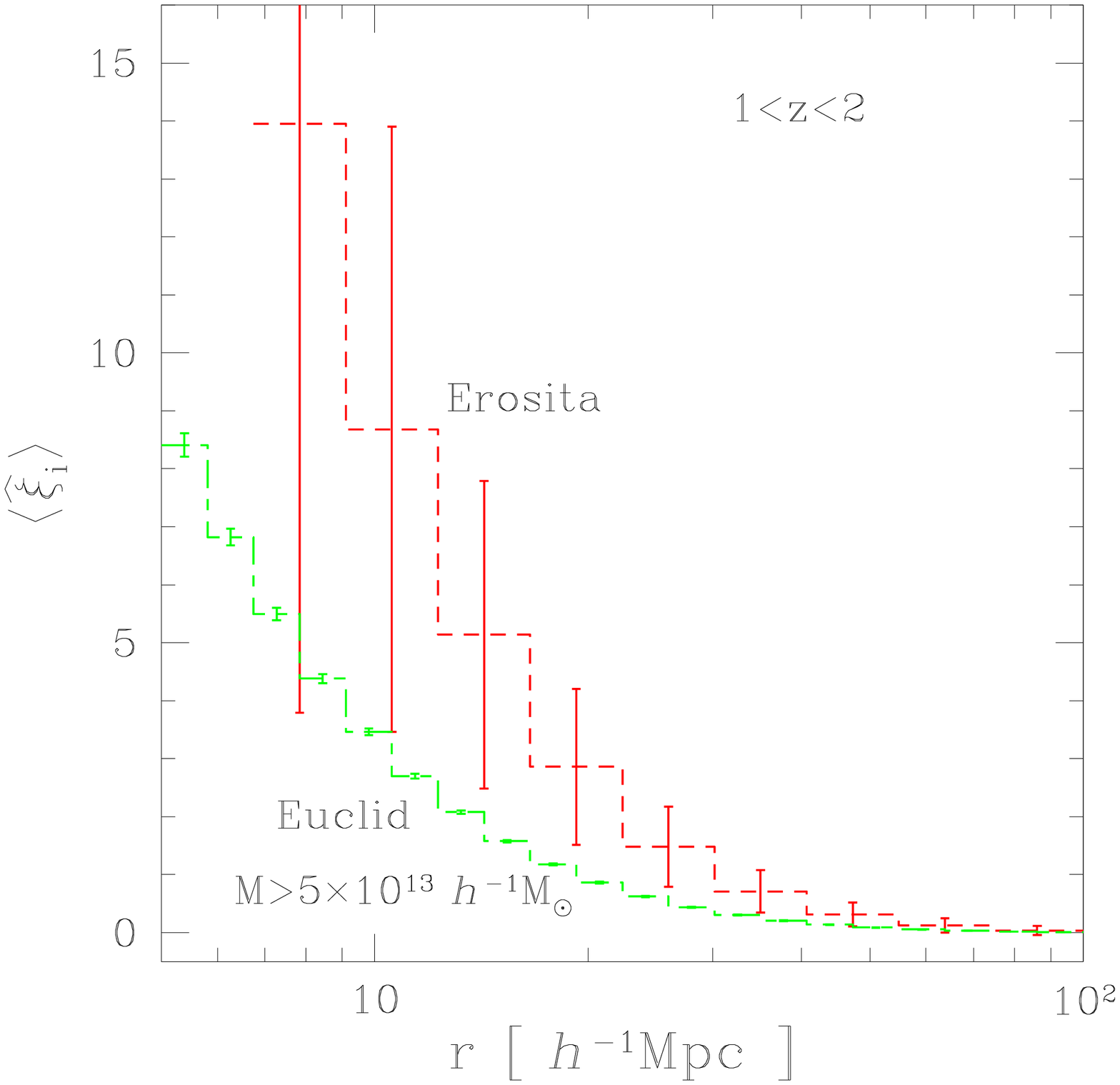}}
\end{center}
\caption{The mean correlation, $\lag\hxi_i\rag$, over the redshift interval $1<z<2$,
for the clusters detected by Erosita (upper curve, with ten distance bins)
and Euclid (lower curve, with twenty distance bins).}
\label{fig_XiR_z1to2_Erositaall}
\end{figure}

Following  PLANCK, space missions will map the entire sky in the X-ray (EROSITA) 
and optical (EUCLID) wavebands at unprecedented depth and angular resolution.
Corresponding selection functions are still at the tentative or predictive level.
It is nevertheless instructive to compare estimates of the statistical significance
of the all-sky cluster catalogs expected from these forthcoming 
surveys\footnote{For all the considered surveys, the $M_{\rm lim}(z)$ curves
were estimated using specific assumptions as to the evolution of the X-ray, optical,
and S-Z properties of the clusters, hence on their detectability. Moreover, the
assumed cosmology was either WMAP5 or WMAP7, to be consistent with
published analysis of each survey.
Therefore, these hypotheses may not be totally self-consistent with respect
to each other, but the main results of the comparison between the expected
signals should remain valid.}.
In practice, the total angular area of such surveys is not really $4\pi$ sterad since
we must remove the galactic plane. In the following, for Planck we consider the
two-sided cone of angle $\theta_s=75$ deg (i.e., $|b|>15$ deg), which yields a total
area $\Delta\Omega\simeq 30576$ deg$^2$.
For Erosita and Euclid we take $\theta_s=59$ deg (i.e., $|b|>31$ deg), which
corresponds to a total area that is about one-half of the full sky,
$\Delta\Omega\simeq 20000$ deg$^2$.

- Planck operates at nine frequencies, enabling an efficient detection of the cluster 
S-Z signature but has a rather large PSF (5'-10'). Some 1625 massive clusters
out to $z=1$ are expected over the whole sky. We assume the selection
function by \citet{Melin2006}, shown in middle panel of
Fig.~\ref{fig_Fs_XXL_Planck_Erosita}.

- For Erosita, a simple flux limit is currently  assumed as an average over the
whole sky: $4~10^{-14}$ \flux\ in the $[0.5-2]$ keV band \citep{Predehl2009}.
The associated selection function is shown in the right hand panel of 
Fig.~\ref{fig_Fs_XXL_Planck_Erosita}.
This would yield $71,907$ clusters out to $z=1$. 

- For Euclid, we follow the prescription of the Euclid Science Book for the cluster
optical selection function and adopt a fixed mass threshold of 
$5\times 10^{13} h^{-1} M_{\odot}$ \citep{Refregier2010}.

We show in Fig.~\ref{fig_lnz_Planckall} the angular number densities per redshift
bin. The error bars contain the shot-noise contribution (\ref{Cij-sn}), as well as the
sample-variance contribution (\ref{Cii-W2}) that holds for any angular
window and does not rely on the flat-sky and Limber's
approximations.\footnote{Because we consider a symmetric two-sided angular window
(i.e., two cones of angle $\theta_s$ around the north and south galactic poles),
the coefficients $\tW_2^{(\ell,m)}$ vanish for nonzero $m$ and for odd $\ell$.
For even $\ell$, they are still given by Eqs.(\ref{W2-l0})-(\ref{W2-00}) where
we substitute $(\Delta\Omega) \rightarrow (\Delta\Omega)_{\rm half}$,
where $(\Delta\Omega)_{\rm half}=(\Delta\Omega)/2$ is the area associated
with a single side (so that the last expression (\ref{W2-00}) still applies).}.
The 3D correlation functions are shown in
Figs.~\ref{fig_XiR_z0to1_Planckall} and \ref{fig_XiR_z1to2_Erositaall}.

As compared with the smaller surveys of Sect.~\ref{limited-areas},
these (almost) full-sky surveys provide much more accurate measures
of the evolution with redshift of cluster abundance, and of two-point
correlation functions, thanks to the greater number of objects.
In particular, thanks to its lower mass threshold, Euclid can probe higher
redshifts, both for number counts and correlation functions.  
Although we have only considered two redshift bins for the two-point
correlation function, $0<z<1$ and $1<z<2$,
Figs.~\ref{fig_XiR_z0to1_Planckall} and \ref{fig_XiR_z1to2_Erositaall}
suggest that for Euclid it should be
possible to introduce a smaller redshift binning, such as $\Delta z=0.5$.
We leave it to future works to estimate which redshift binning is the most
efficient at constraining cosmology.

\subsection{Shot noise versus sample variance}
\label{Shot-noise}

\begin{figure}
\begin{center}
\epsfxsize=8.5 cm \epsfysize=6.5 cm {\epsfbox{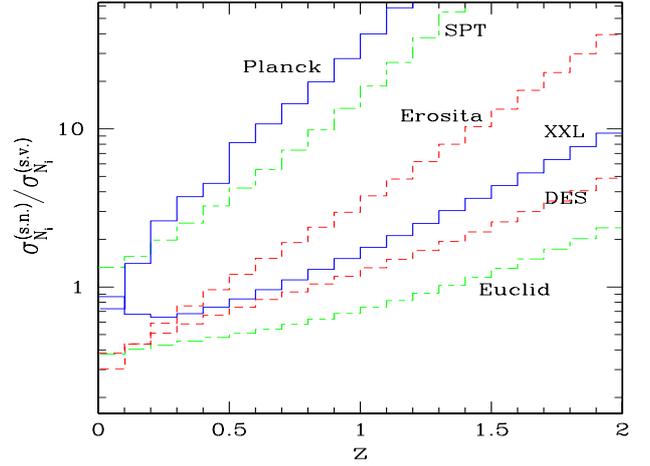}}
\end{center}
\caption{The ratio $\sigma_{N_i}^{(s.n.)}/\sigma_{N_i}^{(s.v.)}$ of the
rms shot-noise contribution $\sigma_{N_i}^{(s.n.)}$ to the rms
sample-variance contribution $\sigma_{N_i}^{(s.v.)}$, of the covariance of the
angular number densities $N_i$ obtained for various surveys.
(For DES and Euclid we only consider the case
$M>5\times 10^{13} h^{-1} M_{\odot}$.)}
\label{fig_Nz_sn_sv}
\end{figure}

\begin{figure}
\begin{center}
\epsfxsize=8.5 cm \epsfysize=6.5 cm {\epsfbox{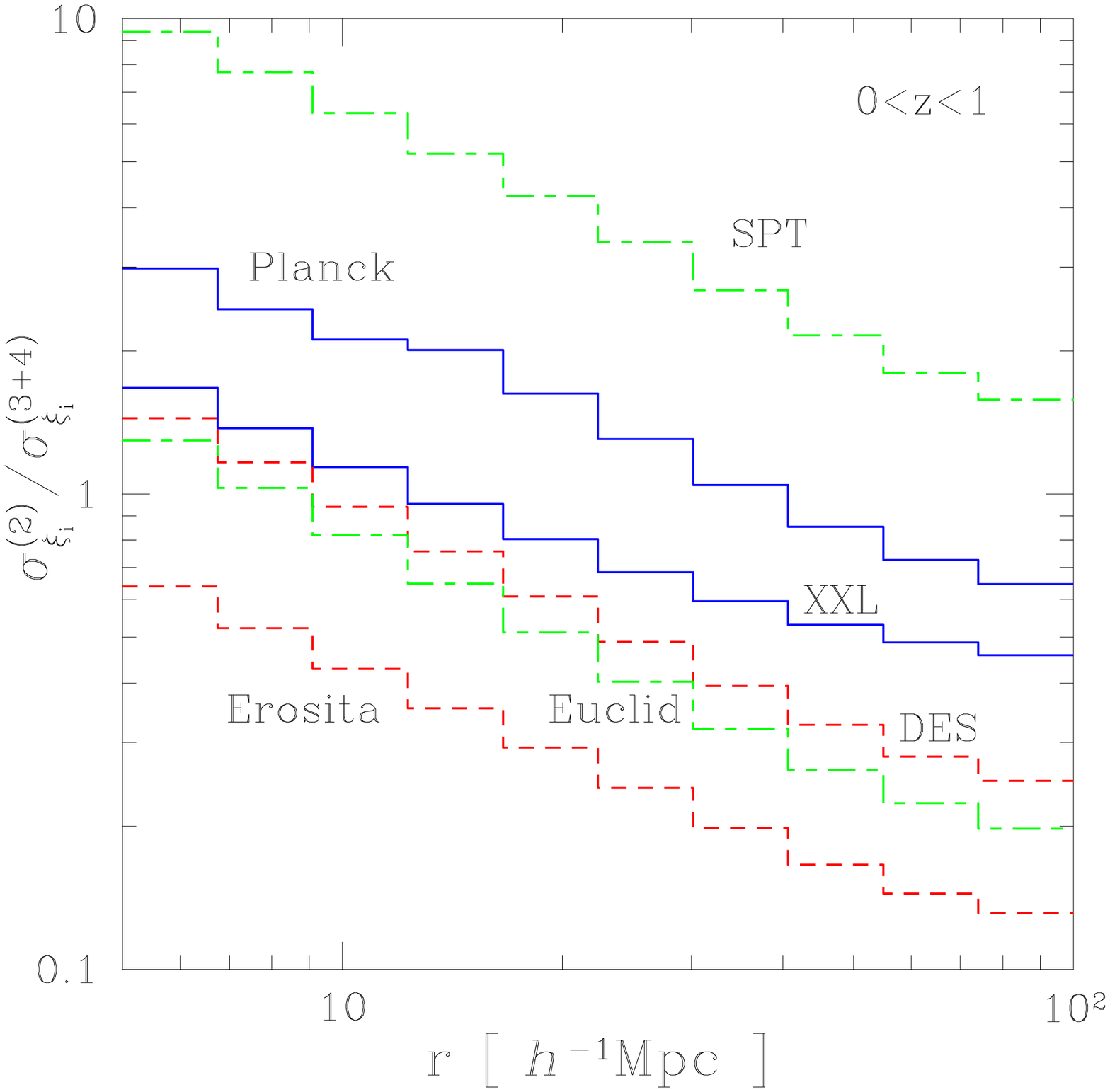}}
\end{center}
\caption{The ratio $\sigma_{\xi_i}^{(2)}/\sigma_{\xi_i}^{(3+4)}$ of the
rms contributions $\sqrt{C^{(2)}}$ and $\sqrt{C^{(3)}+C^{(4)}}$ of the 
covariance matrix of the estimator $\hxiLS_i$. This is a measure of shot-noise
effects.
(For DES and Euclid we only consider the case
$M>5\times 10^{13} h^{-1} M_{\odot}$.)}
\label{fig_XiR_z0to1_sn_sv}
\end{figure}

We show in Fig.~\ref{fig_Nz_sn_sv} the ratio of the shot-noise to sample-variance
contributions to the covariance of number counts, where the rms contributions
$\sigma_{N_i}^{(s.n.)}$ and $\sigma_{N_i}^{(s.v.)}$ are defined by
Eq.(\ref{sigma-Ni-sn-sv-def}).
As expected, shot noise becomes increasingly dominant at higher redshift,
as the number of clusters decreases, and it is smaller for Euclid which has a
wider sky coverage and a lower mass threshold.

We show in Fig.~\ref{fig_XiR_z0to1_sn_sv} the ratio
$\sigma_{\xi_i}^{(2)}/\sigma_{\xi_i}^{(3+4)}$ of the contribution
(\ref{C2-LS-def}) to the sum of contributions (\ref{C3-LS-def}) and
(\ref{C4-LS-def}), to the rms error $\sigma_i=\sqrt{C_{i,i}}$.
In contrast to Fig.~\ref{fig_CiiR_sn_sv} we include the high-order terms
of $C^{(3)}$ and $C^{(4)}$, but the ratio
$\sigma_{\xi_i}^{(2)}/\sigma_{\xi_i}^{(3+4)}$ is again a measure of shot-noise
effects.
As expected, we can see that the contribution $C^{(2)}$ becomes increasingly 
dominant for smaller radial bins since they contain fewer clusters.
We can see that the ordering between the various surveys is not the same
as the one obtained in Fig.~\ref{fig_Nz_sn_sv} for the number counts.
This is because the mass thresholds are not the same (and couplings between
shot-noise and sample-variance effects in the covariance matrix of halo correlations
make the analysis less direct).
Since a higher mass
means both a larger correlation function and larger discreteness effects (because
halos are rarer), it is not always obvious a priori how the relative
importance of shot-noise effects changes from one configuration to another. 

Here we must recall that Fig.~\ref{fig_XiR_z0to1_sn_sv} only shows the diagonal
part of the covariance matrix $C_{i,j}$ and that off-diagonal terms can be
non-negligible, see Sect.~\ref{Two-point-correlation}.

\subsection{High-order and low-order contributions to the sample variance of $\hxi$}
\label{relative-high-order}

\begin{figure}
\begin{center}
\epsfxsize=8.5 cm \epsfysize=6.5 cm {\epsfbox{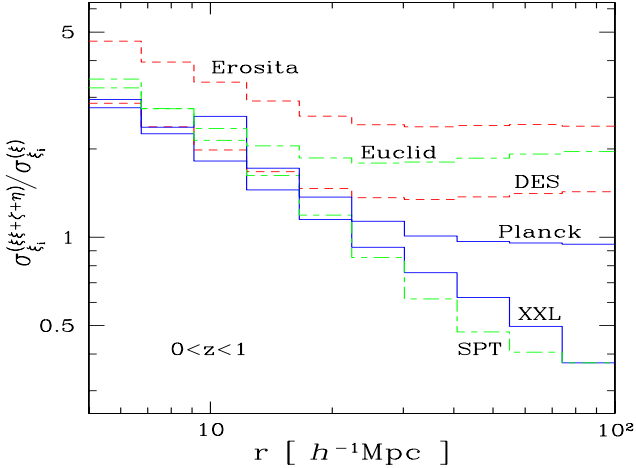}}
\end{center}
\caption{The ratio
$\sigma_{\xi_i}^{(\xi\xi+\zeta+\eta)}/\sigma_{\xi_i}^{(\xi)}$ 
of the rms high-order contribution (\ref{CLS-xixi-1})-(\ref{CLS-eta-1}) to the rms
low-order contribution (second term in Eq.(\ref{Cij-LS-tot})) of the sample variance 
of the correlation $\xi_i$ obtained for various surveys.
(For DES and Euclid we only consider the case $M>5\times 10^{13} h^{-1} M_{\odot}$.)}
\label{fig_XiR_z0to1_sv}
\end{figure}

We show in Fig.~\ref{fig_XiR_z0to1_sv} the ratio of the high-order contributions
to the low-order contribution of the sample variance of the 3D correlation $\xi_i$.
We consider the Landy \& Szalay estimator and we define along the diagonal the rms
contributions as 
$\sigma_{\xi_i}^{(\xi\xi+\zeta+\eta)}=\sqrt{C_{i,i}^{\rm LS (\xi\xi)}
+C_{i,i}^{\rm LS (\zeta)}+C_{i,i}^{\rm LS (\eta)}}$, from
Eqs.(\ref{CLS-xixi-1})-(\ref{CLS-eta-1}) for the high-order term, and
$\sigma_{\xi_i}^{(\xi)}=\sqrt{C_{i,i}^{\rm LS (\xi)}}$ for the low-order term, 
where $C_{i,i}^{\rm LS (\xi)}$ is given by the second term in Eq.(\ref{Cij-LS-tot}).
We do not consider here the shot-noise contribution $\sigma_{\xi_i}^{(2)}$
associated with the first term in Eq.(\ref{Cij-LS-tot}), which was studied in
Fig.~\ref{fig_XiR_z0to1_sn_sv};
however, while $C_{i,i}^{\rm LS (\xi\xi)}$ and $C_{i,i}^{\rm LS (\eta)}$ are
pure sample-variance contributions, $C_{i,i}^{\rm LS (\zeta)}$ and $C_{i,i}^{\rm LS (\xi)}$
are mixed shot-noise and sample-variance contributions. Indeed, they arise from
both the discreteness of the halo population (as shown by the power $\Nb^3$ instead
of $\Nb^4$, which comes from the identification of two objects as explained
in Eq.(\ref{n4-sn-def})) and its large-scale correlations (as shown by the bias
factors $\bb^3$ and $\bb^2$).

As could be expected, we can see in Fig.~\ref{fig_XiR_z0to1_sv} that the relative
importance of high-order terms increases on smaller scales, deeper in the
nonlinear regime where correlations are stronger.
However, in some cases there is a flattening on larger scales because the relative
importance of high-order terms no longer decreases (and could even increase
in the case of low-mass halos as seen in Fig.~\ref{fig_CiiR_NG}).
This is because the low-order contribution $C_{i,i}^{\rm LS (\xi)}$
is actually a mixed ``shot-noise and sample-variance'' contribution, as noticed
above, and shot-noise effects decrease on larger radii (because of the
greater volume), as seen in Fig.~\ref{fig_XiR_z0to1_sn_sv}.
In agreement with this explanation, we can see that this upturn appears earlier
and is greater for the surveys where shot-noise effects are less, that is,
Erosita, Euclid, and DES.

More generally, Fig.~\ref{fig_XiR_z0to1_sv} shows that high-order contributions
to the sample-variance or mixed terms are not negligible (but on small scales
along the diagonal the covariance matrix is often dominated by the
pure shot-noise contribution). For the variety of cases studied in
Fig.~\ref{fig_XiR_z0to1_sv} they do not grow above five times the low-order
contribution along the diagonal, but as shown in Figs.~\ref{fig_Rij_CXiR_NG}
and \ref{fig_Rij_CXiR_Horizon}
their importance can be greater far from the diagonal.
Then, these contributions should be taken into account if one requires accurate
or safe estimates of signal-to-noise ratios.

\subsection{Dependence of the results on cosmology}
\label{cosmology}

We investigate in App.~\ref{app-cosmology} the sensitivity of our results
to the value of the cosmological parameters, by comparing the curves obtained
in the previous sections with those that are obtained when we change either
$h$, $\Omega_{\rm m}$, or $\sigma_8$ by an amount that corresponds to the
current ``$2\!-\!\sigma$'' uncertainty \citep{Komatsu2010}. 
We find that the main features shown in Figs.~\ref{fig_Nz_sn_sv}, 
\ref{fig_XiR_z0to1_sn_sv}, and \ref{fig_XiR_z0to1_sv} remain valid, with modest
quantitative changes (e.g., shot-noise effects become slightly less important, with
respect to sample-variance contributions, when $\sigma_8$ is slightly increased).
Therefore, our results and conclusions are not sensitive to the precise value of the 
cosmological parameters (within their current range of uncertainty).

\section{Conclusion}
\label{Conclusion}

In this paper we have presented a general formalism for obtaining analytical
estimates of the means and covariance matrices of number counts and
correlation functions, for distributions of  cosmological objects such as
clusters of galaxies or galaxies.
To do so, we assumed that the two-point correlation function of these
objects can be factored in terms of a linear bias model, and this simplifies
expressions as spatial and mass (or luminosity, temperature, etc.) integrals
factor. To estimate the high-order contributions to the covariance of
two-point estimators, we also assumed that the three- and four-point
correlations can be described by a hierarchical ansatz, that is, that they can
be written as products of the two-point functions.
This is the simplest model that agrees reasonably well with realistic
distributions (of the dark matter density field, as well as of cosmological
objects such as galaxies or clusters that follow the dark matter density on large
scales). Although this is only an approximate model and it is known that
actual cosmological fields do not exactly obey such a hierarchical clustering,
this allows us to derive explicit expressions that provide a reasonably good
description of covariance matrices.

The main differences or improvements with respect to previous studies are
the following.

- Keeping the application to cluster surveys in mind, rather than the galaxy
surveys that have been the aim of most works, we considered
two-point estimators that involve integrations over broad redshift bins.
Thus, we do not work with local 3D correlations within an homogeneous and
isotropic box at a given redshift, but with averages over a redshift interval
with explicit integration along part of the observational cone, where the
radial direction plays a specific role.

- We took all shot-noise and sample-variance contributions into account,
along with high-order contributions, which in the present case of one-point and
two-point estimators involve products of two two-point correlation functions
and the three- and four-point correlations.

- Within the framework of the simple hierarchical model recalled above,
we gave explicit expressions for all contributions to these means and
covariance matrices. They can be readily used for any population of objects
and any set of cosmological parameters, provided one is able to compute
the mass function (or the luminosity/temperature function), the two-point
correlation and three- and four-point normalization parameters. In practice,
assuming a linear scale-independent bias model (or a uniform scale-dependence 
that can be absorbed into the two-point correlation), it is sufficient to give a
bias $b(M,z)$ in addition to the mass function.

We first studied the number counts per redshift bins, comparing the
relative importance of shot-noise and sample variance contributions and
giving scaling laws obeyed by the signal-to-noise ratios, as a function of
the survey area and the number of fields.
We have explicitly considered the case of large angular windows, and estimated
the angular scale where the flat-sky and Limber's approximations break down,
which occurs at about 10 deg.
We also computed the decay of correlations between distant redshift bins.
In particular, we checked that a redshift binning of width $\Delta z = 0.1$
is broad enough to neglect cross-correlations between different bins.

Next, we studied estimators of the 3D correlation function, averaged over
finite redshift intervals. We compared the Peebles \& Hauser estimator with the
usual Landy \& Szalay estimator, and we evaluated the relative importance of
shot-noise and sample-variance, low-order and high-order, contributions to the
covariance matrix. We also considered the behavior of the off-diagonal terms,
and described how high-order contributions make the covariance matrices
less diagonal as correlations develop between different scales (especially as
one of the scales becomes smaller and more nonlinear).
Then, we performed the same analysis for the 2D angular correlation
function.

Throughout we compared our analytical expressions with results from
numerical simulations and we obtain a reasonably good match. This makes such
analytical results more competitive than simulations, because they are
much faster to compute and allow one to describe rare objects that would have
low-quality statistics in the simulations. 
Finally, we applied our formalism to
several future cluster surveys, and considered both limited-area and full-sky
missions.

We hope our results can help for estimating the signal-to-noise ratio of current and
future surveys. This is useful for comparing the efficiency of different
probes and different survey configurations, such as the choice of redshift
binning, survey area, or number of subfields.

Our study should be extended in several directions. First, it would be interesting
to consider the noise associated with photometric redshifts.
Second, one should include the effect of redshift-space distortions, which are
likely to be important on small scales. Third, the computation of the means
and covariance matrices studied in this paper is only an intermediate tool for
comparing theoretical predictions with observations, and the final goal
is to derive constraints on cosmological parameters or astrophysical processes
(e.g., scaling laws for cluster mass-luminosity-temperature relationships).
Our results may be used to further investigate the cosmological information that
can be extracted from cluster surveys and to optimize observing configurations
so as to improve those constraints.
We leave these tasks to future works.

\acknowledgements

We thank the anonymous referee for comments that helped to improve the
presentation of the paper.
F.P. acknowledges support from Grant No. 50 OR 1003 of the
Deutsches Zemtrum für Luft- und Raumfahrt (DLR) and from the
Transregio Programme TR33 of the Deutsche
Forschungsgemeinschaft (DfG).

\appendix

\section{Mean and covariance of number counts}
\label{Method}

To avoid introducing numerous Dirac factors, owing to the discreteness
of the observed distribution of objects, we follow the simple approach described
in Sect.36 of \citet{Peebles1980} to compute the statistical properties of counts
in cells. We illustrate in this section this method for the computation of the
mean and covariance of number counts within redshift bins.

We divide the ``volume'' of the space $(z,\vOm,\ln M)$, which enters the
expression (\ref{Ni-1}) of the angular number density of observed objects in
redshift bin $i$, over $\cN$ small (infinitesimal) cells labeled by the index
$\alpha$, so that Eq.(\ref{Ni-1}) reads as
\beq
\hN_i = \frac{1}{(\Delta\Omega)} \sum_{\alpha_i} \hn_{\alpha_i} ,
\label{hNi}
\eeq
where subscript $i$ refers to the redshift bin $i$.
Then, since the cell $\alpha_i$ is infinitesimally small it contains at most one
object, whence \citep{Peebles1980}
\beq
\hn_{\alpha_i} =0 \;\; \mbox{or} \;\; 1 , \;\; \mbox{and} \;\; 
\hn_{\alpha_i}^2 = \hn_{\alpha_i} .
\label{hn-alpha}
\eeq
Moreover, by definition its average is given by
\beq
\lag \hn_{\alpha_i} \rag = \dd z \frac{\dd\chi}{\dd z} \, \cD^2 \dd\vOm
\frac{\dd M}{M} \, \frac{\dd n}{\dd\ln M} .
\label{hn-mean}
\eeq
Of course, we recover for the mean number of objects in the redshift bin $i$
the expression (\ref{Ni-3}), which could also be read from Eq.(\ref{Ni-1}) using the
average
\beq
\lag \frac{\dd\hn}{\dd\ln M} \rag = \frac{\dd n}{\dd\ln M} .
\label{hn-n}
\eeq

We now consider the covariance of the angular number densities $\hN_i$. From
Eq.(\ref{hNi}) we have
\beqa
(\Delta\Omega)^2 \lag \hN_i \hN_j \rag & = & \lag \biggl ( \sum_{\alpha_i} 
\hn_{\alpha_i} \biggl ) \biggl ( \sum_{\alpha_j} \hn_{\alpha_j} \biggl ) \rag \\
& = & \delta_{i,j} \sum_{\alpha_i} \lag \hn_{\alpha_i}^2 \rag +
\sum_{\alpha_i \neq \alpha_j}  \lag \hn_{\alpha_i} \hn_{\alpha_j} \rag \\
& = & \delta_{i,j} \sum_{\alpha_i} \lag \hn_{\alpha_i} \rag +
\sum_{\alpha_i \neq \alpha_j}  \lag \hn_{\alpha_i} \hn_{\alpha_j} \rag .
\label{Ni-Nj-1}
\eeqa
In the second line we used the fact that the redshift bins do not overlap,
so that for two ``volumes'' $\alpha_i$ and $\alpha_j$ to coincide, bins
$i$ and $j$ must be the same (and $\delta_{i,j}$ is the Kronecker symbol),
while in the third line we used Eq.(\ref{hn-alpha}).
The first term in Eq.(\ref{Ni-Nj-1}) corresponds to the shot noise, due to the
discreteness of the object distribution. The second term includes the
nonzero-distance correlation between objects, and reads as
(for $\alpha_i\neq\alpha_J$)
\beq
\lag \hn_{\alpha_i} \hn_{\alpha_j} \rag = \lag \hn_{\alpha_i} \rag 
\lag \hn_{\alpha_j} \rag \left[ 1 + \xih_{\alpha_i,\alpha_j} \right] ,
\label{xi-ij}
\eeq
where $\xih_{\alpha_i,\alpha_j}$ is the ``halo'' two-point correlation function
between ``volumes'' $\alpha_i$ and $\alpha_j$, see \citet{Peebles1980}.
This yields
\beqa
\lag \hN_i \hN_j \rag & = & \delta_{i,j} \frac{\lag \hN_i \rag}{(\Delta\Omega)}
+ \int_i\dd \chi_i \, \cD_i^2 \frac{\dd\vOm_i}{(\Delta\Omega)} \frac{\dd M_i}{M_i}
\, \frac{\dd n}{\dd\ln M_i}  \nonumber \\
&& \times \int_j\dd \chi_j \, \cD_j^2 \frac{\dd\vOm_j}{(\Delta\Omega)}
\frac{\dd M_j}{M_j} \, \frac{\dd n}{\dd\ln M_j} \, \left[ 1 + \xih_{i,j} \right] ,
\label{Ni-Nj-2}
\eeqa
using obvious notations where we label the quantities associated with $\hN_i$
and $\hN_j$ by the subscripts $i$ and $j$ and we integrate over the
bins $i$ and $j$.
This could also be directly obtained from Eq.(\ref{Ni-1}) by writing
\beqa
\lag \frac{\dd \hn}{\dd\ln M_i}  \frac{\dd \hn}{\dd\ln M_j} \rag & = &
\frac{\dd n}{\dd\ln M_i} \frac{\dd n}{\dd\ln M_j} \left[ 1 + \xih_{i,j} \right]
\nonumber \\
&& \hspace{-2.9cm} + \frac{M_j}{\cD_j^2} \, \delta_D(\chi_j\!-\!\chi_i)
\delta_D(\vOm_j\!-\!\vOm_i) \delta_D(M_j\!-\!M_i) \frac{\dd n}{\dd\ln M_i} ,
\label{hni-hnj}
\eeqa
where the second term with the Dirac factors gives the shot-noise contribution.

In this derivation we have assumed in Eq.(\ref{hn-alpha}) that space can be
divided into infinitesimal volumes that contain either zero or one object
and that each object only appears in one cell. 
Even though clusters and dark matter halos are actually extended objects, it is
still possible to define a point distribution by associating a single point to each 
cluster or halo, for instance the halo mass center. Thus, this approach,
which follows \citet{Peebles1980}, applies to these cases as well and to
any distribution of discrete objects, as long as we restrict ourselves to
count distributions and do not study the internal structure of these objects.

\section{Finite-size effects}
\label{Finite-size}

\begin{figure}
\begin{center}
\epsfxsize=7.8 cm \epsfysize=3.5 cm {\epsfbox{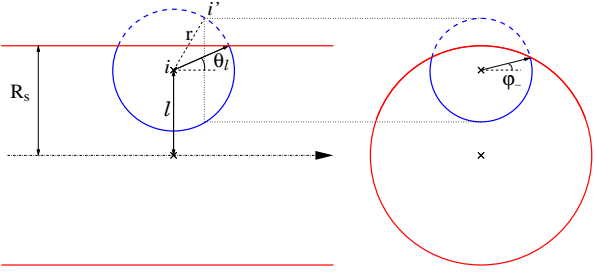}}
\end{center}
\caption{Geometrical illustration of finite-size effects. Close to the survey
boundary, part of the sphere of radius $r$ extends beyond the observational
cone and should not be counted. The left plot is a transverse view, orthogonal
to the central line of sight, whereas the right plot is a view from a point far away
on the line of sight. }
\label{fig-finite}
\end{figure}

As noticed in Sect.~\ref{Two-point-correlation}, in our computations of the
mean and covariance of the estimators $\hxi$ and $\hxiLS$ we neglect
finite-size effects. Indeed, we do not take the fact into account that when
a point $i$ gets close to the survey boundaries the available space for 
points $i'$ located in the distance bin $[\Rim,\Rip]$, with respect to point $i$,
is only a fraction of this spherical shell since a part of it extends beyond the
observational cone. This means that we overestimate the total number of pairs.
This has no impact on the mean, $\lag\hxi_i\rag$, since this effect cancels out
between the numerator and denominator in (\ref{xi-1}), but it means that
we slightly overestimate the signal-to-noise ratio.

To estimate the magnitude of this error, we compute the geometrical factor
illustrated in Fig.~\ref{fig-finite}. 
Approximating the observational cone as a cylinder of radius
$R_s=\cD\theta_s$, a point $i$ at distance $\ell$ from the central line of sight
is the center of a spherical shell of radius $r$, onto which we count all neighbors
$i'$ to estimate the correlation $\xi$ at this distance $r$.
We denote $F(\ell)$ as the fraction of this sphere that is enclosed within the
observational cone. In our computations elsewhere we used the approximation
$F=1$, but for $R_s-r<\ell<R_s$ we actually have $F<1$. 
As in the transverse view shown in the left hand plot of Fig.~\ref{fig-finite},
the angle $\theta_{\ell}$ associated with the farthest point of intersection
between the cylinder and the sphere satisfies $\ell + r \sin\theta_{\ell} = R$,
whence
\beq
R_s\!-\!r\!<\!\ell\!<\!R_s , \;\; 0 \!<\! \theta_{\ell} \!<\! \frac{\pi}{2} : \;\;\;
\sin\theta_{\ell} = \frac{R_s-\ell}{r} .
\label{theta-l}
\eeq
Next, in the plane of each vertical section (i.e., at fixed $\theta$), shown
in the right hand plot of Fig.~\ref{fig-finite} that corresponds to a projection
along the line
of sight, the cylinder appears as a circle of radius $R_s$, whereas the
section of the sphere of center $i$ appears as a circle of radius $r\sin(\theta)$.
Both circles intersect (again for $R_s-r<\ell<R_s$) at the symmetric polar angles
$\varphi_{\pm}$, with
\beq
R_s^2 = \ell^2 + r^2 \sin^2\theta + 2 \ell r \sin\theta \sin\varphi_{\pm} .
\eeq
Then, the surface of the sphere that extends outside of the observational cylinder
writes as
\beq
S^{\rm out} = 4 r^2 \int_{\theta_{\ell}}^{\pi/2} \dd\theta \, \sin\theta
\int_{\varphi_-}^{\pi/2} \dd\varphi .
\label{S-out}
\eeq
Thus, for $R_s-r<\ell<R_s$ the fraction of the sphere that is enclosed within the
observational cylinder reads as
\beqa
F(\ell) & \!\! = \! & 1 - \frac{\cos\theta_{\ell}}{2} + \int_0^{\cos\theta_{\ell}}
\frac{\dd x}{\pi} \, {\rm Arcsin}\! \left( \! \frac{R_s^2\!-\!\ell^2\!-\!r^2(1\!-\!x^2)}
{2\ell r\sqrt{1-x^2}} \! \right)  \nonumber \\
&&
\label{Fl-1}
\eeqa
whereas $F(\ell)=1$ for $0<\ell<R_s-r$. 
Then, integrating the position of the central point $i$ over the cylinder, the fraction
of volume for pairs at distance $r$, with respect to the approximation $F=1$,
writes as
\beqa
\frac{N'}{N} & = & \int_0^{R_s} \frac{\dd\ell}{R_s} \frac{2\ell}{R_s} F(\ell) \\
& = & \left(1-\frac{r}{R_s}\right)^2 + \int_{R_s-r}^{R_s} \frac{\dd\ell}{R_s}
\frac{2\ell}{R_s} F(\ell) .
\label{N'-N}
\eeqa
This gives the ratio of the number of pairs $N'$, which is measured in the survey,
to the number $N$ obtained when we do not take finite-size effects into account.
For instance, at $z=1$, which corresponds to the angular distance
$\cD\simeq 2352 h^{-1}$Mpc, and for a survey angular window of area
$50$ deg$^2$, which corresponds to $\theta_s \simeq 0.0696$ rad,
we have $R_s=\cD\theta_s\simeq 164 h^{-1}$Mpc.
Then, we obtain $N'/N \simeq 0.91$ for a shell at radius $r=30 h^{-1}$Mpc.
This means that the approximation $F=1$ overestimates the number of pairs
by about $10\%$ and the signal-to-noise ratio by $5\%$.

\section{Computation of the mean of the estimators $\hxi$ and $\hxiLS$}
\label{Computation-of-the-mean-of-the-estimator}

Defining the 3D Fourier-space top-hat as
\beq
\tW_3(kR) = \! \int_0^R \!\! \frac{\dd \vr}{4\pi R^3/3} \, e^{\ii\vk\cdot\vr} = 
3 \, \frac{\sin(kR)-kR\cos(kR)}{(kR)^3} ,
\label{W3-def}
\eeq
the 3D Fourier-space window of the $i$-shell reads as
\beqa
\tW^{(3)}_i(k) & = & \int_{\cV_i}\frac{\dd\vr}{\cV_i} \, e^{\ii\vk\cdot\vr}
\nonumber \\
& = & \frac{\Rip^3 \tW_3(k\Rip) - \Rim^3 \tW_3(k\Rim)}{\Rip^3- \Rim^3} ,
\label{W3-D-def}
\eeqa
where the superscript $(3)$ recalls that we consider a 3D radial bin.
Then, writing the two-point correlation function in terms of the power spectrum,
as in Eq.(\ref{xi-Pk}), we obtain for its radial average (\ref{xi-i-i'-def})
\beq
\xir_{i'}(z) = \int_0^{\infty} \frac{\dd k}{k} \, \Delta^2(k,z) \,
\tW^{(3)}_i(k) .
\label{I3-def}
\eeq
(Here $i$ and $i'$ refer to the same radial bin; the prime only recalls that we
are integrating over a neighbor $i'$ within a small radial shell with respect to
another point in the observational cone.)

\section{Derivation of the covariance of the Peebles \& Hauser estimator $\hxi$}
\label{Computation-of-the-covariance-of-the-estimator}

We compute here the covariance of the estimators $\hxi_i$, which is identical
to the covariance of the quantities $(1+\hxi_i)$. To simplify the expressions
we do not consider mass binning here, but it is straightforward to generalize
to the case of several mass bins.
From the definition
(\ref{xi-1}) we can write with obvious notations the second moment as
\beqa
\lag (1+\hxi_i)(1+\hxi_j)\rag & = & \frac{1}{\QQ_i} \int \dd\chi_i \, \cD_i^2
\frac{\dd\vOm_i}{(\Delta\Omega)} \frac{\dd M_i}{M_i} \int \dd\vr_{i'}
\frac{\dd M_{i'}}{M_{i'}} \nonumber \\
&& \hspace{-1.5cm} \times \frac{1}{\QQ_j} \int \dd\chi_j \, \cD_j^2
\frac{\dd\vOm_j}{(\Delta\Omega)} \frac{\dd M_j}{M_j} \int \dd\vr_{j'}
\frac{\dd M_{j'}}{M_{j'}} \nonumber \\
&& \hspace{-1.5cm} \times \, \lag \frac{\dd\hn}{\dd\ln M_i} 
\frac{\dd\hn}{\dd\ln M_{i'}} \frac{\dd\hn}{\dd\ln M_j}
\frac{\dd\hn}{\dd\ln M_{j'}} \rag .
\label{xii-xij-1}
\eeqa
The average in Eq.(\ref{xii-xij-1}) can be written as in Eq.(\ref{hni-hnj}), with
many Dirac factors for the shot-noise contributions. However, as in
App.~\ref{Method}, it may be easier to follow \citet{Peebles1980}
and to divide ``volumes'' over small (infinitesimal) cells that contain
$\hn$ objects, with $\hn=0$ or $1$.
Then, we can split the average $\lag\hn_i\hn_{i'}\hn_j\hn_{j'}\rag$ as
\beqa
\lefteqn{ \lag\hn_i\hn_{i'}\hn_j\hn_{j'}\rag = 
\lag\hn_i\hn_{i'}\hn_j\hn_{j'}\rag^{\sv}
+ \delta_{i,j} \lag\hn_i\hn_{i'}\hn_{j'}\rag^{\sv} }\nonumber \\
&& + \delta_{i,j'} \lag\hn_i\hn_{i'}\hn_{j}\rag^{\sv}
+ \delta_{i',j} \lag\hn_i\hn_{i'}\hn_{j'}\rag^{\sv} \nonumber \\
&& + \delta_{i',j'} \lag\hn_i\hn_{i'}\hn_{j}\rag^{\sv}
+ \delta_{i,j} \delta_{i',j'} \lag\hn_i\hn_{i'}\rag^{\sv} \nonumber \\
&& + \delta_{i,j'} \delta_{i',j} \lag\hn_i\hn_{i'}\rag^{\sv} ,
\label{n4-sn-def}
\eeqa
where we have explicitly written the first ``pure sample-variance'' contribution and
the last six ``shot-noise'' contributions associated with the Kronecker symbols.
The remaining averages with the superscript ``(s.v.)'' denote ``sample-variance''
averages, that is, without further shot-noise terms.
Here we used the fact that the objects $i$ and $i'$ are separated by
the finite distance $r_{i'}$, with $r_{i'}\geq\Rim$, so that the elementary ``cells''
$i$ and $i'$ cannot coincide and there is no shot-noise contribution of the
form $\delta_{i,i'}$. For the same reason there is no term $\delta_{j,j'}$.
Next, the ``sample-variance'' averages of Eq.(\ref{n4-sn-def}) read as
\citep{Peebles1980}
\beqa
\lefteqn{ \lag\hn_i\hn_{i'}\hn_j\hn_{j'}\rag^{\sv} = \lag\hn_i\rag \lag\hn_{i'}\rag
\lag\hn_j\rag \lag\hn_{j'}\rag \left[ 1 + \xih_{i,i'} + \xih_{i,j} + \xih_{i,j'}
\right. } \nonumber \\
&& + \xih_{i',j} + \xih_{i',j'} + \xih_{j,j'} + \zetah_{i',j,j'} +  \zetah_{i,j,j'}
+ \zetah_{i,i',j'} + \zetah_{i,i',j} \nonumber \\
&& \left. + \xih_{i,i'} \xih_{j,j'} + \xih_{i,j} \xih_{i',j'} + \xih_{i,j'} \xih_{i',j}
+ \etah_{i,i',j,j'} \right] ,
\label{n4-def}
\eeqa
\beqa
\lag\hn_i\hn_{i'}\hn_{j'}\rag^{\sv} & = & \lag\hn_i\rag \lag\hn_{i'}\rag
\lag\hn_{j'}\rag \left[ 1 + \xih_{i,i'} +\xih_{i,j'} +\xih_{i',j'} \right. \nonumber \\ 
&& \left. + \zetah_{i,i',j'} \right] ,
\label{n3-def}
\eeqa
\beq
\lag\hn_i\hn_{i'}\rag^{\sv} = \lag\hn_i\rag \lag\hn_{i'}\rag 
\left[ 1 + \xih_{i,i'} \right] ,
\label{n2-def}
\eeq
where $\xih$, $\zetah$, and $\etah$ are the two-point, three-point, and
four-point correlation functions of the objects.
Since we have
\beq
\lag\hxi_i\hxi_j\rag - \lag\hxi_i\rag \lag\hxi_j\rag =
\lag(1\!+\!\hxi_i) (1\!+\!\hxi_j)\rag - \lag 1\!+\!\hxi_i\rag \lag 1\!+\!\hxi_j\rag ,
\eeq
we obtain from Eqs.(\ref{xii-xij-1})-(\ref{n2-def}) the decomposition (\ref{xii-xij-2})
of the covariance matrix, with the explicit expressions (\ref{C2-def})-(\ref{C4-def})
of the various ``sample-variance'' and ``shot-noise'' contributions.
Here we used the symmetries\footnote{The integration over the points
$i$ and $i'$ in Eq.(\ref{xii-xij-1}) should be understood as $\int \dd\chi_i \cD_i^2
\dd\vOm_i \int \dd\chi_{i'} \cD_{i'}^2 \dd\vOm_{i'} \,
\Theta(\Rim\!<\!|\vx_i\!-\!\vx_{i'}|\!<\!\Rip)$, which is more clearly symmetric,
where $\Theta$ is a top-hat window that takes values 0 or 1 with obvious notations.
In practice, we actually slightly ``break'' this symmetry by using the variables
$(\chi_i,\vOm_i;\vr_{i'})$ as in Eq.(\ref{xii-xij-1}) if we do not take
boundary effects into account, as in this paper.}
$\{i\leftrightarrow i'\}$ and $\{j\leftrightarrow j'\}$ of Eq.(\ref{xii-xij-1}).
In Eq.(\ref{C3-def}) the object ``$j'$'' is at the distance $r_{j'}$ from the object
``$i$'', since this shot-noise contribution comes from the case where the
objects $i$ and $j$ are the same object (or from one of the three remaining cases
``$i=j'$'', ``$i'=j$'', or ``$i'=j'$'').
The shot-noise contribution (\ref{C2-def}) comes from the identification ``$i=j$
and $i'=j'$'' (or ``$i=j'$ and $i'=j$'').
This implies that the distances $r_{i'}$ and $r_{j'}$ are equal,
which gives rise to the Kronecker symbol $\delta_{i,j}$ since we consider the
case of nonoverlapping distance bins $[\Rim,\Rip]$.

From Eq.(\ref{xi-2}) the contribution $C_{i,j}^{(2)}$ of Eq.(\ref{C2-def}) also reads as
\beq
C_{i,j}^{(2)} = \delta_{i,j} \frac{2}{(\Delta\Omega)\QQ_i} (1+\lag\hxi_i\rag) .
\label{C2-1}
\eeq
In order to estimate the contributions $C_{ij}^{(3)}$ and $C_{ij}^{(4)}$ we
assume that the radial bins $[\Rim,\Rip]$ are restricted to large enough scales
to neglect three- and four-point correlation functions, as well as products such as
$\xi_{i;j'} \xi_{i';j}$. Thus, we only keep in this Appendix the contributions that
are constant or linear over the two-point correlation function $\xi_{i;j}$ of the
objects, which we recall with the superscripts ``1'' and ``$\xi$'' below.
Moreover, we again assume that the two-point correlation function can be
factored in as in Eq.(\ref{xij-bb}).

The first contribution to $C_{i,j}^{(3)}$, associated with the factor $1$
in the brackets in Eq.(\ref{C3-def}), reads as
\beq
C_{i,j}^{(3,1)} = \frac{4}{(\Delta\Omega)\QQ_i\QQ_j} \int \dd\chi \, \cD^2 \,
\nb^3 \, \cV_i \cV_j .
\label{C3-0-1}
\eeq
The contributions that are linear over $\xi$ sum up as
\beqa
C_{i,j}^{(3,\xi)} & = & \frac{4}{(\Delta\Omega)\QQ_i\QQ_j} \int \dd\chi \, \cD^2 \, \bb^2 \, \nb^3 \, \cV_i \cV_j \nonumber \\
&& \times \, \left[ \overline{\xir_{i'}}  + \overline{\xir_{j'}} 
+ \overline{\xir_{i',j'}} \right] ,
\label{C3-xi-1}
\eeqa
where $\overline{\xir_{i'}}$ and $\overline{\xir_{j'}}$ are
defined as in Eqs.(\ref{xi-i-i'-def}) and (\ref{I3-def}), whereas
$\overline{\xir_{i',j'}}$ is defined in Eq.(\ref{I3-ij-xi-def}) and also writes as
\beq
\overline{\xir_{i',j'}} =  \int_0^{\infty} \frac{\dd k}{k} \, \Delta^2(k,z) \, 
\tW_i^{(3)}(k) \, \tW_j^{(3)}(k) .
\label{I3-ij-def}
\eeq
Next, at this order the contribution (\ref{C4-def}) to the
covariance simplifies as
\beq
C_{i,j}^{(4,\xi)} = \frac{4}{\QQ_i\QQ_j} \int \dd\chi \, \cD^5
\, \bb^2 \nb^4 \, \cV_i \cV_j \, \xicyl ,
\label{C4-xi-2}
\eeq
where $\xicyl$ is Limber's approximation (\ref{I-thetas-def}) to
Eq.(\ref{xib-ij-def}).
Then, collecting all terms, we obtain the expression (\ref{Cij-tot}) for the
covariance .

\section{Derivation of the mean and covariance of the Landy \& Szalay
estimator $\hxiLS$}
\label{Computation-Landy-Szalay}

We can relate the Landy \& Szalay estimator $\hxiLS$ defined by Eq.(\ref{xi-LS-1})
to the Peebles \& Hauser estimator (\ref{xi-1}) by
\beq
\hxiLS_i = \hxi_i - 2 \hxic_i ,
\label{xi-LS-c}
\eeq
where we defined the cross-term $\hxic_i$ by
\beqa
1+\hxic_i & = & \frac{1}{\QQ_i} \int \dd z \, \frac{\dd\chi}{\dd z} \, \cD^2
\frac{\dd\vOm}{(\Delta\Omega)} \frac{\dd M}{M} \int_i \dd\vr' \frac{\dd M'}{M'} \nonumber \\
&& \times \, \frac{\dd\hn}{\dd\ln M} \frac{\dd n}{\dd\ln M'} .
\label{xi-c-1}
\eeqa
We obtain at once, using Eqs.(\ref{hn-n}) and (\ref{QQ-1}),
\beq
\lag \hxic_i \rag =0 ,
\label{hxi-c-2}
\eeq
which leads to Eq.(\ref{xi-LS-2}).

From the relation (\ref{xi-LS-c}) we have for the covariance of the estimator
$\hxiLS$,
\beq
C^{\rm LS}_{i,j} = C_{i,j} - 2 \lag\hxi_i\hxic_j\rag - 2 \lag\hxic_i\hxi_j\rag
+ 4 \lag\hxic_i\hxic_j\rag ,
\label{Cij-LS-1}
\eeq
where $C_{i,j}$ is the covariance of the Peebles \& Hauser estimator $\hxi$, defined in
Eq.(\ref{xii-xij-2}).
To compute the cross-terms in (\ref{Cij-LS-1}) we write as in Eq.(\ref{xii-xij-1}),
\beqa
\lag (1+\hxi_i)(1+\hxic_j)\rag & = & \frac{1}{\QQ_i} \int \dd\chi_i \, \cD_i^2
\frac{\dd\vOm_i}{(\Delta\Omega)} \frac{\dd M_i}{M_i} \int_i \dd\vr_{i'} 
\frac{\dd M_{i'}}{M_{i'}} \nonumber \\
&& \hspace{-1.5cm} \times \frac{1}{\QQ_j} \int \dd\chi_j \, \cD_j^2
\frac{\dd\vOm_j}{(\Delta\Omega)} \frac{\dd M_j}{M_j} \int_j \dd\vr_{j'}
\frac{\dd M_{j'}}{M_{j'}} \nonumber \\
&& \hspace{-1.5cm} \times \, \lag \frac{\dd\hn}{\dd\ln M_i} 
\frac{\dd\hn}{\dd\ln M_{i'}} \frac{\dd\hn}{\dd\ln M_j} \rag \, 
\frac{\dd n}{\dd\ln M_{j'}} .
\label{xii-xicj-1}
\eeqa
Proceeding as in App.~\ref{Computation-of-the-covariance-of-the-estimator},
this gives
\beq
C^{\rm c}_{i,j} = \lag\hxi_i\hxic_j \rag = C_{i,j}^{\rm c (3)} + C_{i,j}^{\rm c (4)} ,
\label{xii-xicj-2}
\eeq
with
\beqa
C_{i,j}^{\rm c (3)} & = & \frac{2}{(\Delta\Omega)\QQ_i\QQ_j} \int \dd\chi_i \, 
\cD_i^2 \frac{\dd\vOm_i}{(\Delta\Omega)} \frac{\dd M_i}{M_i} \int_i \dd\vr_{i'} 
\frac{\dd M_{i'}}{M_{i'}}  \nonumber \\
&& \hspace{-0.7cm} \times \int_j \! \dd\vr_{j'} \frac{\dd M_{j'}}{M_{j'}}
\frac{\dd n}{\dd\ln M_i} \frac{\dd n}{\dd\ln M_{i'}}
\frac{\dd n}{\dd\ln M_{j'}} \left[ 1\!+\! \xih_{i,i'} \right] ,
\label{C2-c-def}
\eeqa
\beqa
C_{i,j}^{\rm c (4)} & = & \frac{1}{\QQ_i\QQ_j} \int \dd\chi_i \cD_i^2
\frac{\dd\vOm_i}{(\Delta\Omega)} \frac{\dd M_i}{M_i} \int_i \dd\vr_{i'} 
\frac{\dd M_{i'}}{M_{i'}} \frac{\dd n}{\dd\ln M_i}  \nonumber \\
&& \hspace{-0.6cm} \times \frac{\dd n}{\dd\ln M_{i'}} \int \dd\chi_j \cD_j^2
\frac{\dd\vOm_j}{(\Delta\Omega)} \frac{\dd M_j}{M_j} \int_j \dd\vr_{j'}
\frac{\dd M_{j'}}{M_{j'}} \frac{\dd n}{\dd\ln M_j} \nonumber \\
&& \hspace{-0.6cm} \times \frac{\dd n}{\dd\ln M_{j'}}
\left[ 2 \xih_{i;j} + \zetah_{i,i';j} \right] .
\label{C3-c-def}
\eeqa
Next, to compute the last term in Eq.(\ref{Cij-LS-1}) we write
\beqa
\lag (1+\hxic_i)(1+\hxic_j)\rag & = & \frac{1}{\QQ_i} \int \dd\chi_i \cD_i^2
\frac{\dd\vOm_i}{(\Delta\Omega)} \frac{\dd M_i}{M_i} \int_i \dd\vr_{i'}
\frac{\dd M_{i'}}{M_{i'}} \nonumber \\
&& \hspace{-1.5cm} \times \frac{1}{\QQ_j} \int \dd\chi_j \cD_j^2
\frac{\dd\vOm_j}{(\Delta\Omega)} \frac{\dd M_j}{M_j} \int_j \dd\vr_{j'}
\frac{\dd M_{j'}}{M_{j'}} \nonumber \\
&& \hspace{-1.5cm} \times \, \lag \frac{\dd\hn}{\dd\ln M_i} 
\frac{\dd\hn}{\dd\ln M_j} \rag \, \frac{\dd n}{\dd\ln M_{i'}} 
\frac{\dd n}{\dd\ln M_{j'}} ,
\label{xici-xicj-1}
\eeqa
whence
\beq
C^{\rm cc}_{i,j} = \lag\hxic_i\hxic_j \rag = C_{i,j}^{\rm cc (3)}
+ C_{i,j}^{\rm cc (4)} ,
\label{xici-xicj-2}
\eeq
with
\beqa
C_{i,j}^{\rm cc (3)} & = & \frac{1}{(\Delta\Omega)\QQ_i\QQ_j} \int \dd\chi_i  
\cD_i^2 \frac{\dd\vOm_i}{(\Delta\Omega)} \frac{\dd M_i}{M_i} \int_i \dd\vr_{i'} 
\frac{\dd M_{i'}}{M_{i'}} \nonumber \\
&& \times \int_j \dd\vr_{j'} \frac{\dd M_{j'}}{M_{j'}} \; \frac{\dd n}{\dd\ln M_i}
\frac{\dd n}{\dd\ln M_{i'}} \frac{\dd n}{\dd\ln M_{j'}} ,
\label{C0-cc-def}
\eeqa
\beqa
C_{i,j}^{\rm cc (4)} & = & \frac{1}{\QQ_i\QQ_j} \int \! \dd\chi_i \cD_i^2
\frac{\dd\vOm_i}{(\Delta\Omega)} \frac{\dd M_i}{M_i} \int_i \! \dd\vr_{i'}
\frac{\dd M_{i'}}{M_{i'}} \frac{\dd n}{\dd\ln M_i}  \nonumber \\
&& \times \frac{\dd n}{\dd\ln M_{i'}} \int \! \dd\chi_j \cD_j^2
\frac{\dd\vOm_j}{(\Delta\Omega)} \frac{\dd M_j}{M_j} \int_j \!
\dd\vr_{j'} \frac{\dd M_{j'}}{M_{j'}} \frac{\dd n}{\dd\ln M_j} \nonumber \\
&& \times \frac{\dd n}{\dd\ln M_{j'}} \; \xih_{i;j} ,
\label{C2-cc-def}
\eeqa
Collecting all terms in Eq.(\ref{Cij-LS-1}), which reads as
$C^{\rm LS}_{i,j} = C_{i,j} - 2 C_{i,j}^{\rm c} - 2 C_{j,i}^{\rm c}
+ 4 C^{\rm cc}_{i,j}$, we obtain the decomposition (\ref{Cij-LS-2})
with the contributions (\ref{C2-LS-def})-(\ref{C4-LS-def}).

\section{Computation of high-order terms for the covariance of $\hxiLS$}
\label{Computation-high-order-terms}

We compute here the high-order terms for the covariance $C_{i,j}^{\rm LS}$
of the Landy-Szalay estimator $\hxiLS_i$ that we had neglected in
Eq.(\ref{Cij-LS-tot}).

For numerical computations, it is often more efficient to express the quantities
that we encounter in this work in terms of the real-space correlation $\xi(x)$,
instead of the Fourier-space power spectra $P(k)$ or $\Delta^2(k)$, 
provided $\xi(x)$ is known (e.g., computed in advance on a fine
grid\footnote{In practice, we compute in advance
$\xi(x,z)$ on a 2D grid, over distance and redshift, using Eq.(\ref{xi-Deltak})
and the nonlinear power spectrum from \citet{Smith2003}. To
obtain meaningful and accurate results, one needs to make sure that 
$\xi(x)$ is accurately computed, especially on large scales where one should recover
linear theory and a smooth two-point correlation.}). Indeed, this
replaces oscillatory integrals by integrals with slowly-varying factors, which allows
faster and more accurate computations.
This comes from mostly considering various kinds of
volume averages of correlation functions, such as Eq.(\ref{xib-ij-def}),
which are more naturally written in configuration space. This yields integrations
over bounded or unbounded domains with typically positive and slowly-varying
kernels. In contrast, the transformation to Fourier space yields highly oscillatory 
kernels as soon as some underlying real-space volumes are finite with a
size much larger than some other scales (see for instance the 2D top-hat
(\ref{W-thetas}) for a window $\theta_s$ that is much broader than the
typical angular scale $1/(k_{\perp}\cD)$).
On the other hand, intermediate analytical computations are often easier
to perform in Fourier space, mostly because of the convolution theorem.
Then, a convenient method is to first write expressions in terms of Fourier-space
power spectra, perform integrations over angles, and finally go back to the
real-space correlation function, using the fact that from Eq.(\ref{xi-Pk}), $\xi(x)$ 
and $\Delta^2(k)$ are related by
\beq
\xi(x) = \int \frac{\dd k}{k} \, \Delta^2(k) \, \frac{\sin(k x)}{k x} ,
\label{xi-Deltak}
\eeq
\beq
\Delta^2(k) = \frac{2}{\pi} \int \frac{\dd x}{x} \, \xi(x) \, (k x)^2 \, \sin(k x) .
\label{Deltak-xi}
\eeq
As shown below, this method also allows partial factorization of most integrals.

A first high-order contribution to the covariance $C_{i,j}^{\rm LS}$ arises from
the product $\xi_{i;j'} \xi_{i';j}$ in Eq.(\ref{C4-LS-def}), which also writes as 
Eq.(\ref{CLS-xixi-1}) where we introduced the quantity
$\overline{\xir_{i;j'} \xir_{i';j}}$ defined by
\beq
\overline{\xir_{i;j'} \xir_{i';j}} = \int \frac{\dd\chi_j}{\cD_i} 
\int\frac{\dd\vOm_i\dd\vOm_j}{(\Delta\Omega)^2} 
\int\frac{\dd\vr_{i'}\dd\vr_{j'}}{\cV_i\cV_j} \, \xi_{i;j'} \xi_{i';j} .
\label{Kij-def}
\eeq
Expressing the two-point correlation functions in terms of the power spectrum,
using the flat-sky (small angle) approximation, as well as Limber's approximation
as we did for Eq.(\ref{Cij-3}), we obtain after integration over angles and
over the two radial shells,
\beqa
\overline{\xir_{i;j'} \xir_{i';j}}  & = & \frac{2\pi}{\cD} \int \dd\vk_1\dd\vk_2 \,
P(k_1) P(k_2) \, \delta_D(k_{1\parallel}+k_{2\parallel}) \nonumber \\ 
&& \hspace{-0.4cm} \times \, \tW_i^{(3)}(k_1) \, \tW_j^{(3)}(k_2) \,
\tW_2(|\vk_{1\perp}+\vk_{2\perp}|\cD\theta_s)^2 .
\label{xi-ij'-i'j-1}
\eeqa
Again, the factor $2\pi\delta_D(k_{1\parallel}\!+\!k_{2\parallel})$ comes from the
integration over $\chi_j$, which suppresses longitudinal wavelengths.
Using the exponential representation of Dirac functions, Eq.(\ref{xi-ij'-i'j-1}) can
be partially factorized as
\beqa
\overline{\xir_{i;j'} \xir_{i';j}} & \!=\!\! & \!\int\!\!\frac{\dd\vr}{\cD(2\pi)^2} 
\!\int\!\!\dd\vk_1\dd\vk_2 \, P(k_1) P(k_2) \, \tW_i^{(3)}(k_1) \,
\tW_j^{(3)}(k_2) \,  \nonumber \\ 
&& \hspace{-1cm} \times \! \int\!\!\dd\vk_{\perp} \tW_2(k_{\perp}\cD\theta_s)^2
e^{\ii r_{\parallel}\cdot(k_{1\parallel}\!+k_{2\parallel})
+\ii\vr_{\perp}\cdot(\vk_{1\perp}\!+\vk_{2\perp}\!-\vk_{\perp})}  ,
\label{Kij-1}
\eeqa
and the integration over angles yields
\beqa
\overline{\xir_{i;j'} \xir_{i';j}} &  = & 2 \int \! \frac{\dd r}{r}  \int
\frac{\dd k_1}{k_1} \frac{\Delta^2(k_1)}{\cD k_1} \sin(k_1 r) \tW_i^{(3)}(k_1)
\nonumber \\
&& \times \int \frac{\dd k_2}{k_2} \frac{\Delta^2(k_2)}{\cD k_2} \sin(k_2 r) 
\tW_j^{(3)}(k_2) \nonumber \\
&& \times \int \frac{\dd k}{k} \cD k \sin(k r) \tW_2(k\cD\theta_s)^2 .
\label{Kij-2}
\eeqa
This also reads as
\beq
\overline{\xir_{i;j'} \xir_{i';j}} = 2\theta_s \! \int\! \frac{\dd r \; r}{(\cD\theta_s)^2} 
\; \cI_i^{(3)}(r) \, \cI_j^{(3)}(r) \, A^{(3)}\!\left(\!\frac{r}{\cD\theta_s}\!\right) ,
\label{Kij-3}
\eeq
where we introduced
\beq
A^{(3)}(y) = \int_0^{\infty} \dd u \, \sin(y u) \, \tW_2(u)^2 ,
\label{A3_y-def}
\eeq
and
\beq
\cI_i^{(3)}(r) = \int \frac{\dd k}{k} \, \Delta^2(k) \, \frac{\sin(kr)}{kr} \, 
\tW_i^{(3)}(k) .
\label{cI-3-def}
\eeq
The function $A^{(3)}(y)$ can be written as
\beqa
0< y< 2 : A^{(3)}(y) & = & \frac{2}{3\pi} \left[ 3\pi y - 2 (4+y^2) {\bf E}(y/2)
\right .\nonumber \\
&& \left. - 2 (-4+y^2) {\bf K}(y/2) \right] , \\
y> 2 : A^{(3)}(y) & = & \frac{2 y}{3\pi} \left[ 3\pi - (4+y^2) {\bf E}(2/y)
\right .\nonumber \\
&& \left. + (-4+y^2) {\bf K}(2/y) \right] , 
\label{A3_y}
\eeqa
where ${\bf K}(k)$ and ${\bf E}(k)$ are the complete elliptic integrals of the
first and second kinds \citep{Gradshteyn}. One can check that $A^{(3)}(y)$
is a positive, nonoscillatory, and continuous function (but not analytic at $y=2$),
with $A^{(3)}(y) \sim 2y$ for $y\rightarrow 0$ and
$A^{(3)}(y) \sim 1/y$ for $y\rightarrow \infty$.

Going back to configuration space, by substituting Eq.(\ref{Deltak-xi}),
the integral (\ref{cI-3-def}) can be written as
\beqa
\cI_i^{(3)}(r) & = & \int \frac{\dd x \; x \, \xi(x)}{r(\Rip^3-\Rim^3)} \,
\left[ \Rip^2 W_3\!\left(\frac{x}{\Rip},\frac{r}{\Rip}\right) \right. \nonumber \\
&& \left.  - \Rim^2 W_3\!\left(\frac{x}{\Rim},\frac{r}{\Rim}\right) \right] 
\label{cI-3-W3}
\eeqa
with
\beq
W_3(a,b) = \frac{2}{\pi} \int_0^{\infty} \dd u \, \sin(a u) \sin(b u) \tW_3(u) ,
\label{W3-ab-def}
\eeq
which for $a>0$ and $b>0$ is given by
\beq
\bea{rl}
|a-b|>1 : & W_3= 0 \\
|a-b|<1, \; a+b<1 : & W_3= 3 a b \\
|a-b|<1, \; a+b>1 : & W_3= \frac{3}{4} [1-(a-b)^2] .
\ea
\eeq
Thus, using Eq.(\ref{cI-3-W3}), the quantity $\overline{\xir_{i;j'} \xir_{i';j}}$ of
Eq.(\ref{Kij-3}) involves slowly varying integrals over real-space variables,
which partially factor as three factors within the integrand of Eq.(\ref{Kij-3}).
This makes it more efficient to use Eq.(\ref{Kij-3}) than the Fourier-space
expressions (\ref{xi-ij'-i'j-1}) or (\ref{Kij-2}).

To evaluate the two remaining contributions, associated with the factors
$\zeta_{i,i',j'}$ in Eq.(\ref{C3-LS-def}) and $\eta_{i,i';j,j'}$ in Eq.(\ref{C4-LS-def}),
we use the model for the three- and four-point correlation functions
described in Sect.~\ref{three-point}.
Thus, using Eq.(\ref{zeta-def}) for the three-point correlation function that enters
Eq.(\ref{C3-LS-def}), this contribution to Eq.(\ref{C3-LS-def}) reads as
\beqa
C_{i,j}^{\rm LS (3,\zeta)} & = & \frac{4}{(\Delta\Omega)\QQ_i\QQ_j} \int \dd\chi \,
\cD^2 \, \bb^3 \, \nb^3 \, \cV_i \cV_j \, \frac{S_3}{3}  \nonumber \\
&& \times \, \left[ \overline{\xir_{i,i'} \xir_{i,j'}} + \overline{\xir_{i',i} \xir_{i',j'}}
+ \overline{\xir_{j',i} \xir_{j',i'}} \right] .
\label{C3-zeta-1}
\eeqa
The first term in the bracket in Eq.(\ref{C3-zeta-1}) is given by
\beq
\overline{\xir_{i,i'} \xir_{i,j'}} = \overline{\xir_{i'}} \, \times \, 
\overline{\xir_{j'}}  ,
\eeq
where $ \overline{\xir_{i'}}$ was defined in Eq.(\ref{xi-i-i'-def}), because the 
integrations over $\vr_{i'}$ and $\vr_{j'}$ are independent. The second term reads as
\beqa
\overline{\xir_{i',i} \xir_{i',j'}} & = & \int\!\dd\vk_1\dd\vk_2 \, P(k_1) P(k_2) \,
\tW_i^{(3)}(|\vk_1+\vk_2|) \nonumber \\
&& \times \,  \tW_j^{(3)}(k_2) ,
\label{xi-i'i-xi-i'j'-1}
\eeqa
which no longer factors.
Introducing an auxiliary wavenumber $\vk$ and the Dirac factor
$\delta_D(\vk_1+\vk_2-\vk)$, which we write under its exponential form
as in Eq.(\ref{Kij-1}), and using the inverse Fourier transform of the 3D
shell (\ref{W3-D-def}),
\beqa
\frac{\Theta(\vr\in\cV_i)}{\cV_i} & = & \int \frac{\dd\vk}{(2\pi)^3} \, 
e^{-\ii\vk\cdot\vr} \, \tW_i^{(3)}(k) \\
& = & \frac{1}{2\pi^2} \int \frac{\dd k}{k} \, k^3 \, \frac{\sin(k r)}{k r} \,
\tW_i^{(3)}(k) ,
\label{W3-inverseFourier}
\eeqa
as well as Eq.(\ref{cI-3-def}), we obtain
\beq
\overline{\xir_{i',i} \xir_{i',j'}} = \int_{\cV_i} \frac{\dd\vr}{\cV_i} \, \xi(r) \, 
\cI_j^{(3)}(r) . 
\label{J3ij-1}
\eeq
The third term in Eq.(\ref{C3-zeta-1}) is obtained from Eq.(\ref{J3ij-1})
by exchanging the labels ``$i$'' and ``$j$''.
 
We now turn to the four-point contribution to Eq.(\ref{C4-LS-def}), using
Eq.(\ref{eta-def}) for the halo four-point correlation function $\etah_{i,i';j,j'}$.
Thanks to the symmetries $\{i\leftrightarrow i'\}$ and $\{j\leftrightarrow j'\}$
we have two different contributions 
(a) and (b) associated with the topology of the left diagram in Fig.~\ref{fig-eta}, 
each with a multiplicity factor $2$, and four different contributions (c), (d), (e), and
(f), associated with the topology of the right diagram, with multiplicity factors
$4,4,2$, and $2$.

The first contribution (a) reads as
\beq
C_{i,j}^{\rm LS (4,a)} = \frac{2}{\QQ_i\QQ_j} \int
\dd\chi \, \cD^5 \, \bb^4 \, \nb^4 \, \cV_i \cV_j \, \frac{S_4}{16} \, 
\overline{\xir_{i,i'}\xi_{i;j}\xir_{i;j'}}
\label{C4-a-1}
\eeq
with
\beqa
\overline{\xir_{i,i'}\xi_{i;j}\xir_{i;j'}} & = & \int \frac{\dd\chi_j}{\cD_i}
\int\frac{\dd\vOm_i\dd\vOm_j}{(\Delta\Omega)^2} 
\int\frac{\dd\vr_{i'}\dd\vr_{j'}}{\cV_i\cV_j} \, \xi_{i,i'} \xi_{i;j} \xi_{i;j'} \nonumber \\
& = & \overline{\xir_{i'}} \times \overline{\xi_{i;j} \xir_{i;j'}} .
\eeqa
Proceeding as for Eq.(\ref{Kij-def}), we obtain
\beq
\overline{\xi_{i;j} \xir_{i;j'}} = 2\theta_s \int \frac{\dd r \; r}{(\cD\theta_s)^2} \;
\xi(r) \, \cI_j^{(3)}(r) \, A^{(3)} \!\left(\frac{r}{\cD\theta_s}\right) ,
\label{Kj-def}
\eeq
The contribution $C_{i,j}^{\rm LS (4,b)}$ is the symmetric one with respect to
$\{i\leftrightarrow j\}$ of Eq.(\ref{C4-a-1}); that is, the product
$\overline{\xir_{i'}} \times \overline{\xi_{i;j} \xir_{i;j'}}$ is replaced by
$\overline{\xir_{j'}} \times \overline{\xi_{i;j} \xir_{j;i'}}$.

Next, the contribution $C_{i,j}^{\rm LS (4,c)}$ reads as
\beq
C_{i,j}^{\rm LS (4,c)} = \frac{4}{\QQ_i\QQ_j} \int
\dd\chi \, \cD^5 \, \bb^4 \, \nb^4 \, \cV_i \cV_j \, \frac{S_4}{16} \, 
\overline{\xir_{i',i}\xi_{i;j}\xir_{j,j'}}
\label{C4-c-1}
\eeq
where the geometrical average writes as
\beq
\overline{\xir_{i',i}\xi_{i;j}\xir_{j,j'}} = \overline{\xir_{i'}} \times \xicyl 
\times \overline{\xir_{j'}} ,
\eeq
since integrals over $\vr_{i'}$ and $\vr_{j'}$ can be factored.

The contribution (d) involves $\overline{\xir_{j';i}\xi_{i;j}\xir_{j;i'}}$ where no
factorization is possible. 
Proceeding as for Eq.(\ref{Kij-def}) we obtain
\beqa
\overline{\xir_{j';i}\xi_{i;j}\xir_{j;i'}} & = & 2\theta_s \int \! 
\frac{\dd r \; r}{(\cD\theta_s)^2} \;
\xi(r) \, \cI_i^{(3)}(r) \, \cI_j^{(3)}(r) \, A^{(3)}
\!\left(\!\frac{r}{\cD\theta_s}\!\right) . \nonumber \\
&&
\label{Lij-def}
\eeqa
The contribution (e) involves $\overline{\xir_{j';i}\xir_{i,i'}\xir_{i';j}}$ that can be
written as
\beqa
\overline{\xir_{j';i}\xir_{i,i'}\xir_{i';j}} & = & 3\theta_s \int 
\frac{\dd r \; r}{(\cD\theta_s)^2} \; A^{(3)}\!\left(\!\frac{r}{\cD\theta_s}\!\right)
\, \cI_j^{(3)}(r) \nonumber \\
&& \hspace{-1.2cm} \times \int_{\Rim}^{\Rip} \!\! 
\frac{\dd r' \, r'^2}{\Rip^3\!-\!\Rim^3} \xi(r') \int_{|r-r'|}^{r+r'} 
\frac{\dd r'' \, r''}{r \, r'} \, \xi(r'') ,
\label{Tij-def}
\eeqa
whereas contribution (f) is obtained from (e) by exchanging the labels ``$i$'' and
``$j$''.

Collecting all terms, the high-order contributions to the covariance matrix
$C_{i,j}^{\rm LS}$ are given by Eqs.(\ref{CLS-xixi-1})-(\ref{CLS-eta-1}).

\section{Computation of the mean of the estimators $\hw$ and $\hwLS$}
\label{Computation-of-the-mean-of-the-estimator-w}

We give here explicit expressions of the average (\ref{w-i-i'-def}) of the correlation
function over an angular ring.
As in Sect.~\ref{Flat-sky-approximation}, using the flat-sky and Limber's
approximations, we obtain
\beq
\overline{\xith_{i'}}(z) = \pi  \int_0^{\infty}
\frac{\dd k}{k} \frac{\Delta^2(k,z)}{\cD k} \tW_i^{(2)}(k\cD) ,
\label{I2i-def}
\eeq
where we introduced the 2D Fourier-space window of the $i$-ring,
\beqa
\tW_i^{(2)}(k_{\perp}\cD) & = & \int_{\cA_i} \, \frac{\dd\vtheta}{\cA_i} \,
e^{\ii \vk_{\perp}\cdot\cD\vtheta} \nonumber \\
&& \hspace{-0.6cm} = \frac{\thetaip^2 \tW_2(k_{\perp}\cD\thetaip) 
- \thetaim^2 \tW_2(k_{\perp}\cD\thetaim)}{\thetaip^2- \thetaim^2} ,
\label{W2i-def}
\eeqa
and $\tW_2$, associated with a full circular window, was defined in
Eq.(\ref{W-thetas}).
In terms of the two-point correlation function, Eq.(\ref{I2i-def}) also writes as
\beqa
\overline{\xith_{i'}}(z) & \!\!=\! & \frac{4}{\thetaip^2 \!-\! \thetaim^2} \biggl\lbrace 
\int_{\cD\thetaim}^{\cD\thetaip} \frac{\dd x \; x^2}{\cD^3} \;
\xi(x) \, \sqrt{1 \!-\! \cD^2\thetaim^2/x^2} \nonumber \\
&&  + \int_{\cD\thetaip}^{\infty} \frac{\dd x \; x^2}{\cD^3} \; \xi(x) \,
\left[ \sqrt{1 \!-\! \cD^2\thetaim^2/x^2} \right . \nonumber \\
&& \left. - \sqrt{1 \!-\! \cD^2\thetaip^2/x^2} \right] \biggl\rbrace ,
\eeqa
which avoids introducing oscillatory kernels.

\section{Computation of the covariance of $\hwLS$}
\label{Computation-high-order-terms-w}

The low-order contribution (\ref{Cij-LS-w-tot}) to the covariance matrix of the
estimator $\hwLS$ involves the angular average (\ref{I2ij-def-xi}).
Using Limber's approximation it also reads as
\beq
\overline{\xith_{i',j'}} =  \pi  \int_0^{\infty}
\frac{\dd k}{k} \frac{\Delta^2(k,z)}{\cD k} \tW_i^{(2)}(k\cD) \tW_j^{(2)}(k\cD) .
\label{I2ij-def}
\eeq

We now compute the high-order terms of the covariance $C_{i,j}^{\rm LS}$,
which are given in Eqs.(\ref{CLS-w-xixi-1})-(\ref{CLS-w-eta-1}).
A first contribution (\ref{CLS-w-xixi-1}) arises from the product
$\xi_{i;j'} \xi_{i';j}$ in Eq.(\ref{C4-LS-w-def}). Using Limber's approximation and
integrating over angles yields
\beqa
C_{i,j}^{\rm LS (4,\xi\xi)} & = & \frac{2(2\pi)^2}{\Nb^4} \int \dd\chi_i \, \cD_i^4 \,
\bb_i^2 \, \nb_i^2 \int \dd\chi_j \, \cD_j^4 \, \bb_j^2 \, \nb_j^2
\nonumber \\
&& \hspace{-0.9cm}  \times \int \dd\vk_{1\perp} \dd\vk_{2\perp} \,
P(k_{1\perp};z_i) P(k_{2\perp};z_j) \, \tW_i^{(2)}(k_{2\perp}\cD_j) \nonumber \\
&& \hspace{-0.9cm}  \times \tW_j^{(2)}(k_{1\perp}\cD_i) 
\tW_2[(\cD_j\vk_{2\perp}-\cD_i\vk_{1\perp})\theta_s]^2 .
\label{Cw-LS-xixi-1}
\eeqa
Introducing a Dirac factor
$\delta_D(\cD_j\vk_{2\perp}-\cD_i\vk_{1\perp}-\vx_{\perp})$, 
which we write with the usual exponential representation in a fashion similar
to Eq.(\ref{Kij-1}), we obtain after integration over angles
\beqa
C_{i,j}^{\rm LS (4,\xi\xi)} & = & \frac{2(2\pi)^4}{\Nb^4} \int \dd\chi_i \, \cD_i^4 \,
\bb_i^2 \, \nb_i^2 \int \dd\chi_j \, \cD_j^4 \, \bb_j^2 \, \nb_j^2
\nonumber \\
&& \hspace{-1.1cm}  \times \int \dd k_1 \, k_1 P(k_1;z_i) \tW_j^{(2)}(k_1\cD_i)
\int \dd k_2 \, k_2 P(k_2;z_j) \nonumber \\
&& \hspace{-1.1cm}  \times \tW_i^{(2)}(k_2\cD_j) \int \dd y \, y J_0(yk_1\cD_i)
J_0(yk_2\cD_j) \nonumber \\
&& \hspace{-1.1cm}  \times \int \dd x \, x J_0(x y) \tW_2(x\theta_s)^2 .
\eeqa
Then, after a rescaling of variables $x$ and $y$, and defining the quantities
\beq
A^{(2)}(y) = \int_0^{\infty} \dd u \, u \, J_0(y u) \, \tW_2(u)^2 ,
\label{A2-def}
\eeq
\beq
B_i^{(2)}(\theta) = \int \dd\chi \, \cD^5 \, \bb^2 \, \nb^2 \, \cI_i^{(2)}(\theta) ,
\label{B2-def}
\eeq
\beq
\cI_i^{(2)}(\theta) = \int \frac{\dd k}{k} \, \frac{\Delta^2(k)}{\cD k}
\, J_0(k\cD\theta) \, \tW_i^{(2)}(k\cD) ,
\label{cI2-def}
\eeq
we obtain the expression (\ref{CLS-w-xixi-1}), using the property $A^{(2)}(y)=0$
for $y>2$.
As compared with Eq.(\ref{Cw-LS-xixi-1}), introducing the Dirac factor
and the two auxiliary variables $x$ and $y$ has allowed us to partly factor in
the integrals, as seen in Eq.(\ref{CLS-w-xixi-1}), which is convenient for numerical
computations.
Again, it is useful to express Eq.(\ref{cI2-def}) in terms of the real-space two-point
correlation function, which yields
\beqa
\cI_i^{(2)}(\theta) & = & \frac{2}{\pi} \int
\frac{\dd x \; \xi(x)}{\cD (\thetaip^2-\thetaim^2)} \left[ \thetaip^2 
W_2\!\left(\frac{\cD\theta}{x},\frac{\cD\thetaip}{x}\right) \right. \nonumber \\
&& \left. - \thetaim^2 W_2\!\left(\frac{\cD\theta}{x},\frac{\cD\thetaim}{x}\right)
\right] ,
\label{cI2-xi}
\eeqa
with
\beq
W_2(a,b) = \int_0^{\infty} \dd u \, \sin(u) \, J_0(a u) \, \tW_2(b u) .
\label{W2-ab-def}
\eeq
Although there is no explicit expression for the integral (\ref{W2-ab-def}) for
arbitrary $(a,b)$, for $|a-b|>1$ we can use the properties
\beq
\bea{rl}
b<a-1 : & W_2= 0 \\
b>a+1 : & W_2= 2/b^2 . \\
\ea
\eeq
In the band $|a-b|<1$ one can check that $W_2(a,b)$ is positive and decays
as $\sim b^{-2}$ for large $b$, so that the real-space expression (\ref{cI2-xi})
is again more convenient than the Fourier-space expression (\ref{cI2-def}).

The second contribution (\ref{CLS-w-zeta-1}) arises from the three-point correlation
$\zeta$ in Eq.(\ref{C3-LS-w-def}). Using Eq.(\ref{zeta-def}) it reads as
\beqa
C_{i,j}^{\rm LS (3,\zeta)} & = & \frac{4}{(\Delta\Omega)\Nb^4} \int \dd\chi \,
\cD^8 \, \bb^3 \, \nb^3 \, \frac{S_3}{3} \nonumber \\
&& \times \, \left[ \overline{\xith_{i,i'} \xith_{i,j'}} 
+ \overline{\xith_{i',i} \xith_{i',j'}} + \overline{\xith_{j',i} \xith_{j',i'}} \right] ,
\label{C3-w-zeta-1}
\eeqa
where the three terms in the brackets, which correspond to the three
diagrams in Fig.~\ref{fig-zeta}, are again geometrical averages along the
lines of sight, which we compute with Limber's approximation.
In particular, the first term factors as
\beq
\overline{\xith_{i,i'} \xith_{i,j'}} = \overline{\xith_{i'}} \times
\overline{\xith_{j'}} ,
\eeq
where $\overline{\xith_{i'}}$ was defined in Eq.(\ref{w-i-i'-def}), while the second
term reads as
\beqa
\overline{\xith_{i',i} \xith_{i',j'}} & = & \frac{(2\pi)^2}{\cD^2} \int\dd\vk_{1\perp}
\dd\vk_{2\perp} \, P(k_{1\perp}) P(k_{2\perp}) \nonumber \\
&& \times \, \tW_i^{(2)}(|\vk_{1\perp}+\vk_{2\perp}|\cD) \,
\tW_j^{(2)}(k_{2\perp}\cD) .
\label{xi-i'i-xi-i'j'-w-1}
\eeqa
With the same factorization method, and using the inverse Fourier transform of
the 2D shell (\ref{W2i-def}),
\beqa
\frac{\Theta(\vtheta\in\cA_i)}{\cA_i} & = & \int 
\frac{\dd\vk_{\perp}\,\cD^2}{(2\pi)^2} \, e^{-\ii\vk_{\perp}\cdot\cD\vtheta} 
\, \tW_i^{(2)}(k_{\perp}\cD) \\
& = & \frac{\cD^2}{2\pi} \int \frac{\dd k}{k} \, k^2 \, J_0(k\cD\theta) \, 
\tW_i^{(2)}(k\cD) ,
\label{W2-inverseFourier}
\eeqa
we obtain
\beq
\overline{\xith_{i',i} \xith_{i',j'}} = 2\pi^2 \int_{\thetaim}^{\thetaip}
\frac{\dd\theta \; \theta}{\thetaip^2-\thetaim^2} \, \xi_{\rm cyl}(\theta) \,
\cI_j^{(2)}(\theta) ,
\label{xi-i'i-xi-i'j'-w-2}
\eeq
where we introduced
\beqa
\xi_{\rm cyl}(\theta) & = & \int\frac{\dd k}{k} \, \frac{\Delta^2(k)}{\cD k} \, 
J_0(k\cD\theta) \\
& = & \frac{2\theta}{\pi} \int_0^1 \frac{\dd u}{u^2\sqrt{1-u^2}} \;
\xi\!\left(\frac{\cD\theta}{u}\right) .
\label{xi-cyl}
\eeqa
The third term in Eq.(\ref{C3-w-zeta-1}) is obtained from Eq.(\ref{xi-i'i-xi-i'j'-w-2})
by exchanging the labels ``$i$'' and ``$j$''.

The third contribution (\ref{CLS-w-eta-1}) arises from the four-point correlation
$\eta$ in Eq.(\ref{C4-LS-w-def}). As in App.~\ref{Computation-high-order-terms},
we must compute the various terms associated with the diagrams of
Fig.~\ref{fig-eta}, with contributions (a) and (b) associated with the left
diagram and contributions (c), (d), (e), and (f) associated with the right diagram.
The first contribution (a) leads to
\beq
C_{i,j}^{\rm LS (4,a)} = \frac{2}{\Nb^4} \int \dd\chi \, \cD^{11} \, 
\bb^4 \, \nb^4 \, \frac{S_4}{16} \, \overline{\xith_{i,i'}\xi_{i;j}\xith_{i;j'}} .
\eeq
As in App.~\ref{Computation-high-order-terms}, this geometrical
average factors as
\beq
\overline{\xith_{i,i'}\xi_{i;j}\xith_{i;j'}} = \overline{\xith_{i'}} \times
\overline{\xi_{i;j} \xith_{i;j'}}  ,
\eeq
with
\beqa
\overline{\xi_{i;j} \xith_{i;j'}} & = & \frac{(2\pi)^2}{\cD^2} \int\dd\vk_{1\perp}
\dd\vk_{2\perp} P(k_{1\perp}) P(k_{2\perp}) \tW_j^{(2)}(k_{1\perp}\cD)
\nonumber \\
&& \times \, \tW_2(|\vk_{1\perp}+\vk_{2\perp}|\cD\theta_s)^2 \\
& = & \pi^2 \int_0^{2\theta_s} \frac{\dd\theta \; \theta}{\theta_s^2}
\, \xi_{\rm cyl}(\theta) \, \cI_j^{(2)}(\theta) \,
A^{(2)}\!\left(\frac{\theta}{\theta_s}\right) .
\label{K2-xi}
\eeqa
Contribution (b) is the symmetric one of (a) with respect to $i\leftrightarrow j$.

Next, contribution (c) involves the geometrical average
$\overline{\xith_{i',i}\xi_{i;j}\xith_{j,j'}}$, which again factors as
\beq
\overline{\xith_{i',i}\xi_{i;j}\xith_{j,j'}} = \overline{\xith_{i'}} \times
\xicyl \times \overline{\xith_{j'}} .
\eeq

The contribution (d) involves the average
\beqa
\overline{\xith_{j';i}\xi_{i;j}\xith_{j;i'}} & = & \int
\frac{\dd\chi_{i'}\dd\chi_j\dd\chi_{j'}}{\cD^3} \int 
\frac{\dd\vOm_i\dd\vOm_j}{(\Delta\Omega)^2} 
\int\frac{\dd\vtheta_{i'}\dd\vtheta_{j'}}{\cA_i\cA_j} \nonumber \\
&& \times \, \xi_{j';i}\xi_{i;j}\xi_{j;i'} ,
\eeqa
which also writes as
\beqa
\overline{\xith_{j';i}\xi_{i;j}\xith_{j;i'}}  & \! = \! & \pi^3 \! \int_0^{2\theta_s} \! 
\frac{\dd\theta \; \theta}{\theta_s^2} \, \xi_{\rm cyl}(\theta) \, 
\cI_i^{(2)}(\theta) \, \cI_j^{(2)}(\theta) \,
A^{(2)}\!\left(\frac{\theta}{\theta_s}\right) . \nonumber \\
&&
\label{L2-xi}
\eeqa
Contribution (e) involves $\overline{\xith_{j';i}\xith_{i,i'}\xith_{i';j}}$, which
reads as
\beqa
\overline{\xith_{j';i}\xith_{i,i'}\xith_{i';j}} & \!\! = \! & 2\pi^2 \!\!
\int_0^{2\theta_s} \! \frac{\dd\theta\;\theta}{\theta_s^2} \, A^{(2)} \!
\left(\frac{\theta}{\theta_s}\right) \cI_j^{(2)}(\theta) 
\int_{\thetaim}^{\thetaip} \!\!\!\!
\frac{\dd\theta' \, \theta'}{\thetaip^2\!-\!\thetaim^2} \nonumber \\
&& \hspace{-0.9cm} \times \xi_{\rm cyl}(\theta') \int_0^{\pi}\! \dd\varphi \;
\xi_{\rm cyl}(\sqrt{\theta^2\!+\!\theta'^2\!+\!2\theta\theta'\cos\varphi}) ,
\eeqa
whereas contribution (f) is obtained from (e) by exchanging the labels ``$i$'' and
``$j$''.

Collecting all terms, the high-order contributions to the covariance matrix
$C_{i,j}^{\rm LS}$ are given by Eqs.(\ref{CLS-w-xixi-1})-(\ref{CLS-w-eta-1}).

\section{Scaling of the number counts signal-to-noise in simulations}
\label{scaling-horizon}

\begin{figure}
\begin{center}
\epsfxsize=9.5 cm \epsfysize=6.5 cm {\epsfbox{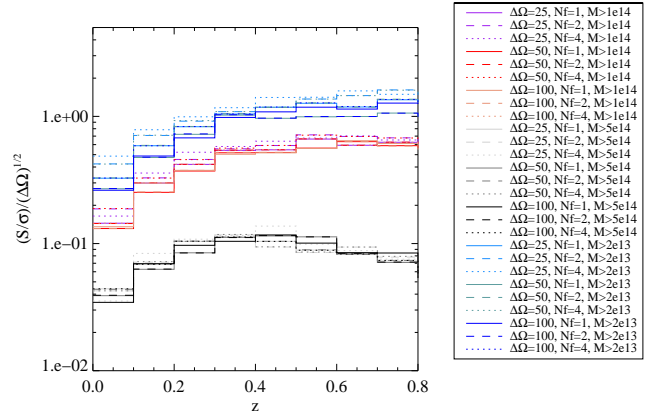}}
\end{center}
\caption{Scaling of the number-counts signal-to-noise ratio by $\sqrt{\Delta\Omega}$
as computed in the Horizon simulation, see Sect.~\ref{simulations}.
Different configurations are displayed according to the total surveyed area
$\Delta\Omega$, the number of subfields $N_f=\cN$, and the mass limit. 
In the right caption, $\Delta\Omega$ is expressed 
in deg$^2$ and the mass unit is $h^{-1} M_{\odot}$.}
\label{dndz-horizon-scaling-domega}
\end{figure}

\begin{figure}
\begin{center}
\epsfxsize=9.5 cm \epsfysize=6.5 cm {\epsfbox{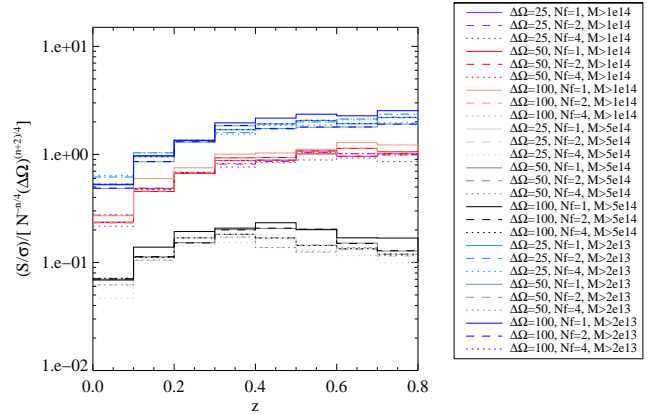}}
\end{center}
\caption{Same as Fig.~\ref{dndz-horizon-scaling-domega} but with a scaling that
depends on the number of subfields: $\cN^{-n/4} \, (\Delta\Omega)^{(n+2)/4}$,
with $n=-0.6$}
\label{dndz-horizon-scaling-n06}
\end{figure}

We present here the result of scaling the number counts with the total surveyed area
$\Delta\Omega$ and the number of subfields $\cN$.
Figures~\ref{dndz-horizon-scaling-domega} and \ref{dndz-horizon-scaling-n06} show
the scalings expected from Eq.(\ref{SN-Ntot}) in the shot-noise and sample-variance
dominated regimes. Multiple survey configurations are explored by varying the total
surveyed area ($\Delta\Omega=25, 50$, and $100$ deg$^2$), the number
of subfields ($\cN=1, 2$, and $4$), and the mass threshold
($M > 2\times 10^{13}, 10^{14}$, and $5 \times 10^{14}h^{-1} M_{\odot}$). 

The weak scatter in those plots shows that (\ref{SN-Ntot}) provides a valid
approximation of the signal-to-noise scaling with respect to $\Delta\Omega$ and
$\cN$.
In agreement with the discussion in Sect.~\ref{Scalings-Nz} and
Fig.~\ref{fig_SN_Nz}, at high redshift and for high mass, the scaling
$\sqrt{(\Delta\Omega)}$ shown in Fig.~\ref{dndz-horizon-scaling-domega} is
best, as expected for the shot-noise dominated regime, whereas at low redshift and
for low mass the scaling $\cN^{-n/4} \, (\Delta\Omega)^{(n+2)/4}$
shown in Fig.~\ref{dndz-horizon-scaling-n06} (with $n=-0.6$) is best,
as expected for the sample-variance dominated regime.

\section{Selection functions used for various surveys}
\label{selection}

\begin{figure*}[htb]
\begin{center}
\epsfxsize=6 cm \epsfysize=5 cm {\epsfbox{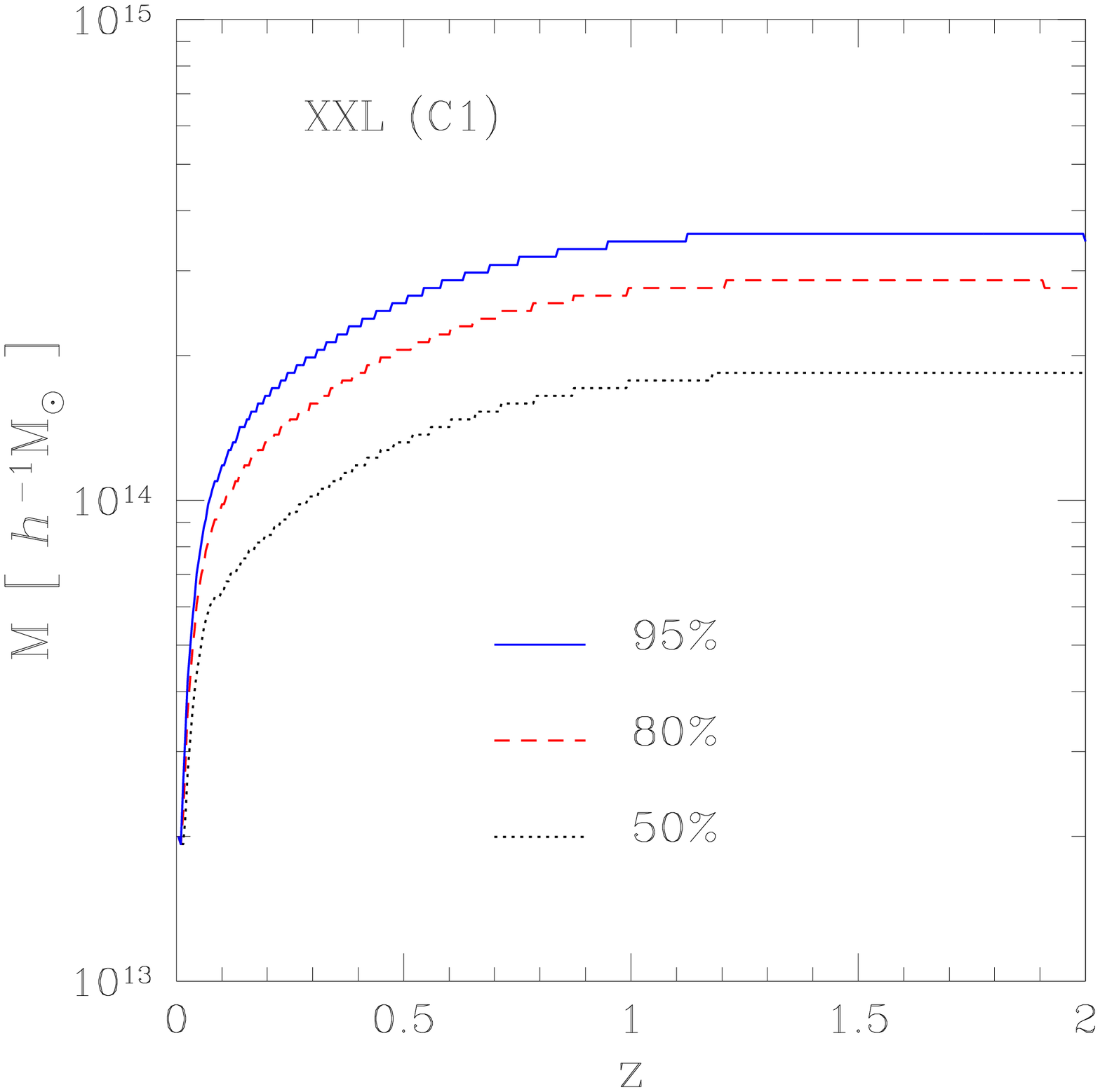}}
\epsfxsize=6 cm \epsfysize=5 cm {\epsfbox{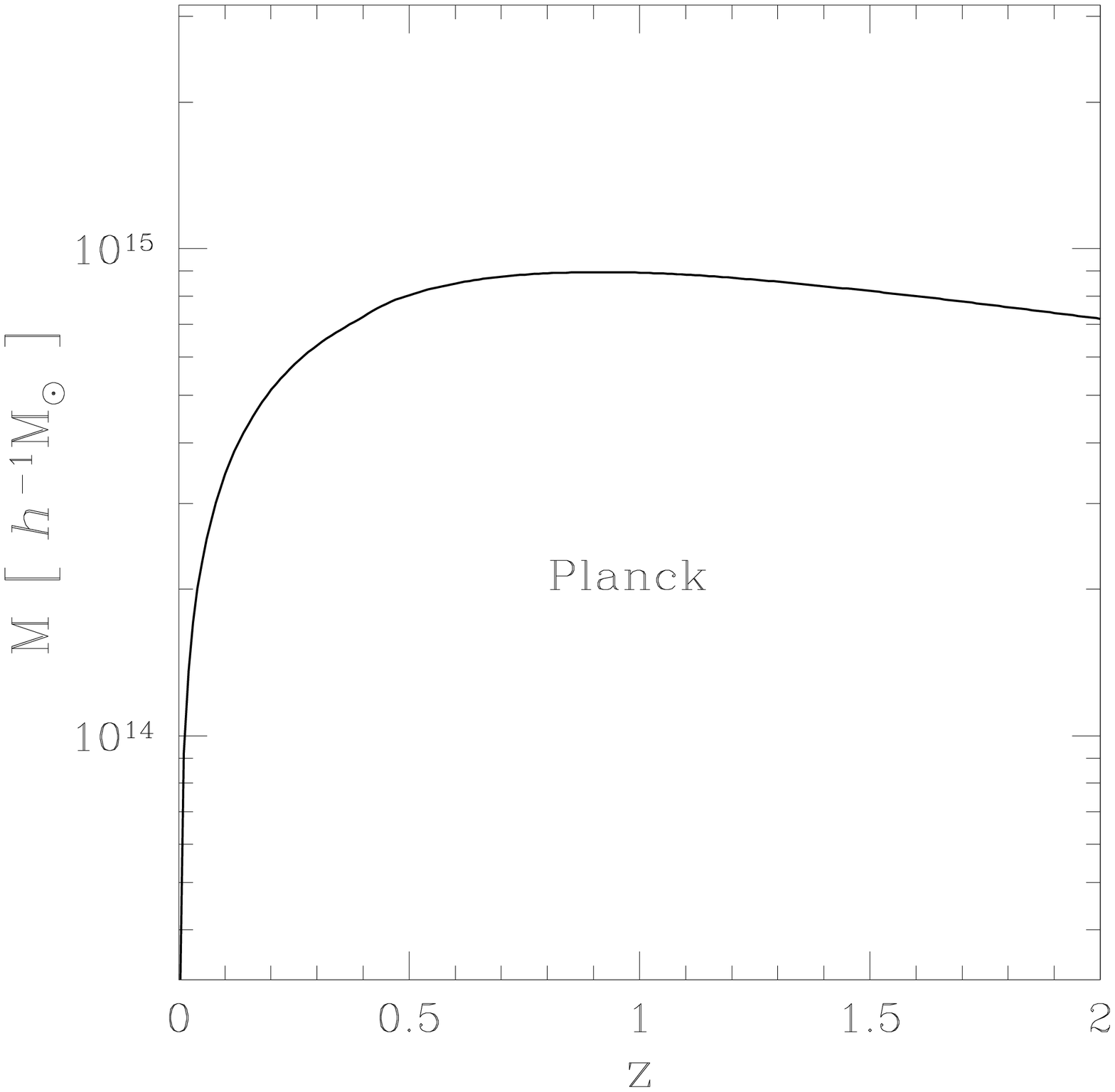}}
\epsfxsize=6 cm \epsfysize=5 cm {\epsfbox{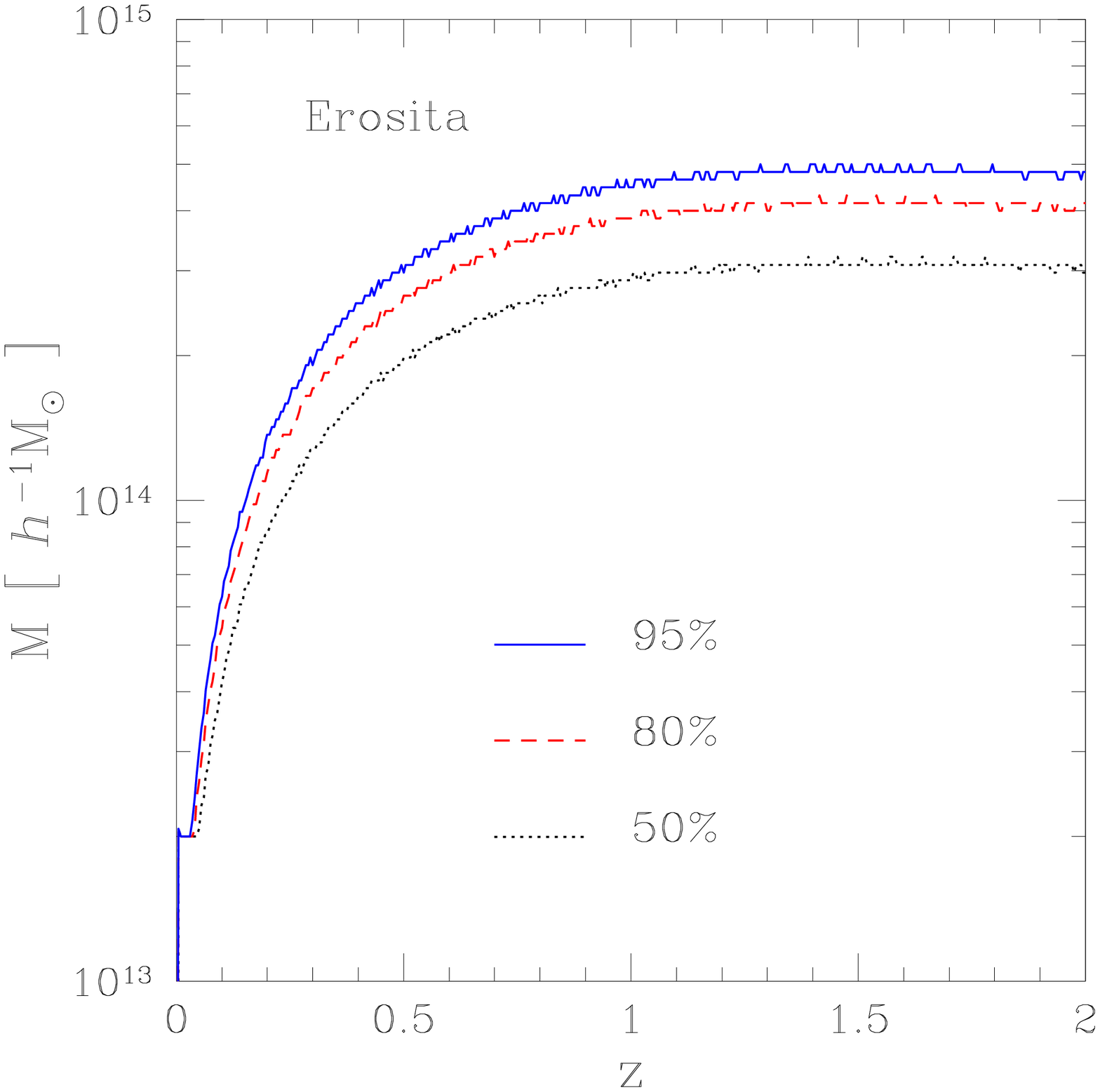}}
\end{center}
\caption{{\it Left panel:} cluster mass associated with a $50\%$, $80\%$, or $95\%$
detection probability (from bottom to top), for the XXL selection function C1,
as a function of redshift. {\it Middle panel:} minimum detectable cluster mass, as a function of redshift, for the Planck space mission. {\it Right panel:} cluster mass associated with a $50\%$, $80\%$, or $95\%$ detection probability (from bottom to top), 
for the Erosita selection function as a function of redshift (we consider a flux limit of 
$4\times 10^{-14}$ \flux\ in the $[0.5-2]$ keV band).}
\label{fig_Fs_XXL_Planck_Erosita}
\end{figure*}

We give in Fig.~\ref{fig_Fs_XXL_Planck_Erosita} the selection functions that we
use for several cluster surveys investigated in Sect.~\ref{Applications}.
For Planck, the curves shown in the middle panel corresponds to a $100\%$
detection probability.

For the other surveys studied in Sect.~\ref{Applications} we consider
simple mass thresholds, rather than detailed selection functions.
More precisely, we consider halos above the two thresholds
$5\times 10^{13}h^{-1} M_{\odot}$ and $5\times 10^{14}h^{-1} M_{\odot}$
for DES and Euclid, and above $5\times 10^{14}h^{-1} M_{\odot}$ for SPT.

\section{Dependence on cosmology}
\label{app-cosmology}

\begin{table}
\begin{center}
\begin{tabular}{c||c|c|c}
 & $h$ & $\Omega_{\rm m}$ & $\sigma_8$ \\ \hline\hline
WMAP7 mean values & 0.702 & 0.274 & 0.816 \rule[-0.3cm]{0cm}{0.7cm}  \\
$2\!-\!\sigma$ deviations & 0.73 & 0.289 & 0.864 \\ 
line style & dashed & dot-dashed & dotted
\end{tabular}
\end{center}
\caption{Three alternative cosmologies.}
%\caption{{\it First line:} means values of the cosmological parameters
%$h$, $\Omega_{\rm m}$, and $\sigma_8$, that we used in Sect.~\ref{Applications} 
%for the applications to real survey cases, from WMAP7 \citep{Komatsu2010}. 
%{\it Second line:} values that we also consider in this appendix, which correspond
%to $2\!-\!\sigma$ deviations. (When we vary $\Omega_{\rm m}$ we keep a flat
%$\Lambda$CDM universe and we change $\Omega_{\rm de}$ according to
%$\Omega_{\rm de}=1-\Omega_{\rm m}$.)
%{\it Third line:} line-style used in Figs.~\ref{fig_Nz_sn_sv_cosmo}, 
%\ref{fig_XiR_z0to1_sn_sv_cosmo}, and \ref{fig_XiR_z0to1_sv_cosmo}, for the
%three cosmologies where we change one of these parameters from its WMAP7 
%mean value.}
\label{Table_cosmo}
\end{table}

In this appendix we investigate the dependence of the results obtained in
Sect.~\ref{Applications} on the value of the cosmological parameters.
Thus, in addition to the WMAP7 cosmology recalled in the first line of
Table~\ref{Table_cosmo}, which was used in Sect.~\ref{Applications}, we also
consider the three modified cosmologies where one among the three parameters
$h$, $\Omega_{\rm m}$, and $\sigma_8$ is changed to the values shown in the
second line of Table~\ref{Table_cosmo}. They correspond to ``$2\!-\!\sigma$''
deviations from WMAP7 \citep{Komatsu2010} and describe current uncertainties.
(When we vary $\Omega_{\rm m}$ we keep a flat
$\Lambda$CDM universe and we change $\Omega_{\rm de}$ according to
$\Omega_{\rm de}=1-\Omega_{\rm m}$.)

\begin{figure}
\begin{center}
\epsfxsize=8.5 cm \epsfysize=6.5 cm {\epsfbox{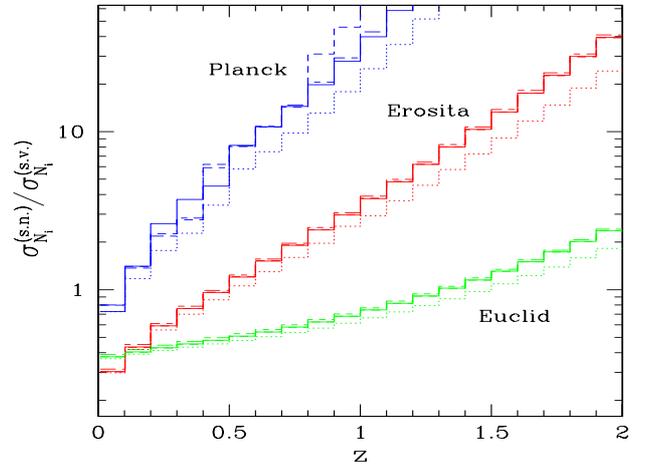}}
\end{center}
\caption{The ratio $\sigma_{N_i}^{(s.n.)}/\sigma_{N_i}^{(s.v.)}$ of the
rms shot-noise contribution $\sigma_{N_i}^{(s.n.)}$ to the rms
sample-variance contribution $\sigma_{N_i}^{(s.v.)}$, of the covariance of the
angular number densities $N_i$, as in Fig.~\ref{fig_Nz_sn_sv}. The fiducial curve
that was shown in Fig.~\ref{fig_Nz_sn_sv} is the solid line (mean WMAP7 cosmology),
whereas the dashed, dot-dashed, and dotted lines correspond to the three 
cosmologies where either $h$, $\Omega_{\rm m}$, or $\sigma_8$, is changed
to the value given in the second line of Table~\ref{Table_cosmo}.}
\label{fig_Nz_sn_sv_cosmo}
\end{figure}

\begin{figure}
\begin{center}
\epsfxsize=8.5 cm \epsfysize=6.5 cm {\epsfbox{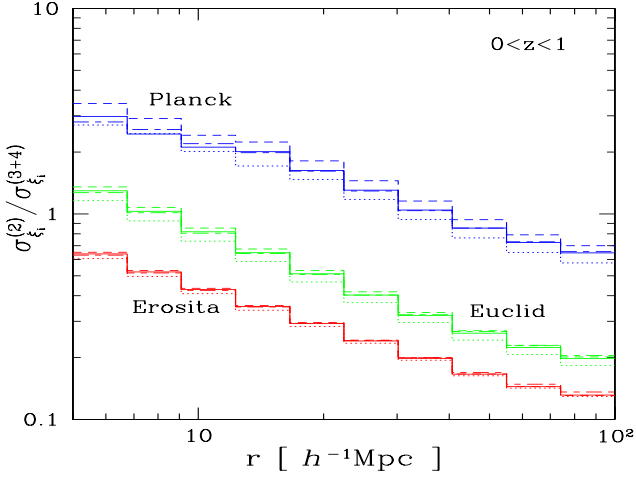}}
\end{center}
\caption{The ratio $\sigma_{\xi_i}^{(2)}/\sigma_{\xi_i}^{(3+4)}$ of the
rms contributions $\sqrt{C^{(2)}}$ and $\sqrt{C^{(3)}+C^{(4)}}$ of the 
covariance matrix of the estimator $\hxiLS_i$, as in
Fig.~\ref{fig_XiR_z0to1_sn_sv}. The line styles are as in
Fig.~\ref{fig_Nz_sn_sv_cosmo} and Table~\ref{Table_cosmo}.}
\label{fig_XiR_z0to1_sn_sv_cosmo}
\end{figure}

\begin{figure}
\begin{center}
\epsfxsize=8.5 cm \epsfysize=6.5 cm {\epsfbox{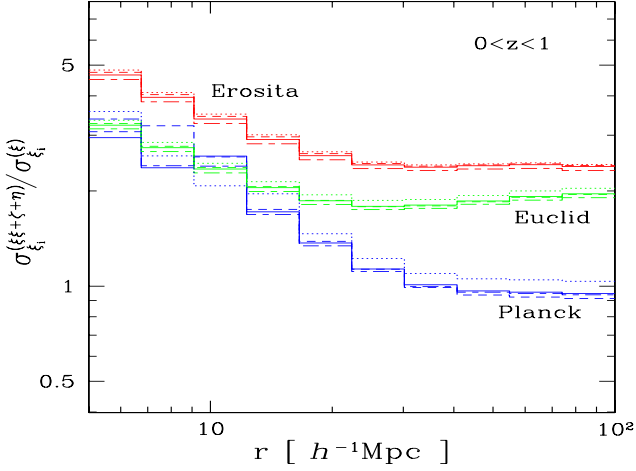}}
\end{center}
\caption{The ratio
$\sigma_{\xi_i}^{(\xi\xi+\zeta+\eta)}/\sigma_{\xi_i}^{(\xi)}$ 
of the rms high-order contribution (\ref{CLS-xixi-1})-(\ref{CLS-eta-1}) to the rms
low-order contribution (second term in Eq.(\ref{Cij-LS-tot})) of the sample variance 
of the correlation $\xi_i$, as in Fig.~\ref{fig_XiR_z0to1_sv}. The line styles are as in
Fig.~\ref{fig_Nz_sn_sv_cosmo} and Table~\ref{Table_cosmo}.}
\label{fig_XiR_z0to1_sv_cosmo}
\end{figure}

Thus, we compare in Figs.~\ref{fig_Nz_sn_sv_cosmo}, 
\ref{fig_XiR_z0to1_sn_sv_cosmo}, and \ref{fig_XiR_z0to1_sv_cosmo}, the
three curves obtained for these three alternative cosmologies with the curve that
was obtained in Sect.~\ref{Applications} for the fiducial WMAP7 cosmology.
To avoid overcrowding the figures we only consider the all-sky surveys,
Planck, Erosita, and Euclid.
We can see that the main features of these figures are not modified when
we consider these alternative cosmologies, so our results and conclusions are
not sensitive to the precise value of the cosmological parameters.
As expected, we can also check that shot-noise effects become less important, with
respect to sample-variance contributions, when $\sigma_8$ is increased.

\bibliographystyle{aa} % style aa.bst
\bibliography{ref1}

\end{document}